\documentclass[twocolumn]{aastex61}

\received{Oct 1, 2018}
\revised{Nov 1, 2018}
\accepted{Dec 1, 2018}
\submitjournal{ApJ}

\shorttitle{SN~2017eaw}
\shortauthors{Szalai et al.}

\usepackage{graphicx}

\begin{document}

\title{The Type II-P Supernova 2017eaw: from explosion to the nebular phase}

\correspondingauthor{Tam\'as Szalai}
\email{szaszi@titan.physx.u-szeged.hu}

\author[0000-0003-4610-1117]{Tam\'as Szalai}
\affiliation{Department of Optics and Quantum Electronics, University of Szeged, D\'om t\'er 9, Szeged, 6720 Hungary}
\affiliation{Konkoly Observatory, MTA CSFK, Konkoly-Thege M. \'ut 15-17, Budapest, 1121, Hungary}

\author{J{\'o}zsef Vink{\'o}}
\affiliation{Konkoly Observatory, MTA CSFK, Konkoly-Thege M. \'ut 15-17, Budapest, 1121, Hungary}
\affiliation{Department of Optics and Quantum Electronics, University of Szeged, D\'om t\'er 9, Szeged, 6720 Hungary}
\affiliation{Department of Astronomy, University of Texas at Austin, Austin, TX 78712, USA}

\author{R\'eka K\"onyves-T\'oth}
\affiliation{Konkoly Observatory, MTA CSFK, Konkoly-Thege M. \'ut 15-17, Budapest, 1121, Hungary}

\author{Andrea P. Nagy}
\affiliation{Department of Optics and Quantum Electronics, University of Szeged, D\'om t\'er 9, Szeged, 6720 Hungary}
\affiliation{Konkoly Observatory, MTA CSFK, Konkoly-Thege M. \'ut 15-17, Budapest, 1121, Hungary}

\author{K. Azalee Bostroem}
\affiliation{Department of Physics, University of California, Davis, CA 95616, USA}


\author{Kriszti\'an S\'arneczky}
\affiliation{Konkoly Observatory, MTA CSFK, Konkoly-Thege M. \'ut 15-17, Budapest, 1121, Hungary}

\author{Peter J. Brown}
\affiliation{George P. and Cynthia Woods Mitchell Institute for Fundamental Physics and Astronomy, Texas A. \& M. University, Department of Physics and Astronomy, 4242 TAMU, College Station, TX 77843, USA}

\author{Ondrej Pejcha}
\affiliation{Institute of Theoretical Physics, Faculty of Mathematics and Physics, Charles University in Prague, Czech Republic}

\nocollaboration

\author{Attila B\'odi}
\affiliation{Konkoly Observatory, MTA CSFK, Konkoly-Thege M. \'ut 15-17, Budapest, 1121, Hungary}

\author{Borb\'ala Cseh}
\affiliation{Konkoly Observatory, MTA CSFK, Konkoly-Thege M. \'ut 15-17, Budapest, 1121, Hungary}

\author{G\'eza Cs\"ornyei}
\affiliation{Konkoly Observatory, MTA CSFK, Konkoly-Thege M. \'ut 15-17, Budapest, 1121, Hungary}

\author{Zolt\'an Dencs}
\affiliation{Konkoly Observatory, MTA CSFK, Konkoly-Thege M. \'ut 15-17, Budapest, 1121, Hungary}

\author{Ott\'o Hanyecz}
\affiliation{Konkoly Observatory, MTA CSFK, Konkoly-Thege M. \'ut 15-17, Budapest, 1121, Hungary}

\author{Bernadett Ign\'acz}
\affiliation{Konkoly Observatory, MTA CSFK, Konkoly-Thege M. \'ut 15-17, Budapest, 1121, Hungary}

\author{Csilla Kalup}
\affiliation{Konkoly Observatory, MTA CSFK, Konkoly-Thege M. \'ut 15-17, Budapest, 1121, Hungary}

\author{Levente Kriskovics}
\affiliation{Konkoly Observatory, MTA CSFK, Konkoly-Thege M. \'ut 15-17, Budapest, 1121, Hungary}

\author{Andr\'as Ordasi}
\affiliation{Konkoly Observatory, MTA CSFK, Konkoly-Thege M. \'ut 15-17, Budapest, 1121, Hungary}

\author{Andr\'as P\'al}
\affiliation{Konkoly Observatory, MTA CSFK, Konkoly-Thege M. \'ut 15-17, Budapest, 1121, Hungary}

\author{B\'alint Seli}
\affiliation{Konkoly Observatory, MTA CSFK, Konkoly-Thege M. \'ut 15-17, Budapest, 1121, Hungary}

\author{\'Ad\'am S\'odor}
\affiliation{Konkoly Observatory, MTA CSFK, Konkoly-Thege M. \'ut 15-17, Budapest, 1121, Hungary}

\author{R\'obert Szak\'ats}
\affiliation{Konkoly Observatory, MTA CSFK, Konkoly-Thege M. \'ut 15-17, Budapest, 1121, Hungary}

\author{Kriszti\'an Vida}
\affiliation{Konkoly Observatory, MTA CSFK, Konkoly-Thege M. \'ut 15-17, Budapest, 1121, Hungary}

\author{Gabriella Zsidi}
\affiliation{Konkoly Observatory, MTA CSFK, Konkoly-Thege M. \'ut 15-17, Budapest, 1121, Hungary}

\collaboration{Konkoly team}

\author{Iair Arcavi}
\affiliation{The School of Physics and Astronomy, Tel Aviv University, Tel Aviv 69978, Israel}

\author{Chris Ashall}
\affiliation{Astrophysics Research Institute, Liverpool John Moores University, IC2, Liverpool Science Park, 146 Brownlow Hill, Liverpool L3 5RF, UK}
\affiliation{Department of Physics, Florida State University, 77 Chieftan Way, Tallahassee, FL 32306, USA}

\author{Jamison Burke}
\affiliation{Las Cumbres Observatory, 6740 Cortona Drive, Suite 102, Goleta, CA 93117-5575, USA}
\affiliation{Department of Physics, University of California, Santa Barbara, CA 93106-9530, USA}

\author{Llu\'is Galbany}
\affiliation{PITT PACC, Department of Physics and Astronomy, University of Pittsburgh, Pittsburgh, PA 15260, USA}

\author{Daichi Hiramatsu}
\affiliation{Las Cumbres Observatory, 6740 Cortona Drive, Suite 102, Goleta, CA 93117-5575, USA}
\affiliation{Department of Physics, University of California, Santa Barbara, CA 93106-9530, USA}

\author{Griffin Hosseinzadeh}
\affiliation{Center for Astrophysics $\mid$ {Harvard \& Smithsonian, 60 Garden St., Cambridge, MA 02138, USA}}
\affiliation{Las Cumbres Observatory, 6740 Cortona Drive, Suite 102, Goleta, CA 93117-5575, USA}
\affiliation{Department of Physics, University of California, Santa Barbara, CA 93106-9530, USA}

\author{Eric Y. Hsiao}
\affiliation{Department of Physics, Florida State University, 77 Chieftan Way, Tallahassee, FL 32306, USA}

\author{D. Andrew Howell}
\affiliation{Las Cumbres Observatory, 6740 Cortona Drive, Suite 102, Goleta, CA 93117-5575, USA}
\affiliation{Department of Physics, University of California, Santa Barbara, CA 93106-9530, USA}

\author{Curtis McCully}
\affiliation{Las Cumbres Observatory, 6740 Cortona Drive, Suite 102, Goleta, CA 93117-5575, USA}
\affiliation{Department of Physics, University of California, Santa Barbara, CA 93106-9530, USA}

\author{Shane Moran}
\affiliation{Tuorla Observatory, Department of Physics and Astronomy, University of Turku, V\"ais\"al\"antie 20, FI-21500 Piikki\"o, Finland}

\author{Jeonghee Rho}
\affiliation{SETI Institute, 189 N. Bernardo Ave, Suite 200, Mountain View, CA 94043, USA}

\author{David J. Sand}
\affiliation{Department of Astronomy/Steward Observatory, 933 North Cherry Avenue, Room N204, Tucson, AZ 85721-0065, USA}

\author{Melissa Shahbandeh}
\affiliation{Department of Physics, Florida State University, 77 Chieftan Way, Tallahassee, FL 32306, USA}

\author{Stefano Valenti}
\affiliation{Department of Physics, University of California, Davis, CA 95616, USA}

\author{Xiaofeng Wang}
\affiliation{Physics Department and Tsinghua Center for Astrophysics (THCA), Tsinghua University, Beijing, 100084, China}

\author{J.\ Craig Wheeler}
\affiliation{Department of Astronomy, University of Texas at Austin, Austin, Texas}

\collaboration{Global Supernova Project}

\begin{abstract}

The nearby SN~2017eaw is a Type II-P (``plateau'') supernova showing early-time, moderate CSM interaction.  We present a comprehensive study of this SN including the analysis of high-quality optical photometry and spectroscopy covering the very early epochs up to the nebular phase, as well as near-UV and near-infrared spectra, and early-time X-ray and radio data. The combined data of SNe 2017eaw and 2004et allow us to get an improved distance to the host galaxy, NGC~6946, as $D \sim 6.85$~$\pm 0.63$ Mpc; this fits in recent independent results on the distance of the host and disfavors the previously derived (30 \% shorter) distances based on SN 2004et. From modeling the nebular spectra and the quasi-bolometric light curve, we estimate the progenitor mass and some basic physical parameters for the explosion and the ejecta. Our results agree well with previous reports on a RSG progenitor star with a mass of $\sim15-16$ M$_\odot$.
Our estimation on the pre-explosion mass-loss rate ($\dot{M} \sim3 \times 10^{-7} -$~$1\times 10^{-6} M_{\odot}$ yr$^{-1}$) agrees well with previous results based on the opacity of the dust shell enshrouding the progenitor, but it is orders of magnitude lower than previous estimates based on general light-curve modeling of Type II-P SNe. Combining late-time optical and mid-infrared data, a clear excess at 4.5 $\mu$m can be seen, supporting the previous statements on the (moderate) dust formation in the vicinity of SN 2017eaw.

\end{abstract}


\section{Introduction} \label{sec:intro}

%
Recently, the growing number of well-observed (i.e. having high signal-to-noise, high cadence data spanning a wide wavelength range) Type II supernovae (SNe) revealed important new details about their progenitors, explosion mechanisms and diversity \citep[see e.g.][and references therein]{Valenti16}. For example, photometry and spectroscopy taken at the earliest phases, during and after shock breakout, turned out to be especially useful for constraining the progenitor radii and/or probing the nearby circumstellar matter \citep{Garnavich16, Khazov16}.  

SN~2017eaw was discovered by P. Wiggins on UT 2017 May 14.238 at a brightness of 12.8 mag \citep{Wiggins17}. Within a few hours the presence of the new transient was confirmed by \citet{Dong17} based on images taken by the Las Cumbres Observatory (LCO) 1m telescope at McDonald Observatory, Texas. The object was classified first as a young Type II SN \citep{Cheng17,Tomasella17}, while, soon after, \citet{Xiang17} found that early spectra of SN~2017eaw matches well with that of young Type II-P explosions; later, the classification has been confirmed by photometry.
 
SN~2017eaw appeared in NGC~6946, at 61.0$''$ west and 143.0$''$ north from the center of the galaxy. The host is a nearby, face-on spiral galaxy, which has produced around a dozen known SNe and other luminous transients including the Type II-L SN 1980K, Type II-P SNe 1948B, 2002hh and 2004et, and the SN impostor 2008S.
The first precise astrometric position of SN~2017eaw, based on ground-based imaging, has been given by \citet{S17}: $\alpha$ = 20$^h$34$^m$44$^s$.238, $\delta$ = $+$60$^d$11$'$36$\arcsec$.00 (with RMS uncertainties of $\Delta \alpha$ = 0.08\arcsec and $\Delta \delta$ = 0.09\arcsec). Later, \citet{Kilpatrick18} determined a very similar, even more precise position of the object on post-explosion Hubble Space Telescope (HST) images: $\alpha$ = 20$^h$34$^m$44$^s$.272, $\delta$ = $+$60$^d$11$'$36$\arcsec$.008 (with an uncertainty of $\sim$0.002-0.003\arcsec in both $\alpha$ and $\delta$).
 
Since SN~2017eaw is one of the closest CC SNe to date and it appeared in a host galaxy that is being monitored almost continuously, the search for the potential progenitor in archival imaging data was started right after the announcement of discovery. The first hints of a possible progenitor were reported by \citet[][based on mid-infrared Spitzer Space Telescope images]{Khan17}, \citet[][based on optical Catalina Sky Survey images]{Drake17}, and \citet[][using HST ACS/WFC F814W images]{VD17}; all of these findings suggested the presence of a red supergiant (RSG) star at the position of the SN. The detailed analysis by \citet{Kilpatrick18}, based on archival HST and {\it Spitzer} images, confirmed the RSG progenitor ($log(L/L_{\odot}$) = 4.9, $T_{\text{eff}}$ = 3350K, M$_{\text{ini}}$ = 13 M$_{\odot}$) obscured by a dust shell; these results agree well with that of \citet{Rui19} based on a similar analysis. In the frameworks of comprehensive studies, two other groups also published their findings on the progenitor candidate of SN~2017eaw. \citet{Williams18} carried out an age-dating study of the surrounding stellar populations of nearby (historic) CC SNe on new HST images; based on that they derived a somewhat smaller mass for the assumed progenitor of SN~2017eaw ($\sim$9 M$_{\odot}$). \citet{Johnson18} presented the results of a long-term multi-channel optical monitoring of the progenitors of Type II SNe; they found a general brightness variability smaller than 5-10$\%$, which is in agreement with the known properties of RSG stars.
 
SN~2017eaw has been the target of several multi-wavelength observing campaigns since its early phase; however, only a few datasets have been published to date. \citet{Tsvetkov18} presented the results of their BVRI photometric campaign covering the first $\sim$200 days. \citet{Rho18} obtained and analyzed near-infrared (near-IR) spectra spanning the time interval 22$-$205 days after discovery, while \citet{Kilpatrick18} published only a single optical spectrum taken during the photospheric phase. Radio and X-ray (non)detections have been also reported (see Section \ref{csm}).

In this paper we present a comprehensive study on the early- and late-time properties of SN~2017eaw. First, we present our ground-based spectroscopic and photometric observations in Section \ref{sec:obs}. After that we give the details on the comparison of the light curves and spectra with those of other SNe~II-P, the extraction of physical parameters from bolometric light curve modeling, and the estimation of the distance of the host galaxy based on the combined data of SNe~2017eaw and 2004et. We also present our findings from the analysis of early-time radio and X-ray (non)detections and of late-time mid-infrared data of SN~2017eaw, and interpret these as potential signs of circumstellar interaction and dust formation in the vicinity of the explosion site, respectively.
In Section \ref{disc}, we discuss our results and, finally, in Section \ref{sec:concl}, we present our conclusions.

\section{Observations} \label{sec:obs}

This section contains the description of the  observational data on SN~2017eaw collected with various ground- and space-based instruments. All data will be publicly released via WIseREP\footnote{https://wiserep.weizmann.ac.il}

\subsection{Photometry}\label{obs_phot}

\begin{figure*}
\begin{center}
\leavevmode
\includegraphics[width=.8\textwidth]{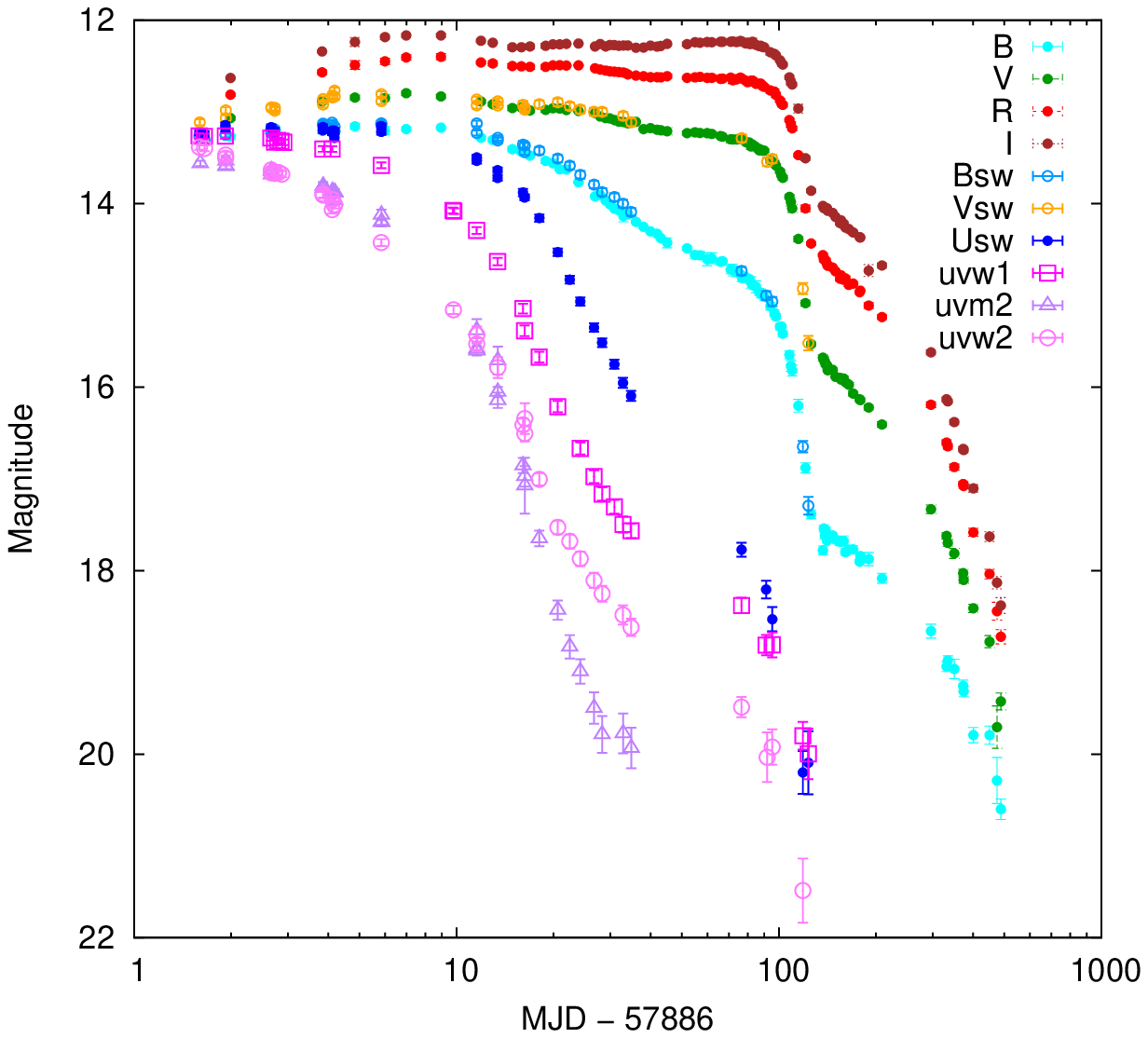}
\includegraphics[width=.45\textwidth]{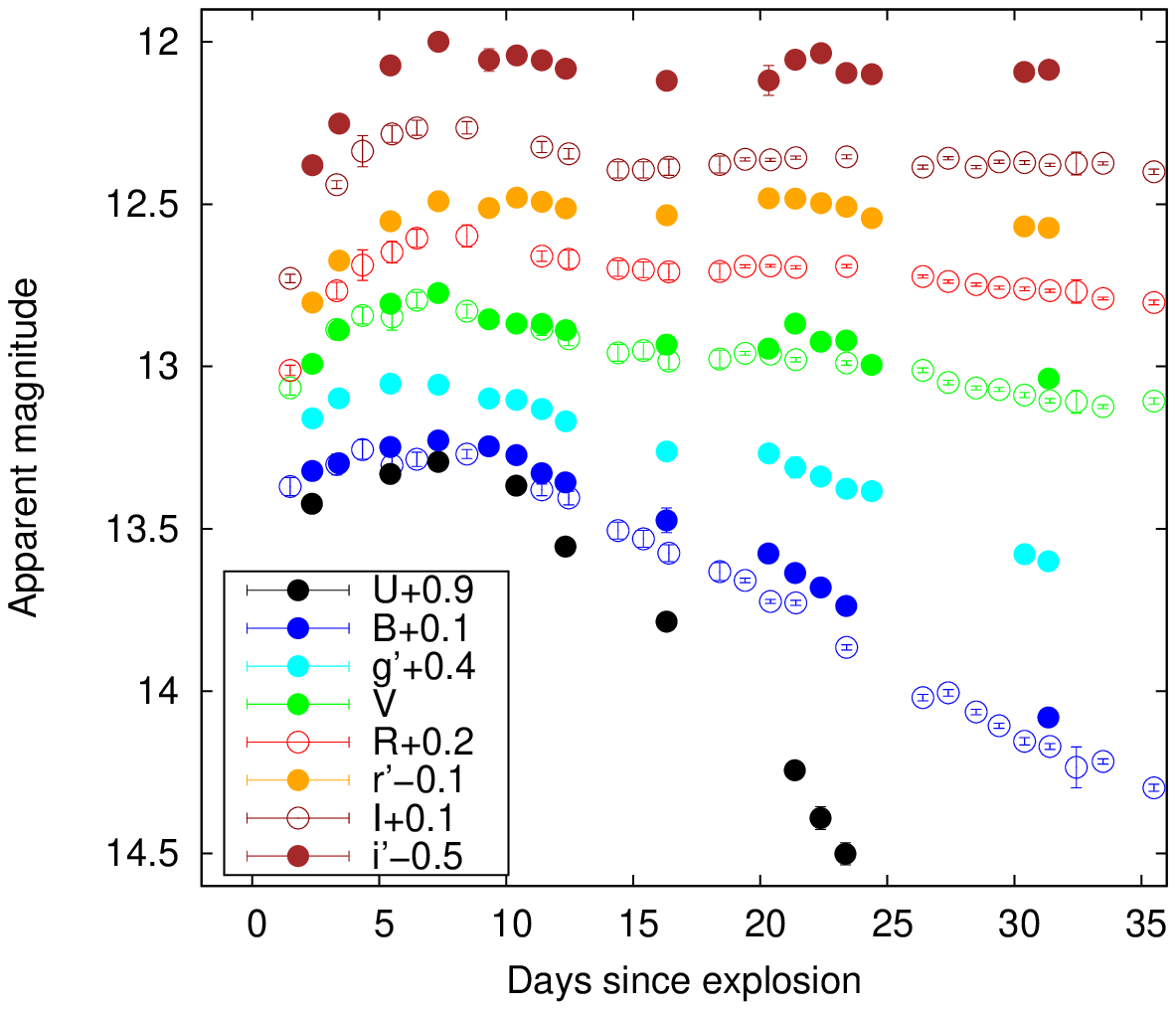} \hspace{5mm}
\includegraphics[width=.45\textwidth]{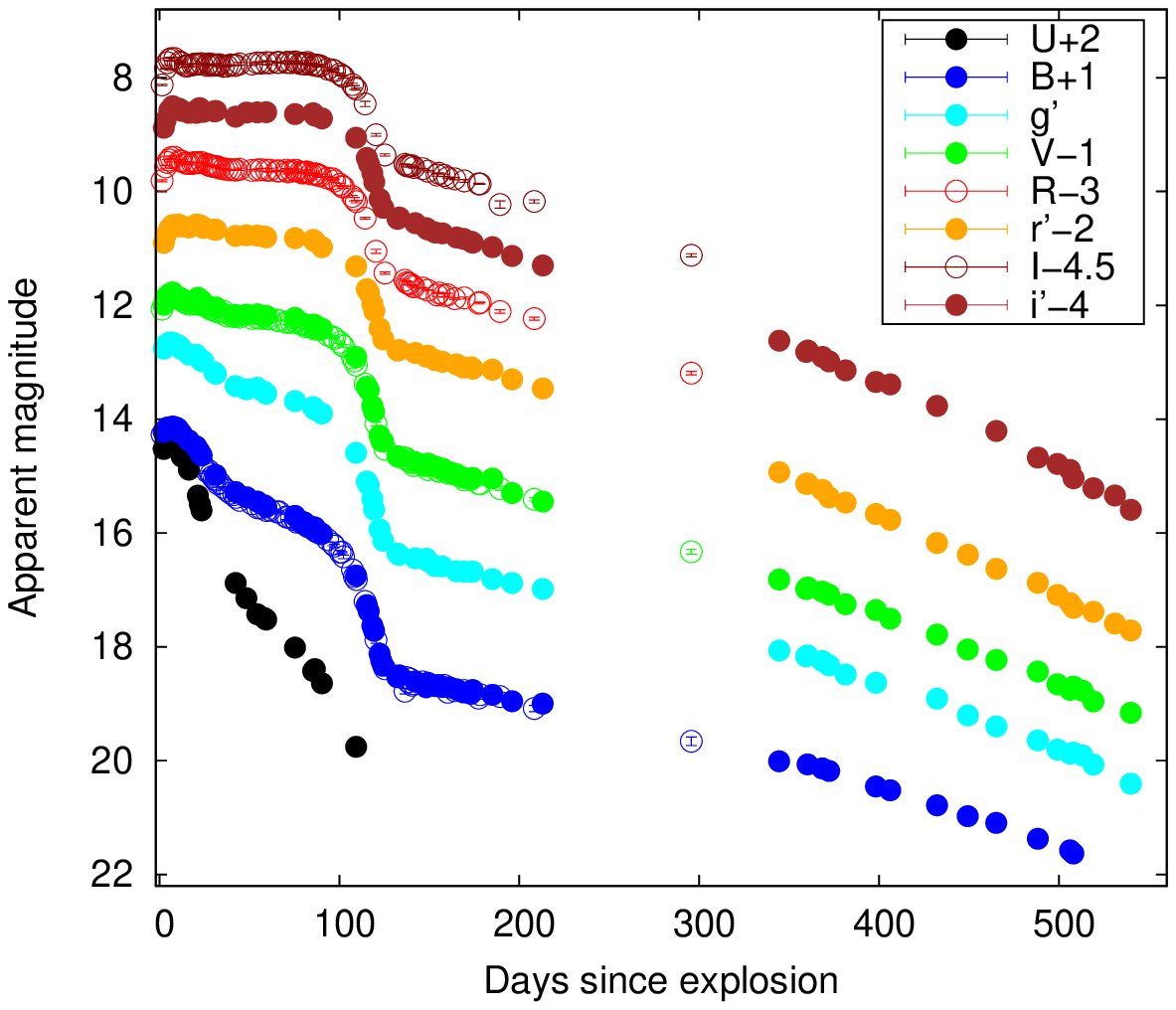}
\end{center}
\caption{Top panel: Multi-colour light curves of SN~2017eaw from Konkoly and {\it Swift}/UVOT. The {\it Swift} UV data have been shifted by +1 magnitude for better visibility. 
Bottom panel: The Konkoly BVRI (open circles) and LCO UBV$g'r'i'$ (filled circles) photometric data during the first month (bottom left) and up to +545 days (bottom right panel).}
\label{fig:photall}
\end{figure*}


Ground-based photometric observations for SN 2017eaw were obtained from the Piszk\'{e}stet\H{o} Mountain Station of Konkoly Observatory, Hungary. We used the 0.6/0.9 m Schmidt-telescope with the attached liquid-cooled FLI Proline PL16801 4096 $\times$ 4096 CCD (FoV 70 $\times$ 70 arcmin$^2$) equipped with Bessell BVRI filters. The CCD frames were bias-, dark- and flatfield-corrected by applying standard IRAF\footnote{IRAF is distributed by the National Optical Astronomy Observatories, which are operated by the Association of Universities for Research in Astronomy, Inc., under cooperative agreement with the National Science Foundation.} routines.
To obtain the Konkoly {\it BVRI} magnitudes, we carried out PSF photometry on the SN and 5 local comparison (tertiary standard) stars using the {\tt allstar} task in IRAF. We applied an aperture radius of {6}\arcsec and a background annulus from 7\arcsec to {10}\arcsec for SN~2017eaw as well as for the local comparison stars.
The magnitudes of the local comparison stars were determined from their PS1-photometry after transforming the PS1 gri magnitudes to the Johnson-Cousins BVRI system.

Long-term photometric data were collected as part of the Global Supernova Project by the Las Cumbres Observatory (LCO). Using {\tt lcogtsnpipe} \citep{Valenti16}, a PyRAF-based photometric reduction pipeline, we measured PSF photometry of the SN. Because the SN is well-separated from its host galaxy, image subtraction is not required. Local sequence stars were calibrated to $g'r'i'$ AB magnitudes from the APASS catalog \citep{Henden16} and to standard fields (e.g. L113) observed on the same night at the same observatory site using UBV magnitudes from \citet{Landolt92}. 

The ground-based optical observations were supplemented by the available Neil Gehrels Swift Observatory \citep[hereafter {\it Swift},][]{gehrels04, burrows05} data taken with the Ultraviolet-Optical Telescope (UVOT) \citep{roming05} and reduced using standard HEAsoft tasks. Individual frames were summed with the {\it uvotimsum} task. Magnitudes were determined via aperture photometry using the task {\it uvotsource} and adopting the most recent zero points \citep{breeveld11}. 

The results of our LCO UBV$g'r'i'$ (in Vega-magnitudes for UBV and AB-magnitudes for $g'r'i'$ bands), Konkoly BVRI, and {\it Swift} photometry (both in Vega-magnitudes) are shown in Fig. \ref{fig:photall}; the data are also presented in Tables \ref{tab:phot}, \ref{tab:lcophot}, and \ref{tab:swphot}, respectively. Intrinsic photometric errors are typically below 0.05 mag, while the overlapping photometric datasets -- LCO/Konkoly/{\it Swift} BV magnitudes, as well as our BVRI data and that of \citet{Tsvetkov18} -- are generally consistent within $\sim$0.1 mag.

\subsection{Spectroscopy}\label{obs_sp}


\begin{figure*}
\begin{center}
\leavevmode
\includegraphics[width=.8\textwidth]{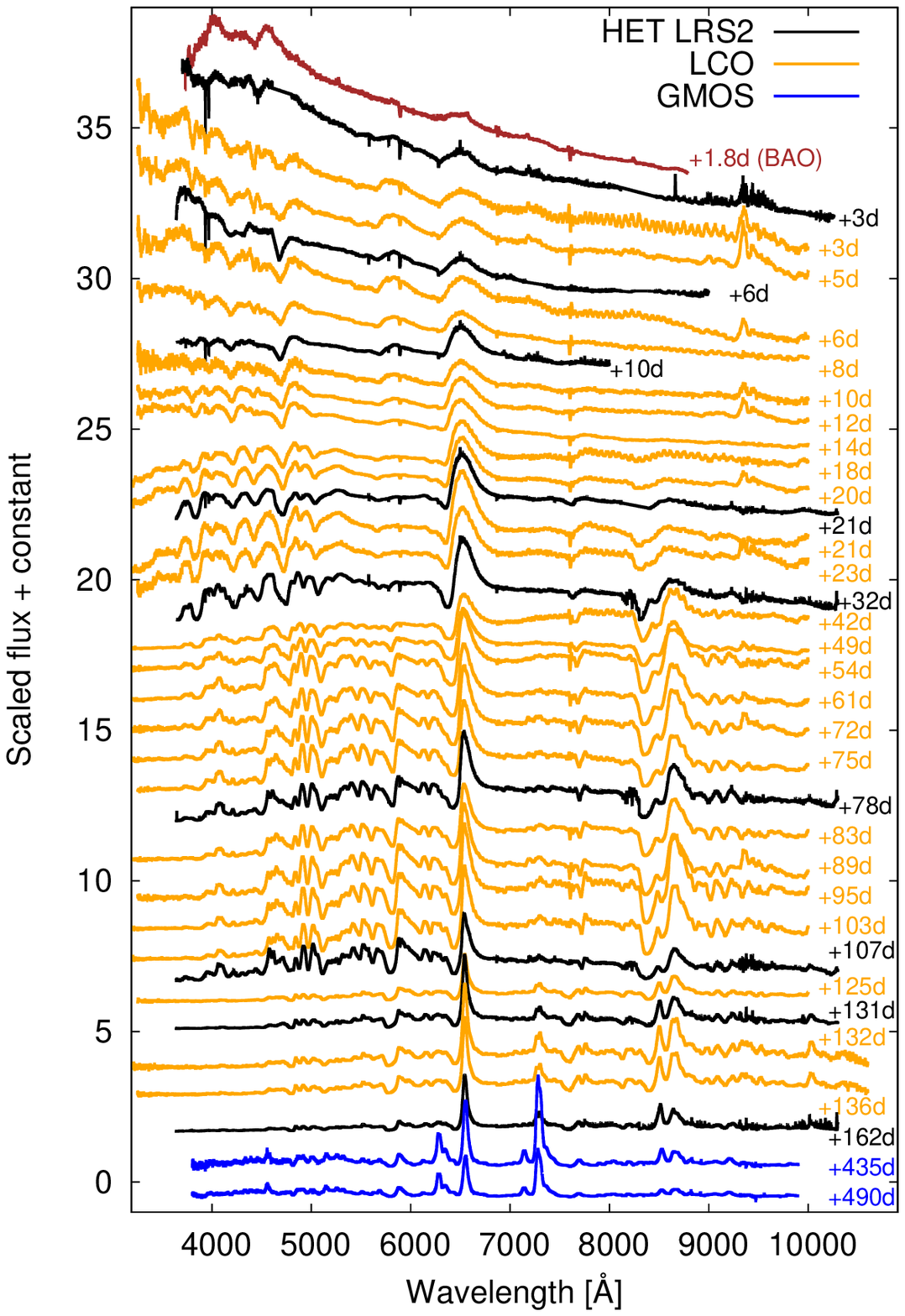}
\end{center}
\caption{Optical spectra of SN~2017eaw obtained with HET LRS2 (black) and Gemini-North GMOS-N (blue), and from LCO sites (orange). An additional public spectrum\footnote{downloaded from WiSERep {\tt https://wiserep.weizmann.ac.il}} obtained at May 14 from the Beijing Astronomical Observatory (BAO) is also shown (brown).}
\label{fig:hetsp}
\end{figure*}

A number of low-resolution optical spectra (R$\sim$400$-$700, in the 370$-$1050 nm range) were collected at LCO sites using the FLOYDS instruments.
Additionally, a sequence of optical spectra on SN 2017eaw was taken with the Low Resolution Spectrograph 2 (LRS2) mounted on the 10m {\sl Hobby-Eberly Telescope}. LRS2 consists of two dual-arm spectrographs covering the 370$-$470 and 460$-$700 nm region (LRS2-B UV- and Orange arm) and the 650$-$842 and 818$-$1050 nm region (LRS2-R Red- and IR-arm), respectively, with an average spectral resolution of about 1500 \citep{2016SPIE.9908E..4CC}. Both arms are fiber-fed by their own Integral Field Unit (IFU) having 280 fibers packed densely to fully cover a $12 \times 6$ arcsec$^2$ field of view. For further details on the instrument and the data reduction see e.g. \citet{Li18}. 

{\it Swift} took two near-ultraviolet (near-UV) spectra with {\it UVOT/UGRISM}
covering the 200$-$500 nm regime. These data were downloaded from the {\it Swift Data Archive}\footnote{\tt https://swift.gsfc.nasa.gov/archive/} and extracted using the HEAsoft task {\it uvotimgrism}. 

Furthermore, five near-IR spectra were taken with the 3.0-m NASA Infrared Telescope Facility (IRTF) and the SpeX spectrograph \citep{Rayner03}. The data were taken in ``SXD" mode, with wavelength coverage from $\sim$0.8-2.4 $\mu$m, cross-dispersed into six orders. The observations were taken using the classic ABBA technique for improved sky-subtraction, and an A0V star was observed for telluric correction. For further details of the observational setup and execution, see \citet{Hsiao19}. The data were reduced using the publicly available {\sc Spextool} software package \citep{Cushing04}, and the telluric corrections were performed with the {\sc XTELLCOR} software suite \citep{Vacca03}.

Optical spectra were obtained during the nebular phase on day 220, 435, and 491.
The first spectrum was taken on 2017-12-19.23 UT with the Low Resolution Imaging Spectrometer (LRIS) \citep{1995oke,1998mccarthy,2010rockosi} at the 10m W. M. Keck Observatory (2017B, project code U109, PI: Valenti).
The spectrum was taken using a 1\arcsec aperture with the 560 dichroic to split the beam between the 600/4000 grism on the blue side and the 400/8500 grating on the red side.
Taken together, the merged spectrum spans $\sim$3200$-$10,200\AA.
Data were reduced in a standard way using the {\tt LPIPE} pipeline \footnote{http://www.astro.caltech.edu/ dperley/programs/lpipe.html}. 
The remaining spectra were observed on 2018-07-22.52 UT and 2018-09-16.31 UT with the Gemini Multi-Object Spectrograph (GMOS) \citep{2004hook,2016gimeno} mounted on the 8m Frederick C. Gillett Gemini North telescope (program ID: GN-2018B-Q-204, PI: Bostroem).
Observations were taken using a 1\arcsec aperture utilizing a red setup and a blue setup to obtain wavelength coverage from 3450$-$9900\AA.
The red setup observations were taken with the R400 grating and the OG515 blocking filter with a resolution of R$\sim$1918. 
The blue setup observations were taken with the B600 grating with a resolution of R$\sim$1688.
The spectra were reduced using a combination of the {\tt Gemini iraf} package and custom Python scripts \footnote{https://github.com/cmccully/lcogtgemini}. 
Extracted spectra were scaled to photometry interpolated or extrapolated to the date of observation.

The journal of all spectroscopic observations is given in Table \ref{tab:spec}. The sequence of optical spectra is plotted in Fig \ref{fig:hetsp}, while the detailed analysis of all optical, near-UV and near-IR spectra are presented in Section \ref{sec:ana}.

\section{Analysis and results} \label{sec:ana}

\begin{table*}
\begin{center}
\caption{Basic data of Type II-P SNe used for comparison}
\label{tab:sne}
\begin{tabular}{ccccccc}
\hline
\hline
Name & Host galaxy & Date of explosion & $z$ & $D$ (Mpc) & $E(B-V)_{\text{tot}}$ & Source \\
~ & ~ & (JD$-$2,400,000) & ~ & ~ & (mag) & ~ \\
\hline
SN 2017eaw & NGC~6946 & {\bf 57886.5 $\pm$ 1.0} & 0.00013 & {\bf 6.85 $\pm$ 0.63} & {\bf 0.41} & 1, 2 \\
SN~2004et & NGC~6946 & 53270.5 $\pm$ 1.0 & 0.00013 & {\bf 6.85 $\pm$ 0.63} & 0.41 & 1, 2 \\
SN~2012aw & NGC~3351 & 56002.6 & 0.00260 &  9.9 $\pm$ 0.1 & 0.07 & 3 \\
SN~2013fs & NGC~7610 & 56571.1 & 0.01190 & 51.0 $\pm$ 3.0 & 0.05 & 4 \\
SN~2016X & UGC~08041 & 57405.9 & 0.00441 & 15.2 $\pm$ 2.0 & 0.04 & 5 \\
SN~2016bkv & NGC~3184 & 57467.5 & 0.00198 & 14.4 $\pm$ 0.3 & 0.01 & 6, 7 \\
\hline
\end{tabular}
\end{center}
\smallskip
{\bf Notes.} Parameters marked with boldface have been determined in this work. References: 1) This work; 2) \citet{Maguire10}; 3) \citet{Bose13}; 4) \citet{Yaron17}; 5) \citet{Huang18}; 6) \citet{Hosseinzadeh18}; 7) \citet{Nakaoka18}. Redshifts are adopted from NASA/IPAC Extragalactic Database (NED, https://ned.ipac.caltech.edu).
\end{table*}

First, we estimate some basic parameters of SN 2017eaw: the moment of explosion ($t_0$), the interstellar extinction toward the SN, and the distance to the host galaxy. 

We adopt $t_0$ = 2,457,886.5$\pm 1.0$ JD (May 13.0$\pm$1.0 UT) as the moment of explosion of SN~2017eaw. This value is strengthened by our distance estimation analysis (see Section \ref{epm}), and suits well to both the date of discovery (May 14.2, 2017) and the epoch of last non-detection (May 12.2, 2017).

Finding the true value of the total extinction in the line-of-sight of SN~2017eaw does not seem to be trivial. Using the reddening map of \citet{SF11}, we get $E(B-V)_{\text{gal}}$=0.30 mag for the Galactic extinction.
For the total extinction, several estimates based on empirical relations between the total reddening and equivalent widths (EWs) of \ion{Na}{1} D lines exist in the literature: for example, \citet{Tomasella17} derived $E(B-V)_{\text{tot}}$=0.22 mag for the total (Galactic+host) extinction using the formulae by \citet{Turatto03}, which is lower than the Galactic component given above, while \citet{Kilpatrick18} determined $E(B-V)_{\text{tot}}$=0.34 mag following the method by \citet{Poznanski12}. 

The $\sim 0.1$ mag difference between these two estimates illustrate the issue that these empirical relations may suffer from relatively high systematic errors \citep[see e.g.][]{Blondin09,Poznanski11,Faran14}. This belief is confirmed by our own analysis. 
Based on our HET spectra, we also determined the EWs of \ion{Na}{1} D1 and D2 features as well as that of the combined line profile (D1+D2) as 0.8, 1.1, and 1.7\AA, respectively.
Because of the very low redshift of SN~2017eaw, the \ion{Na}{1} D doublet at 5890$-$5895\AA\ originating from the Milky Way may be blended with the same features formed in the interstellar medium (ISM) of the host galaxy (and maybe in the CSM around the SN site). In any case, such high EW values would imply $E(B-V) >1$ mag according to the empirical relations 
given by \citet{Poznanski12}. Since the EW(\ion{Na}{1})$-E(B-V)$ relation is suspected to saturate at $E(B-V) \gtrsim 0.2$ mag, these measurements probably overestimate the total reddening toward SN~2017eaw.  

\begin{figure}
\begin{center}
\leavevmode
\includegraphics[width=.45\textwidth]{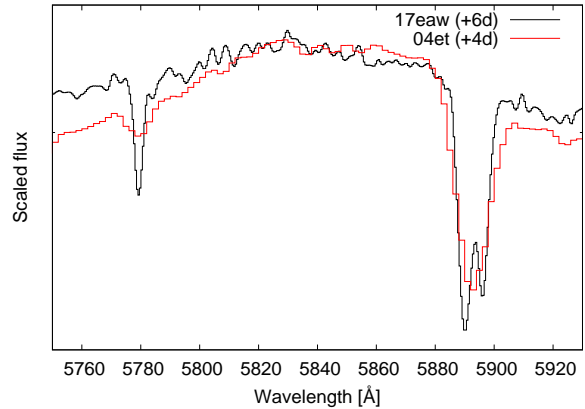}
\end{center}
\caption{Region of DIB 5780\AA\ and \ion{Na}{1} D1 \& D2 lines on both the +6d HET spectrum of SN~2017eaw and the +4d spectrum of SN~2004et (see the text regarding the estimation of total extinction in Section \ref{sec:ana}.)}
\label{fig:sp_ew}
\end{figure}

Diffuse Interstellar Band (DIB) features offer an independent and sometimes more reliable way to estimate the interstellar reddening. 
In the same HET spectrum as above we measured the EW of the unresolved blend of Galactic and host DIB 5780\AA\ feature and got $\sim 0.31$\AA. Repeating the same measurement but using a public spectrum of SN~2004et, a Type II-P SN that occurred in the same host galaxy, resulted in EW(DIB) $\sim 0.19$\AA\ (see Fig. \ref{fig:sp_ew}). These values correspond to $E(B-V)$ $\sim$0.52 and $\sim$0.32 mag, respectively, following \citet{Phillips13}, who applied the
method by \citet{Friedman11}.

For SN~2004et, \citet{Zwitter04} determined $E(B-V)_{\text{tot}}$=0.41 mag based on the method of \citet{MZ97}, which was also adopted by \citet{Maguire10}. Since the optical spectra of SNe 2017eaw and 2004et appear to be very similar (see Section \ref{spec}), including Na D profiles, and this $E(B-V)$ value is close to the mean of the results from the various estimates detailed above, in the rest of this paper we adopt and use $E(B-V)_{\text{tot}}$=0.41 mag as the total reddening toward SN~2017eaw, but note that the uncertainty of this value is at least $\pm 0.1$ mag as explained above.


Similarly, we use $D$ = 6.85 $\pm$ 0.63 Mpc for the distance of the host galaxy that comes from our own detailed analysis using various methods and the combination of other recently published distances to NGC~6946 (see Section \ref{epm}). 
 
In the followings we present a detailed photometric and spectroscopic study of SN~2017eaw, comparing the results with those of several other Type II-P SNe (see Table \ref{tab:sne}). All the fluxes were dereddened using the Galactic reddening law parametrized by \citet{FM07} assuming $R_{\text{V}}$ = 3.1.

\subsection{Photometric comparison}\label{lc}

\begin{figure*}
\begin{center}
\leavevmode
\includegraphics[width=.45\textwidth]{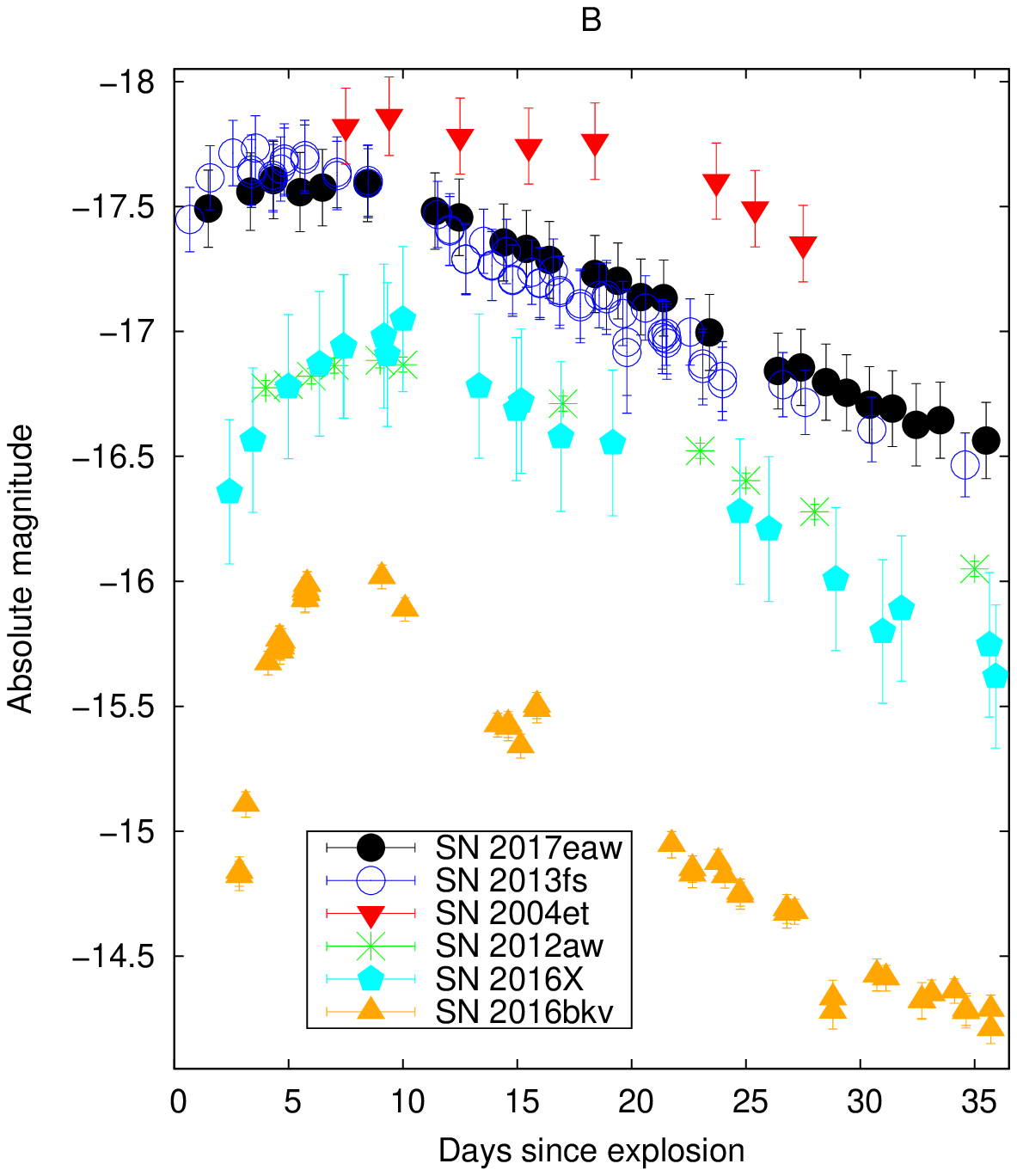} \hspace{5mm}
\includegraphics[width=.45\textwidth]{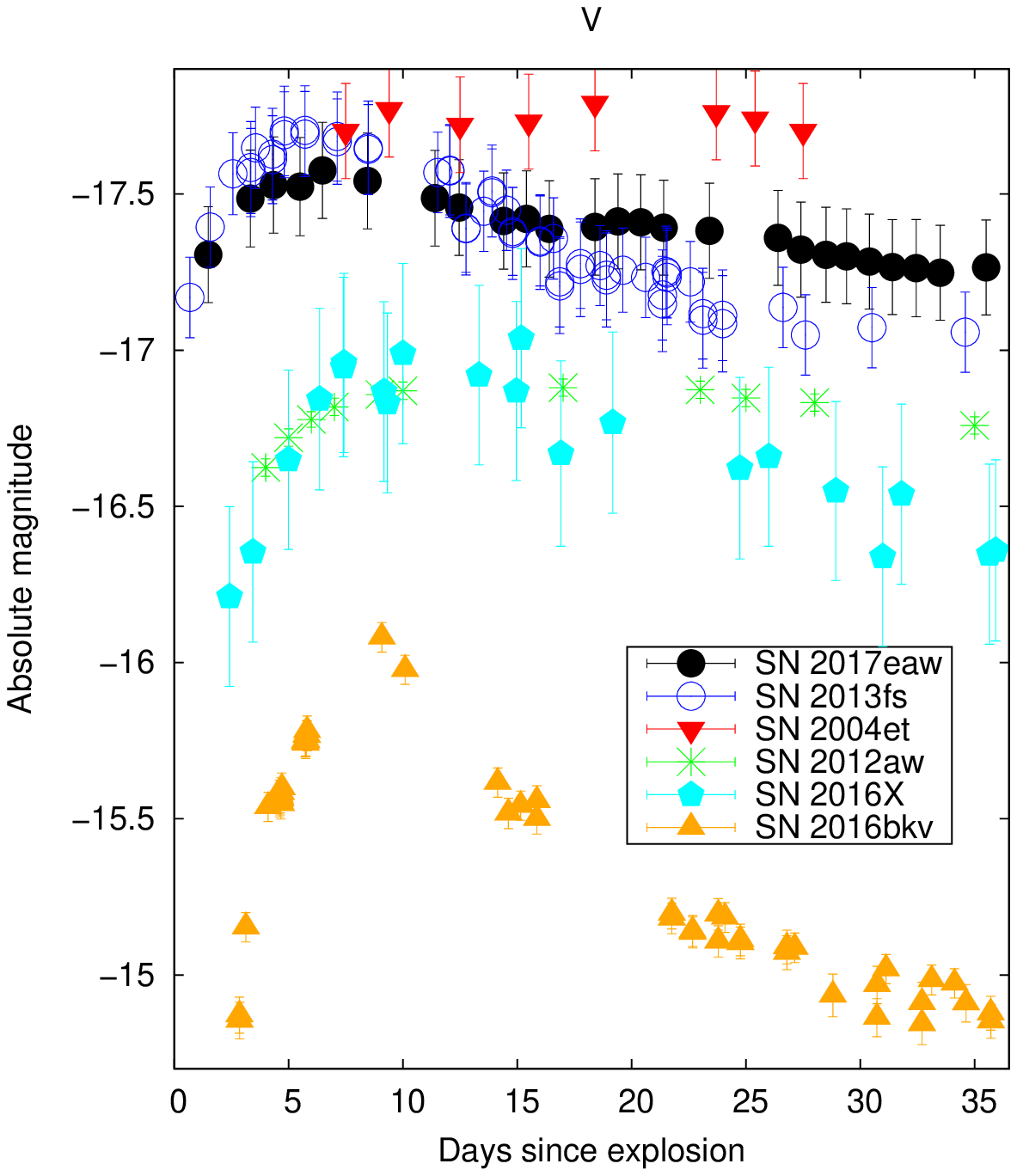}
\includegraphics[width=.45\textwidth]{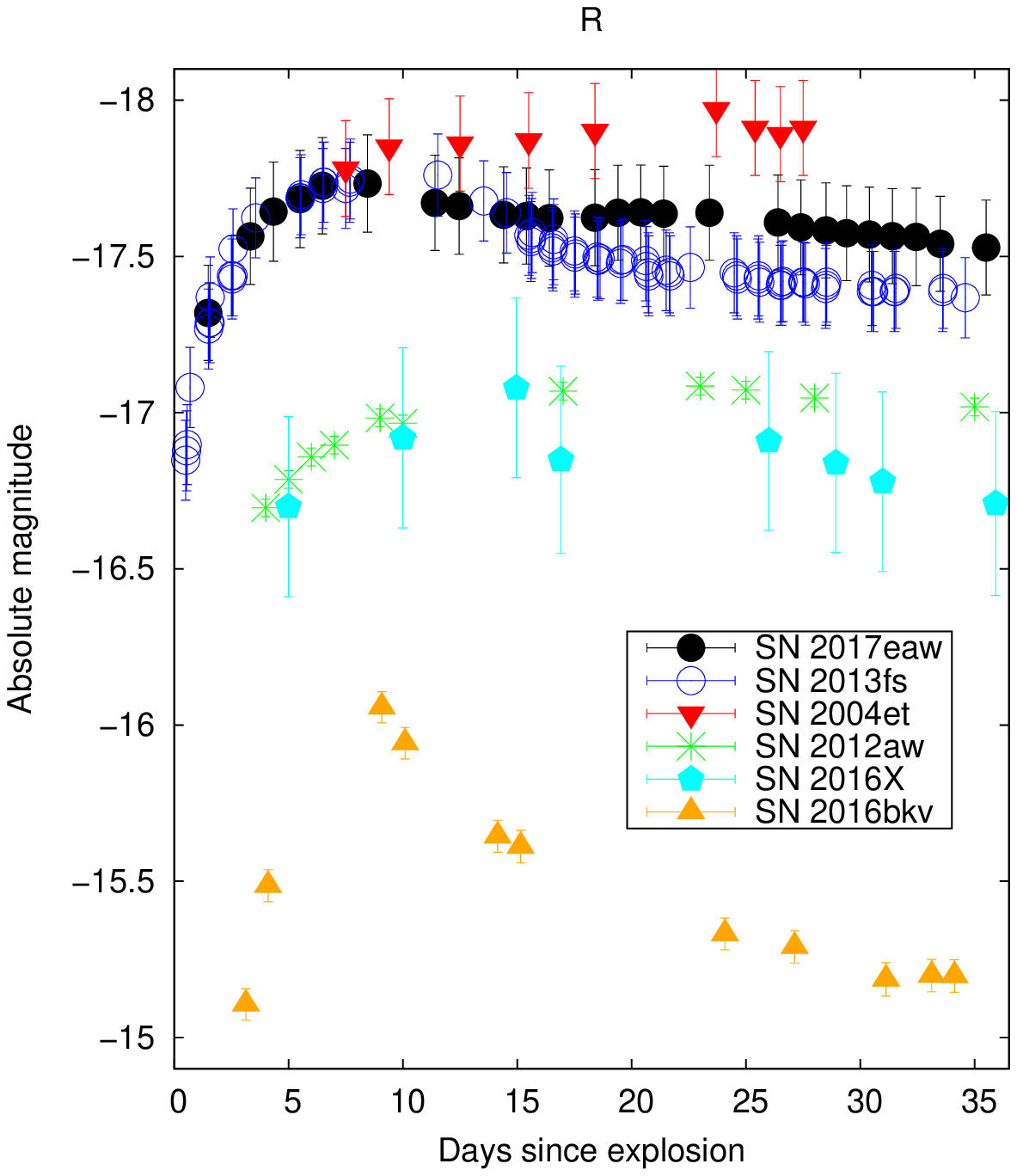} \hspace{5mm}
\includegraphics[width=.45\textwidth]{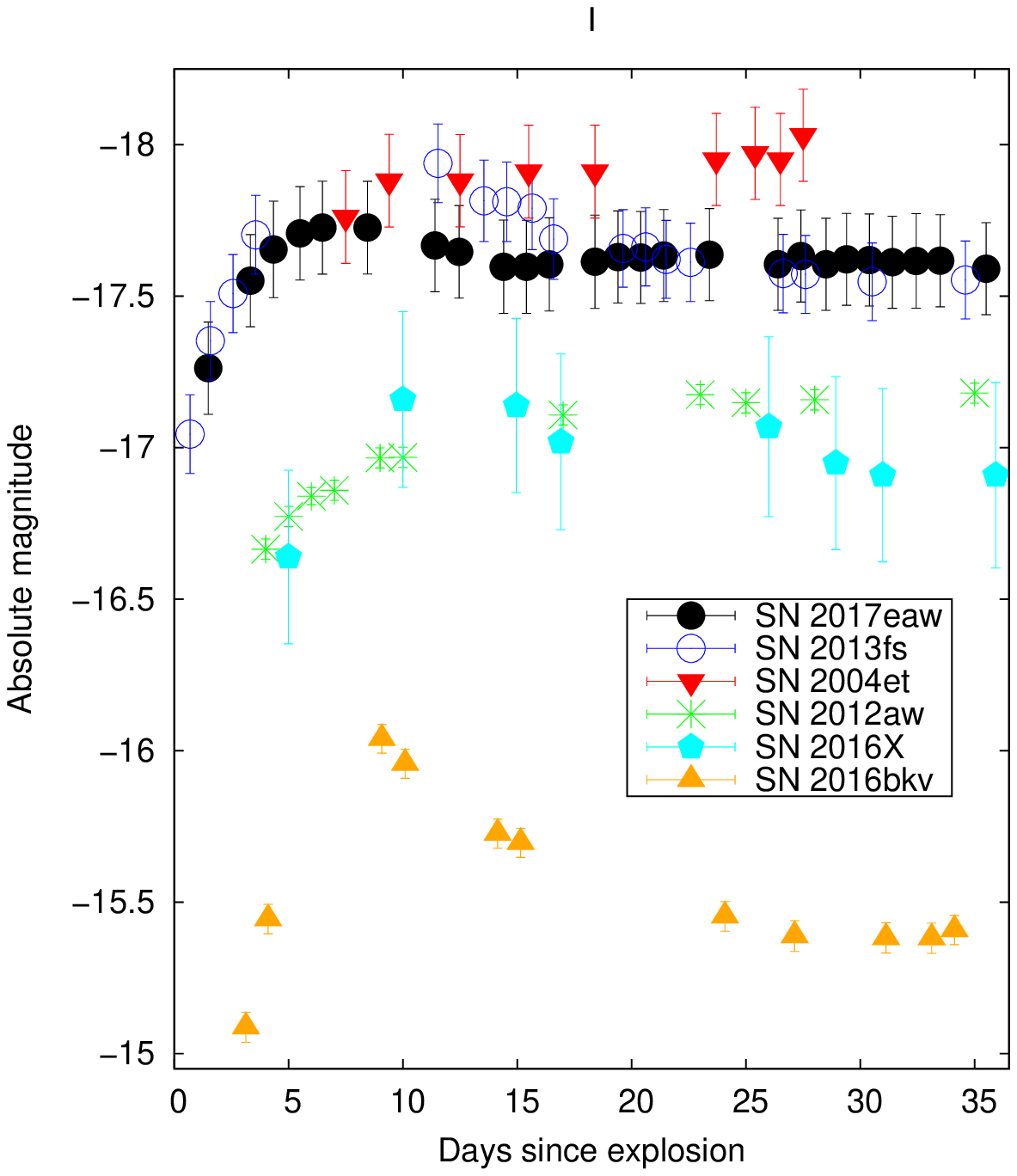}
\end{center}
\caption{Early optical light curves of SN~2017eaw compared to those of other normal or interacting Type II-P SNe.}
\label{fig:phot_comp1}
\end{figure*}

\begin{figure*}
\begin{center}
\leavevmode
\includegraphics[width=.45\textwidth]{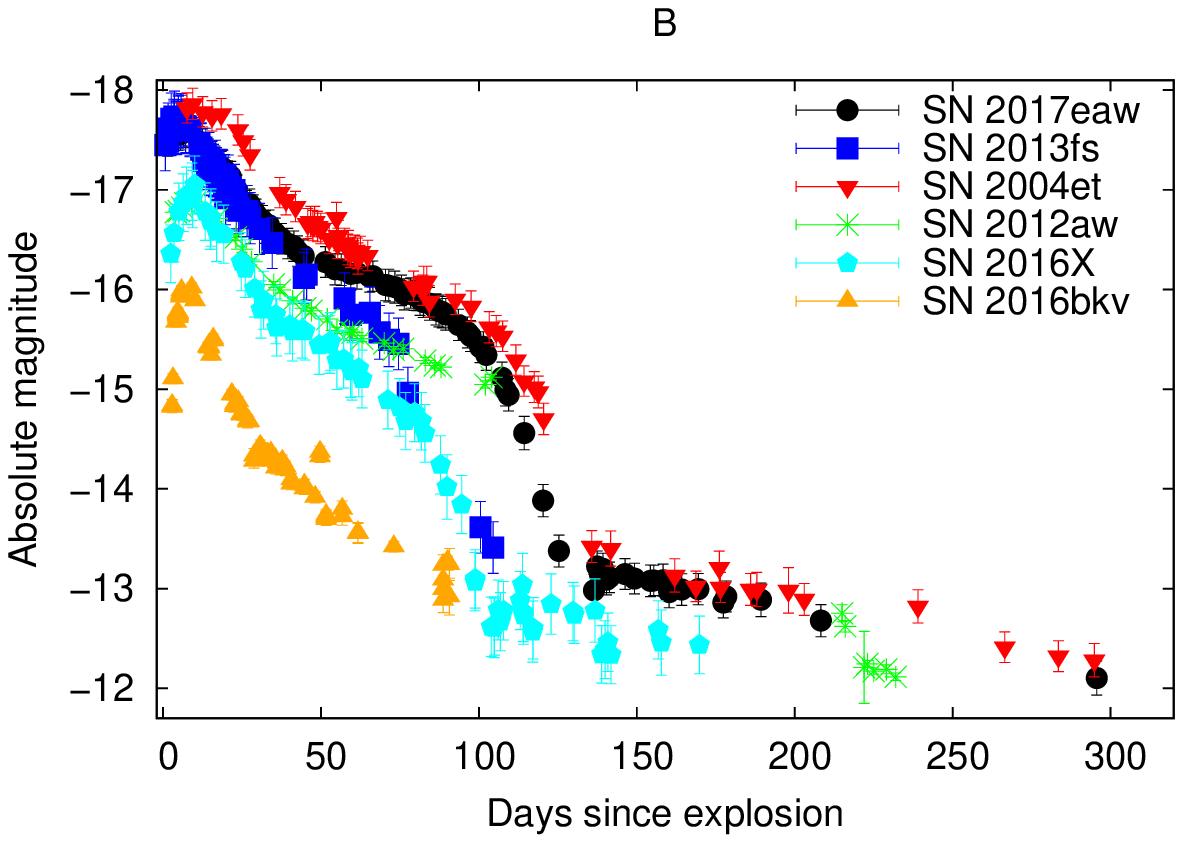} \hspace{5mm}
\includegraphics[width=.45\textwidth]{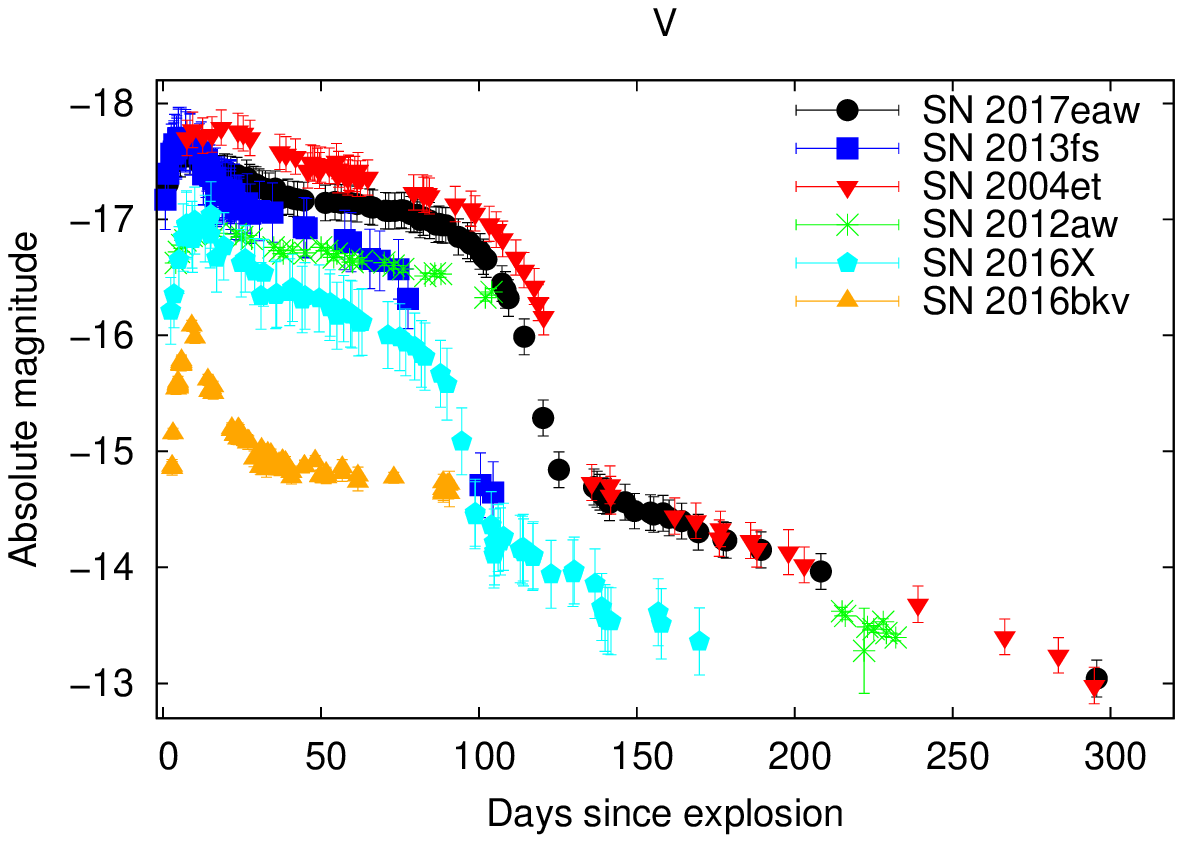}
\includegraphics[width=.45\textwidth]{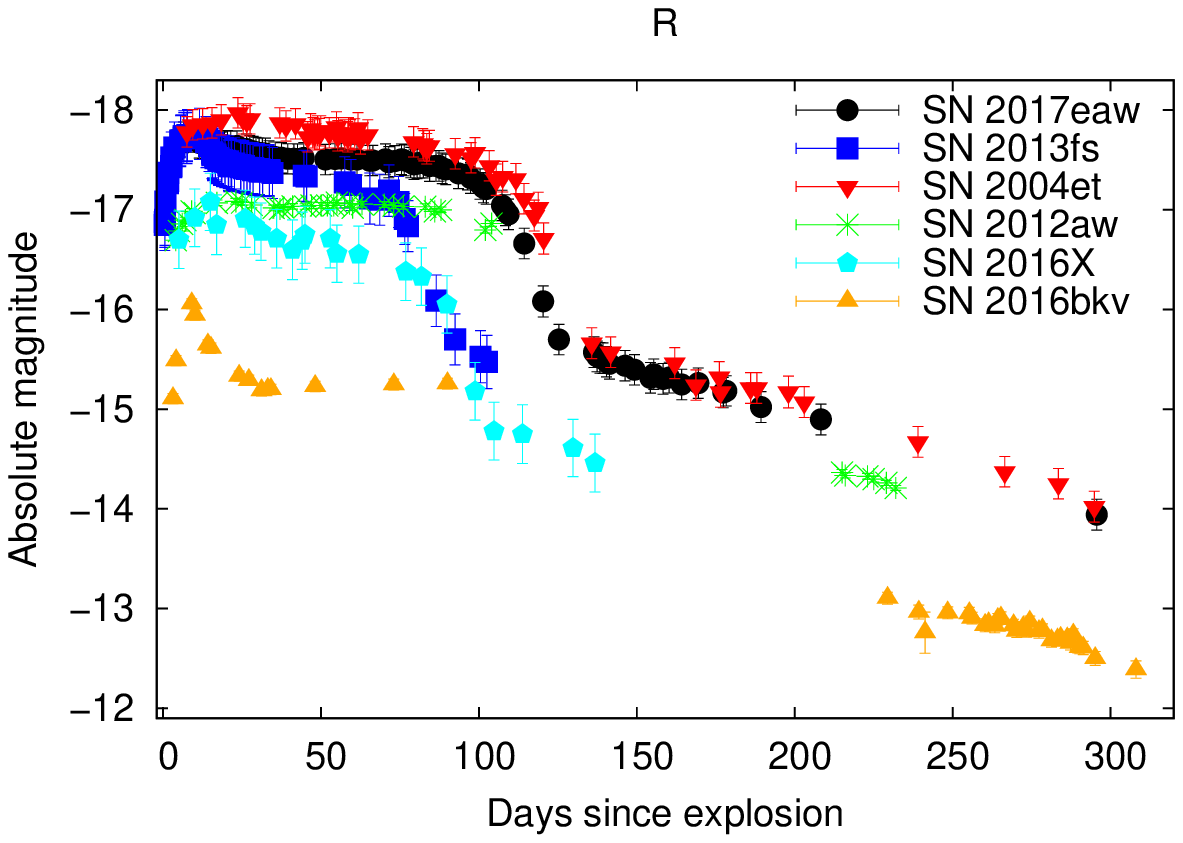} \hspace{5mm}
\includegraphics[width=.45\textwidth]{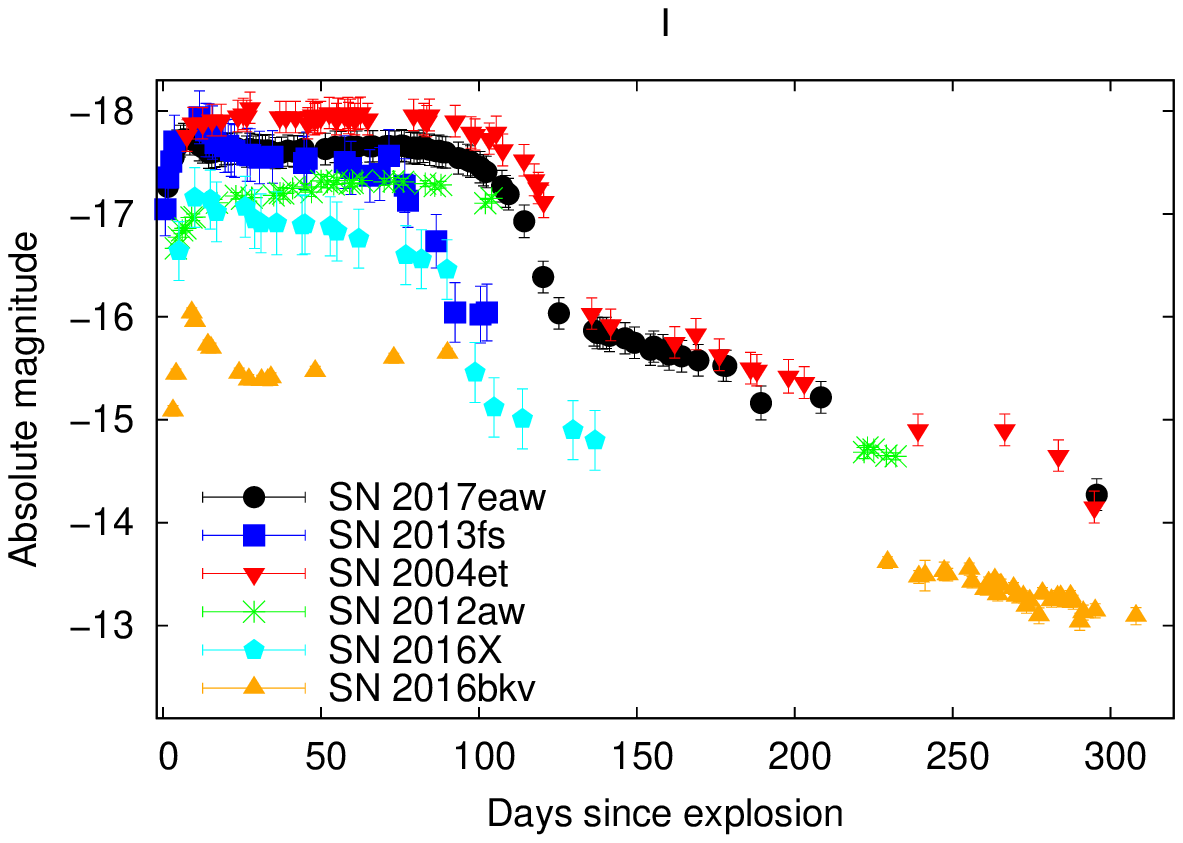}
\end{center}
\caption{Full optical light curves of SN~2017eaw compared to those of other normal or interacting Type II-P SNe.}
\label{fig:phot_comp2}
\end{figure*}

\begin{figure}
\begin{center}
\leavevmode
\includegraphics[width=.5\textwidth]{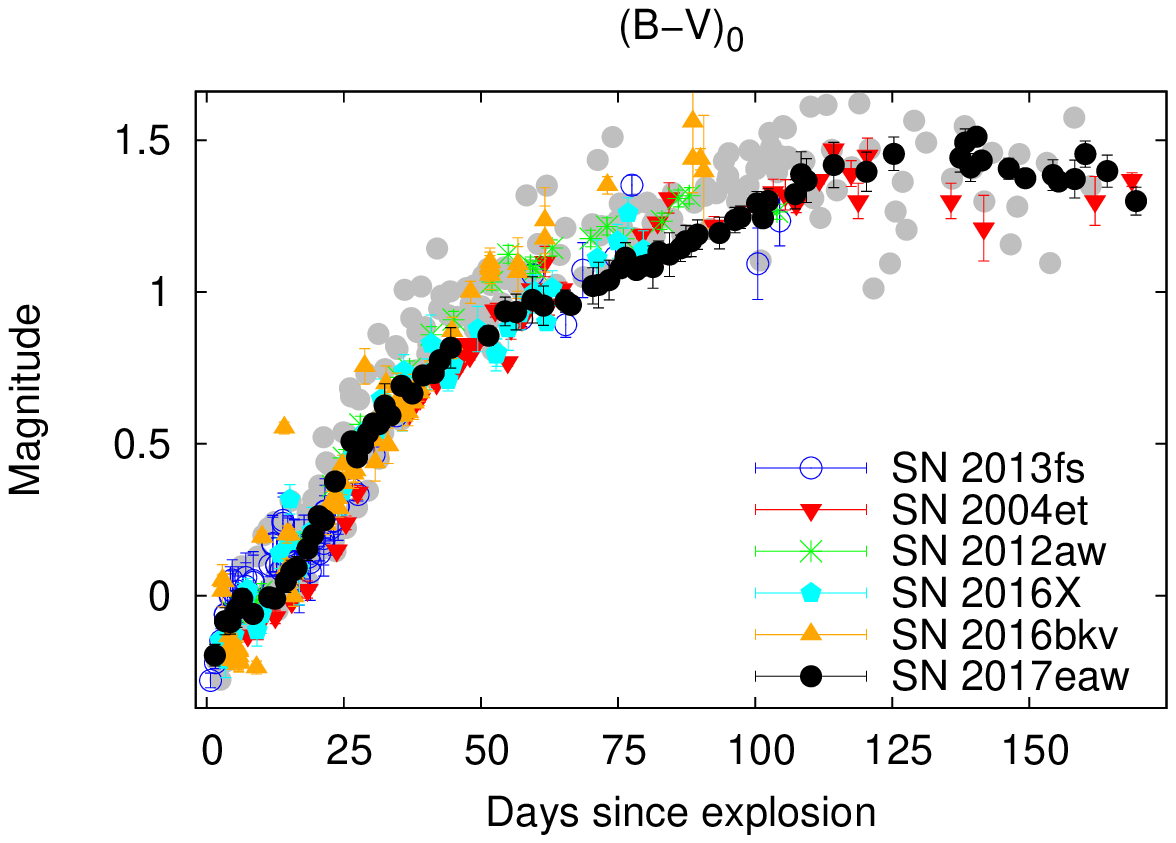} \vspace{2mm}
\includegraphics[width=.5\textwidth]{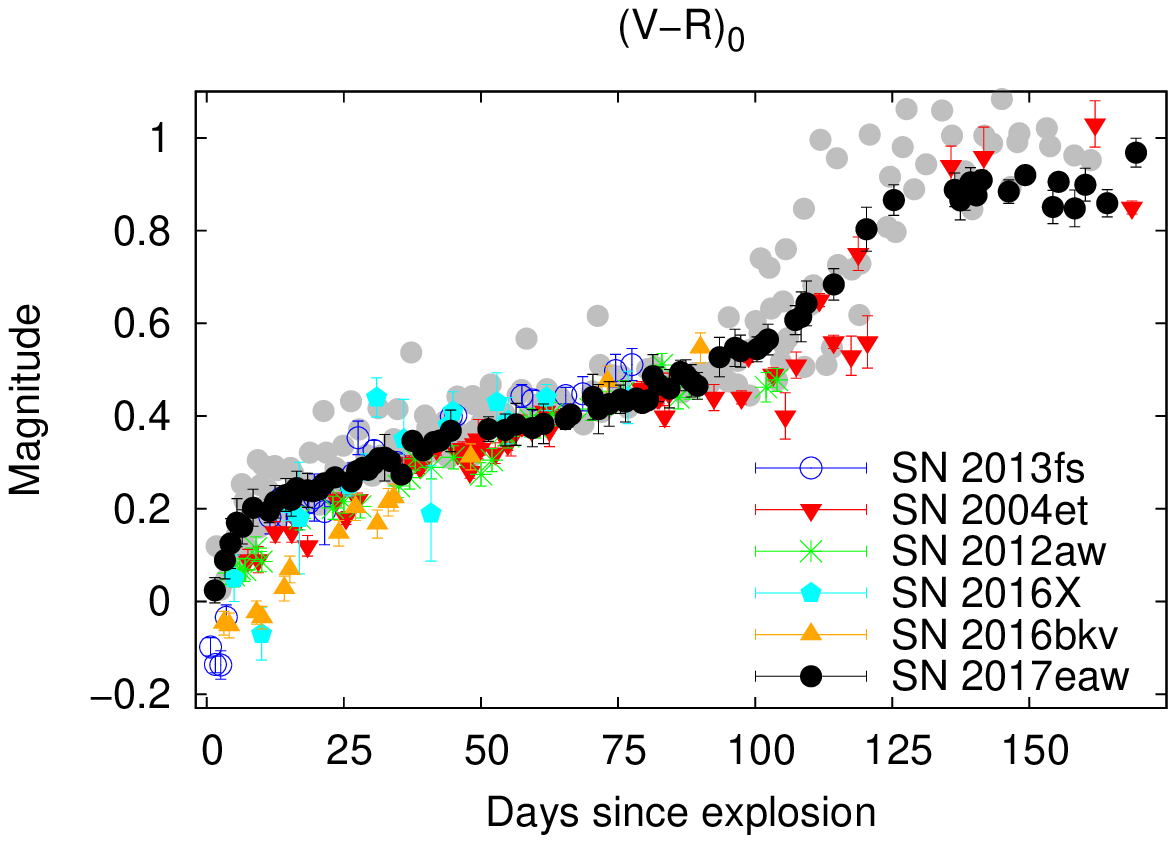} \vspace{2mm}
\includegraphics[width=.5\textwidth]{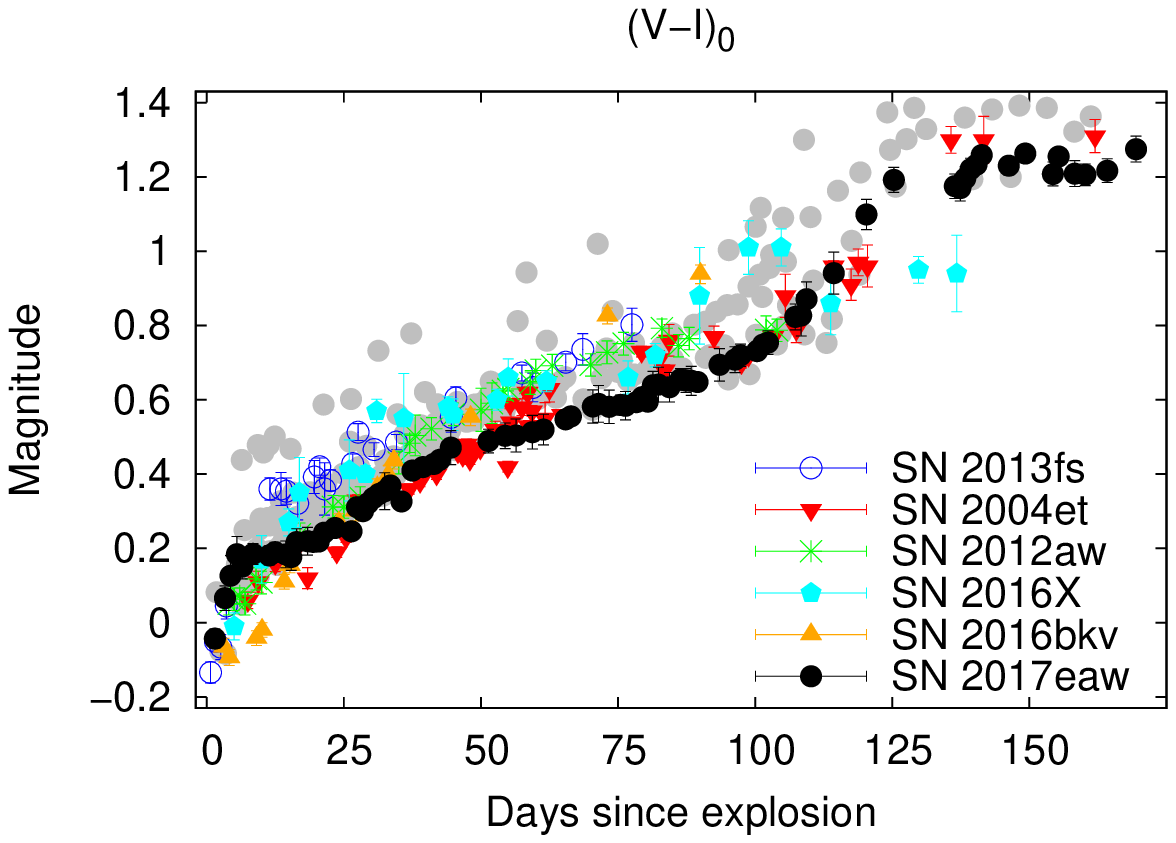}
\end{center}
\caption{Dereddened colour curves of SN~2017eaw compared to those of other normal or interacting Type II-P SNe. Data marked with coloured symbols are adopted from sources given in Table \ref{tab:sne}, while all other data (grey circles) are from \citet{Faran14}.}
\label{fig:colour}
\end{figure}

We have selected several recent, well-observed Type II-P SNe for both photometric and spectroscopic comparison, including ``normal'' (but slightly superluminous) Type II-P SN~2004et (that appeared in the same host as SN~2017eaw), ``normal'' (but slightly subluminous) SN~2012aw, early-caught and strongly interacting SN~2013fs, early-caught and slightly subluminous SN~2016X, and early-time interacting, subluminous SN~2016bkv. Table \ref{tab:sne} lists the basic data of the selected objects as well as their references.

Figs. \ref{fig:phot_comp1} and \ref{fig:phot_comp2} show the early-time and long-term absolute BVRI light curves (LCs) of SN~2017eaw together with that of the other selected SNe, respectively (absolute magnitudes were calculated using the distances and reddening values presented in Table \ref{tab:sne}). As can be seen in Fig. \ref{fig:phot_comp1}, SN~2017eaw shows a small, early bump peaking at $\sim$6-7 days after explosion in all optical channels. This behaviour resembles quite well that of SN~2013fs, and supposed to be the sign of early-time circumstellar interaction \citep{Yaron17,Morozova17,Morozova18,Bullivant18}; this topic is further analyzed in Section \ref{csm}. 

After the early, small bump, SN~2017eaw shows a long plateau up to $\sim$100 days just as ``normal'' SNe II-P (e.g. 2004et), which probably indicates that the masses of the H-envelopes are similar in these cases (unlike SN~2013fs, whose plateau drops at $\sim$25 days earlier).

In Fig. \ref{fig:colour} we present the reddening-corrected colour curves of a sample of SNe II-P. Basically, SN~2017eaw seems to follow the colour evolution of other II-P ones. The $(B-V)_0$ colour is quite blue in the early phases, but evolves relatively rapidly towards redder colours as the ejecta expands and cools; at $\sim$125 days, there is a transition peak after which the $(B-V)_0$ colour becomes gradually bluer. $(V-R)_0$ and $(V-I)_0$ evolve more slowly, and these curves become relatively flat after $\sim$125 days \citep[in part because in this phase the SN II photometric evolution, which depends on the $^{56}$Co decay, is approximately the same in all bands, see e.g.][]{Galbany16}. Nevertheless, we note that, on one hand, the colour data are quite uncertain for most of the selected SNe after $\sim$100 days (sometimes there are no data at all), and, on the other hand, the reddening of several SNe -- including 2017eaw and 2004et -- are somewhat uncertain (as we described above).


\subsection{Spectroscopic comparison}\label{spec}

Based on our observational dataset on SN~2017eaw and published data on other Type II-P SNe listed in Table \ref{tab:sne}, we carried out a detailed comparative spectroscopic analysis.
First, all observed spectra were corrected for the recession velocity of their host galaxies and for the total reddening/extinction listed in Table \ref{tab:sne}.

\subsubsection{Optical spectra}\label{spec_ev}

\begin{figure*}
\begin{center}
\leavevmode
\includegraphics[width=.45\textwidth]{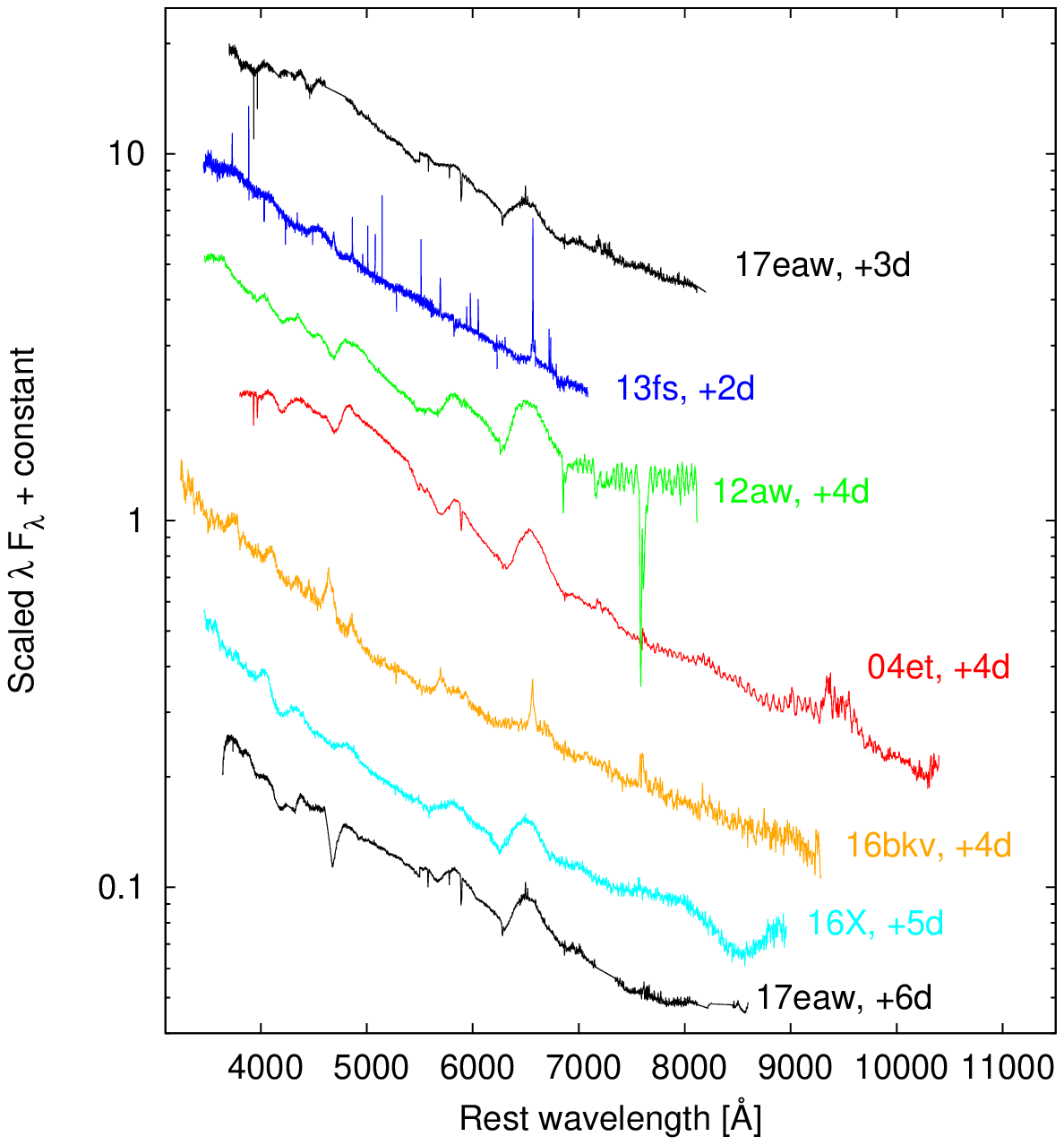} \hspace{5mm}
\includegraphics[width=.45\textwidth]{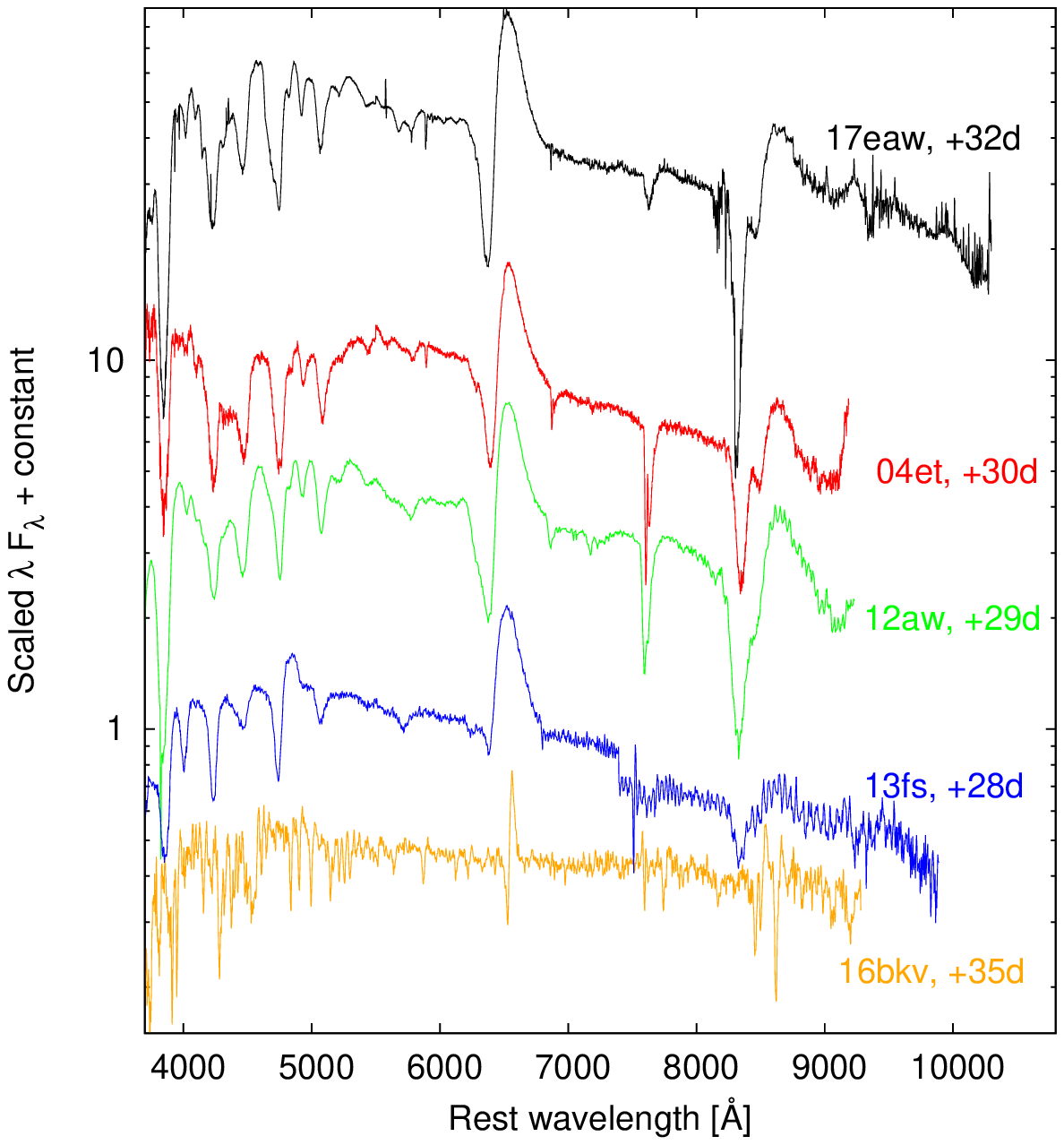}
\includegraphics[width=.45\textwidth]{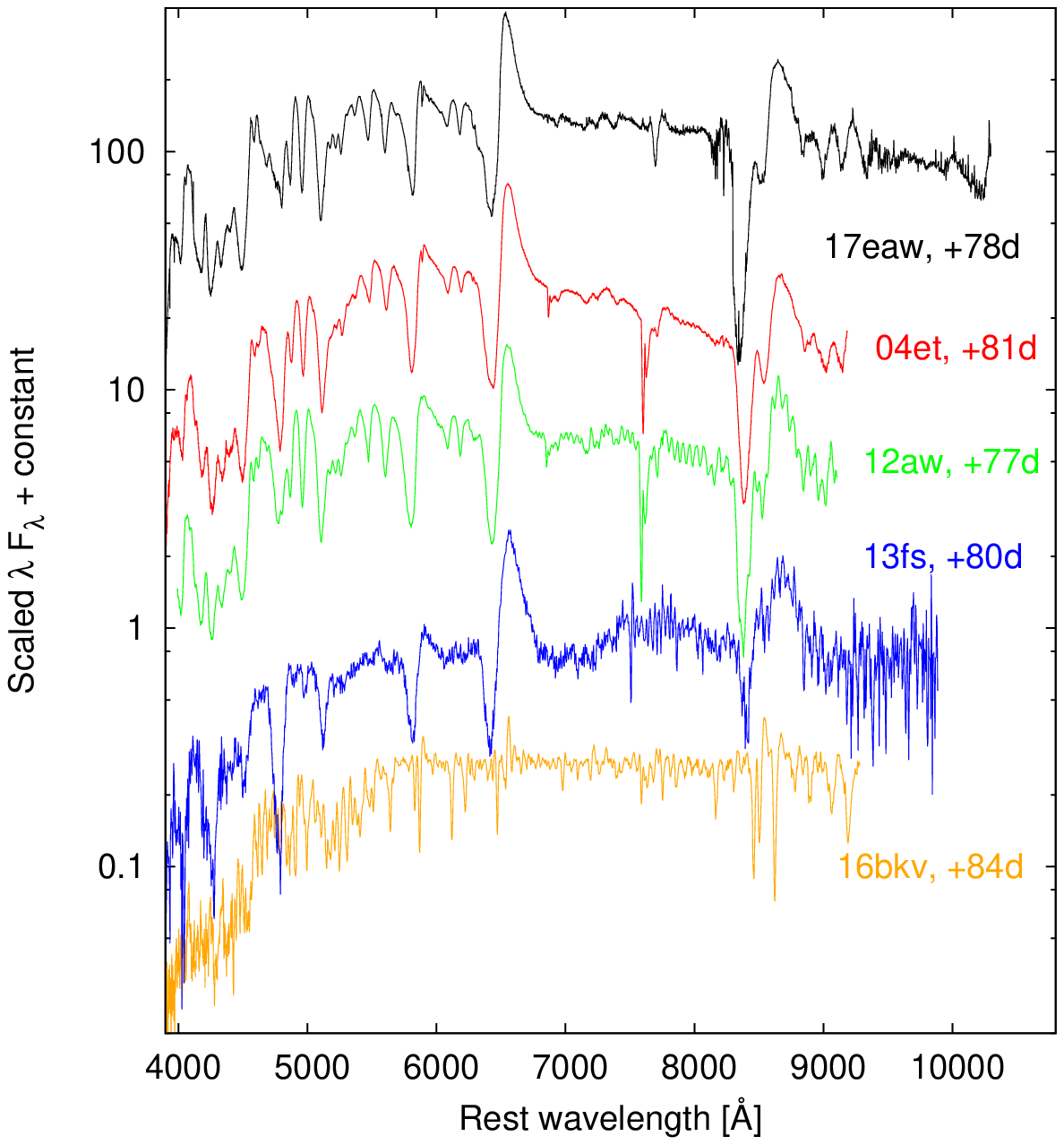}
\end{center}
\caption{Optical spectra of SN~2017eaw compared to those of other normal or interacting Type II-P SNe. All spectra are corrected for redshift and extinction.}
\label{fig:sp_comp}
\end{figure*}

\begin{figure*}
\begin{center}
\leavevmode
\includegraphics[width=.8\textwidth]{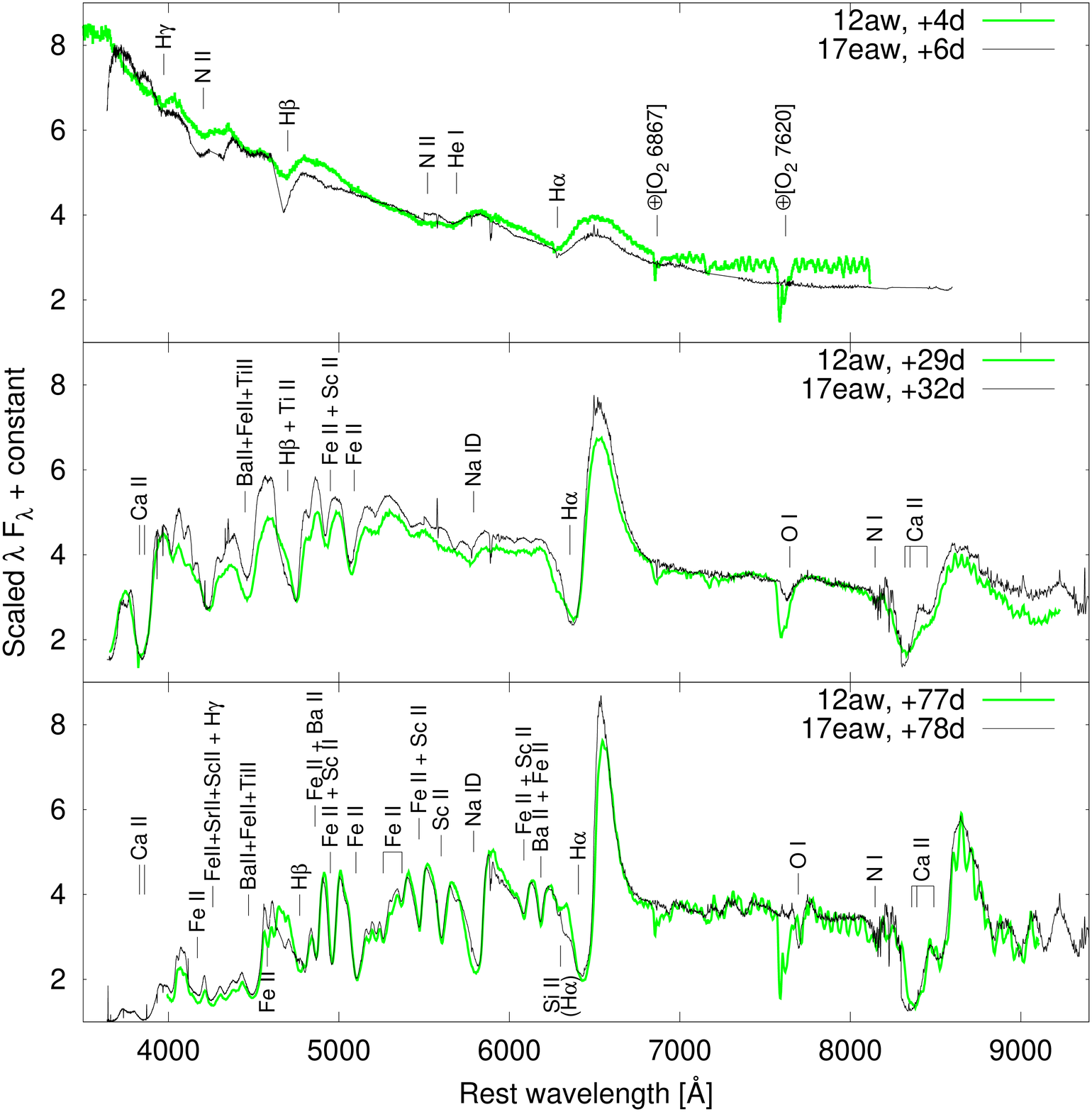}
\end{center}
\caption{Line identification based on the SYNOW modeling of SN~2012aw \citep{Bose13}. Note that at days +77/78, the small absorption component on the blue side of the H$\alpha$ P-Cygni profile may be a high velocity H feature instead of \ion{Si}{2} 6355\AA, see \citet{Guti17}.}
\label{fig:splines}
\end{figure*}

In Fig. \ref{fig:sp_comp}, we present the comparison of the optical spectra of SN~2017eaw with those of other SNe used above for the photometric comparison, selecting three ranges of epochs: 2-6 days, 28-35 days, and 78-84 days (upper left, upper right, and bottom panels, respectively; data sources are listed in Table \ref{tab:sne}). In general, the spectral evolution of SN~2017eaw follows the same trend that can be usually seen in Type II-P SNe; however, minor differences can also be found among the spectra. 

The differences are most apparent during the early days. Both SNe~2013fs \citep{Yaron17,Bullivant18} and 2016bkv \citep{Hosseinzadeh18,Nakaoka18} have been found to exhibit a short-lived but intense circumstellar interaction: they showed numerous narrow emission lines during the first few days after explosion. At the same time, neither SN~2016X \citep{Huang18} nor SN~2017eaw showed any similar phenomenon, even though an early-time moderate CSM interaction may have taken place in 2017eaw (see also Section \ref{csm}), similar to the ``normal'' Type II-P SNe~2004et and 2012aw \citep[we note that, based on a very recent paper of][a weak narrow H$\alpha$ line was observed in the 1.4d spectrum of SN~2017eaw]{Rui19}. 
At $\sim$3 months, all spectra look very similar to each other except those of the low-luminosity SN~2016bkv, which shows much weaker H$\alpha$ and \ion{Ca}{2} features and some further, incredibly narrow features compared to other objects. Note that while SNe 2017eaw, 2004et, and 2012aw are all in the plateau phase at this time, SN~2013fs is already in the declining phase, while in the case of SN~2016bkv, LC sampling is too poor to observe the transition (see Fig. \ref{fig:phot_comp2}).

The spectral similarity between SN~2017eaw, SN~2004et and 2012aw is even more pronounced when their spectrum models, computed with SYNOW, are compared. SN~2012aw is selected as a reference because SYNOW models for this SN are available \citep{Bose13}. We adopted this model sequence for identifying the main features in the spectra of SN~2017eaw at three selected epochs (see Fig. \ref{fig:splines}). As can be seen, all the key spectral features appear with very similar line strengths in both spectra, except maybe H$\beta$ at the earliest epoch, and \ion{Si}{2} 6355\AA\ that seems to be somewhat stronger and at higher velocity in the +78d spectrum of SN~2017eaw than in SN~2012aw.

\subsubsection{Velocity determination}\label{spec_vel}

\begin{figure*}[!h]
\begin{center}
\leavevmode
\includegraphics[width=.45\textwidth]{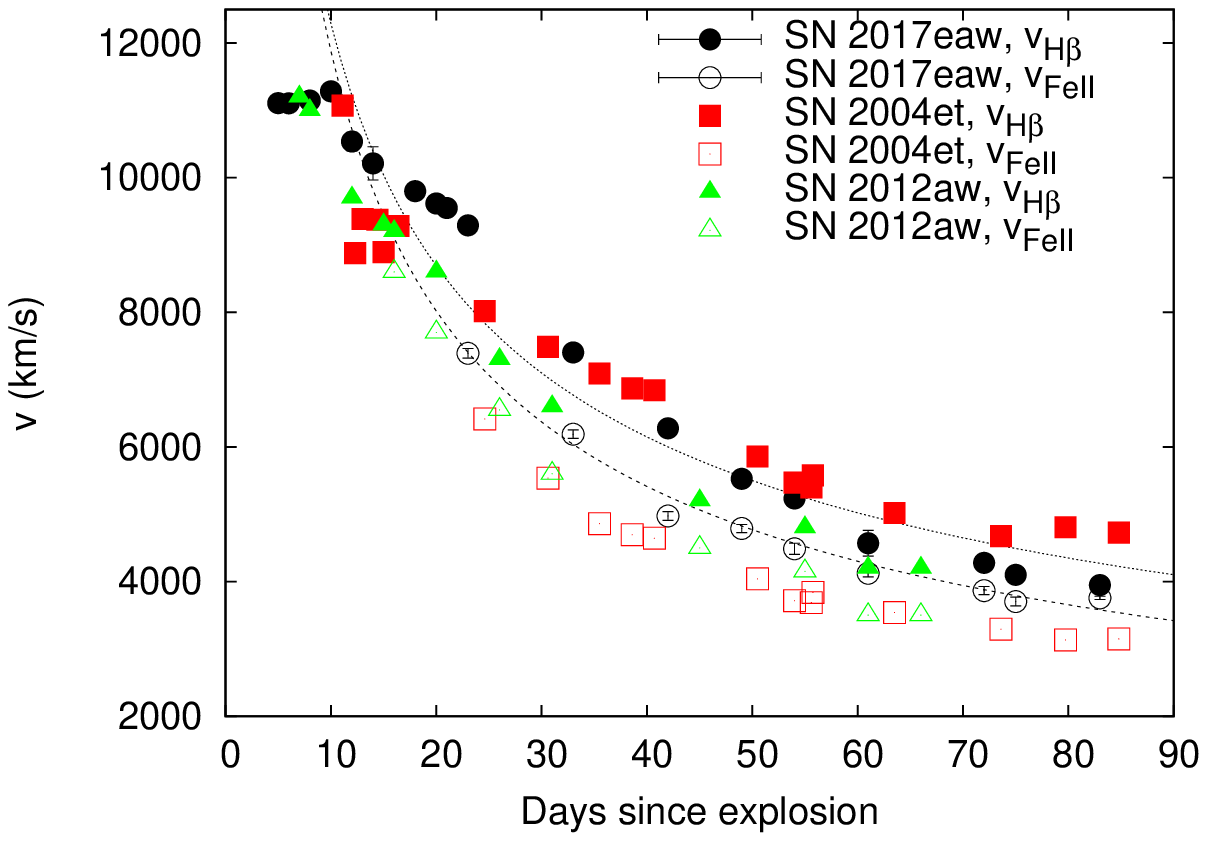}
\includegraphics[width=.45\textwidth]{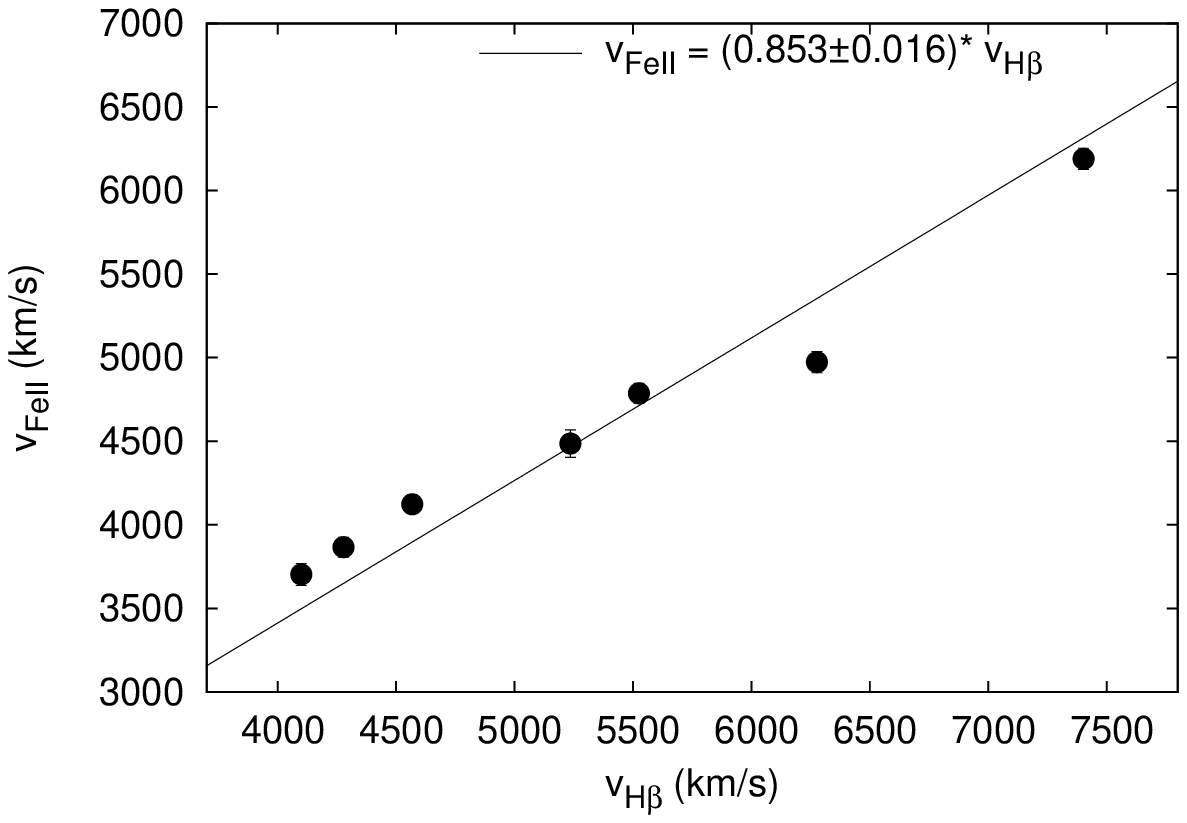}
\end{center}
\caption{ Left: velocity curves of SN~2017eaw compared to those of Type II-P SNe~2004et \citep{TV12} and 2012aw \citep{Bose13}. Open and filled symbols denote velocities calculated from the Doppler-shift of the absorption minima of \ion{Fe}{2} 5169\AA\, and H$\beta$ lines, respectively. Dotted and dashed lines show the fitted curves of $v_{\text{FeII}}$(t)/$v_{\text{FeII}}$(50)=(t/50)$^{-0.567 \pm 0.021}$ and $v_{\text{H}\beta}$(t)/$v_{\text{H}\beta}$(50)=(t/50)$^{-0.499 \pm 0.020}$, respectively (see text for details). Right: The relation between the measured $v_{\text{FeII}}$ and $v_{\text{H}\beta}$ values.}
\label{fig:vel}
\end{figure*}

Using our optical spectra, we also determined the H$\beta$ and \ion{Fe}{2} 5169\AA\ line velocities for SN~2017eaw ($v_{\text{H}\beta}$ and $v_{\text{FeII}}$, respectively), up to +85 days. For calculating $v_{\text{H}\beta}$ and $v_{\text{FeII}}$ values, taking the advantage of the adequate signal-to-noise ratio of both HET and LCO spectra, we simply fitted single Gaussian profiles to the regions of the absorption minima of the two lines.  

 Before +20 days, $H\beta$ is the most appropriate feature for velocity determination; later $v_{\text{H}\beta}$ and its uncertainties become higher because of the increasing optical depth of $H\beta$ as well as the blending with \ion{Ti}{2}, \ion{Fe}{2} and \ion{Ba}{2} features.

After +20 days, $v_{\text{FeII}}$ is thought to be a good indicator of the photospheric velocity ($v_{\text{phot}}$), since the minimum of the \ion{Fe}{2} 5169\AA\ absorption profile tends to form near the photosphere \citep[see][]{Branch03}; however, the detailed investigation of \citet{TV12} showed that the true $v_{\text{phot}}$ may significantly differ from single line velocities. 
Nevertheless, in the case of SN~2012aw, the spectral modeling obtained by \citet{Bose13} shows that $v_{\text{phot}}$ can be well estimated with $v_{\text{H}\beta}$ and $v_{\text{FeII}}$ before and after +20 days, respectively. Thus, based on the high spectral similarity of the two objects, we assume that this estimation is also feasible in the case of SN~2017eaw.

Fig. \ref{fig:vel} shows the results compared to the line velocities of SNe~2004et \citep{TV12} and 2012aw \citep{Bose13}. It is interesting that, despite the spectral similarities mentioned above, SN~2017eaw seems to have a $\lesssim 1000$ km~s$^{-1}$ systematically higher $v_{\text{FeII}}$ than either SN~2004et or SN~2012aw. On the other hand, the $v_{\text{H}\beta}$ velocities for SN~2017eaw are lower than those of SN~2004et after +25 days. As showed in previous studies \citep[see e.g.][and references therein]{TV12,Faran14}, \ion{Fe}{2} 5169\AA\ and H line velocities evolve as  $v$(t)/$v$(50)=(t/50)$^{-\beta}$ in SNe II-P. Repeating this fitting to SN~2017eaw, we get $\beta=0.567 \pm 0.021$ and $0.499 \pm 0.020$ for $v_{\text{FeII}}$ and $v_{\text{H}\beta}$, respectively, which are in good agreements with previous results. Moreover, we also plotted $v_{\text{FeII}}$ against $v_{\text{H}\beta}$ (Fig. \ref{fig:vel}, right panel) and get a linear relation with a slope of 0.853 $\pm$0.016, which agrees also well with that of other SNe II-P \citep[see e.g.][]{Poznanski10,TV12, Faran14, Gall18}.

\subsubsection{Near-UV and near-IR spectra}\label{spec_uvir}

\begin{figure*}
\begin{center}
\leavevmode
\includegraphics[width=.45\textwidth]{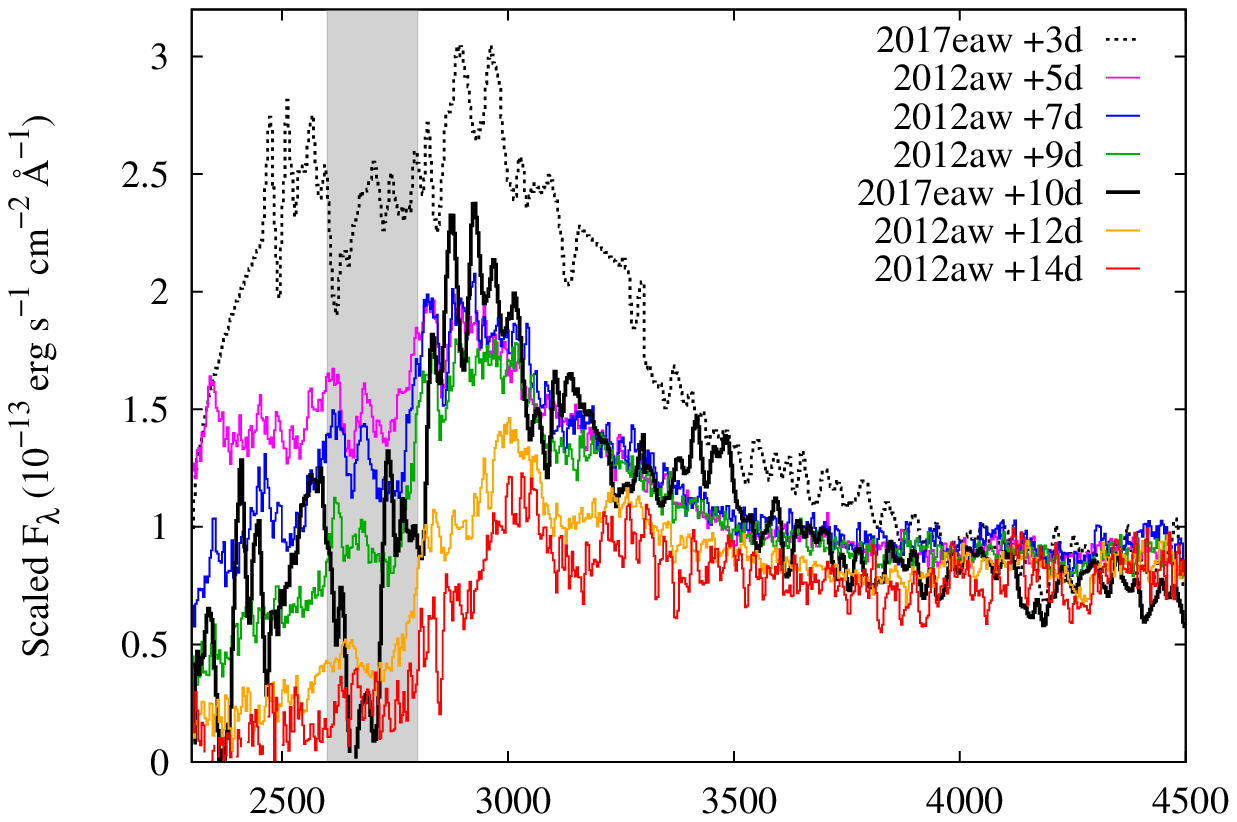}
\includegraphics[width=.45\textwidth]{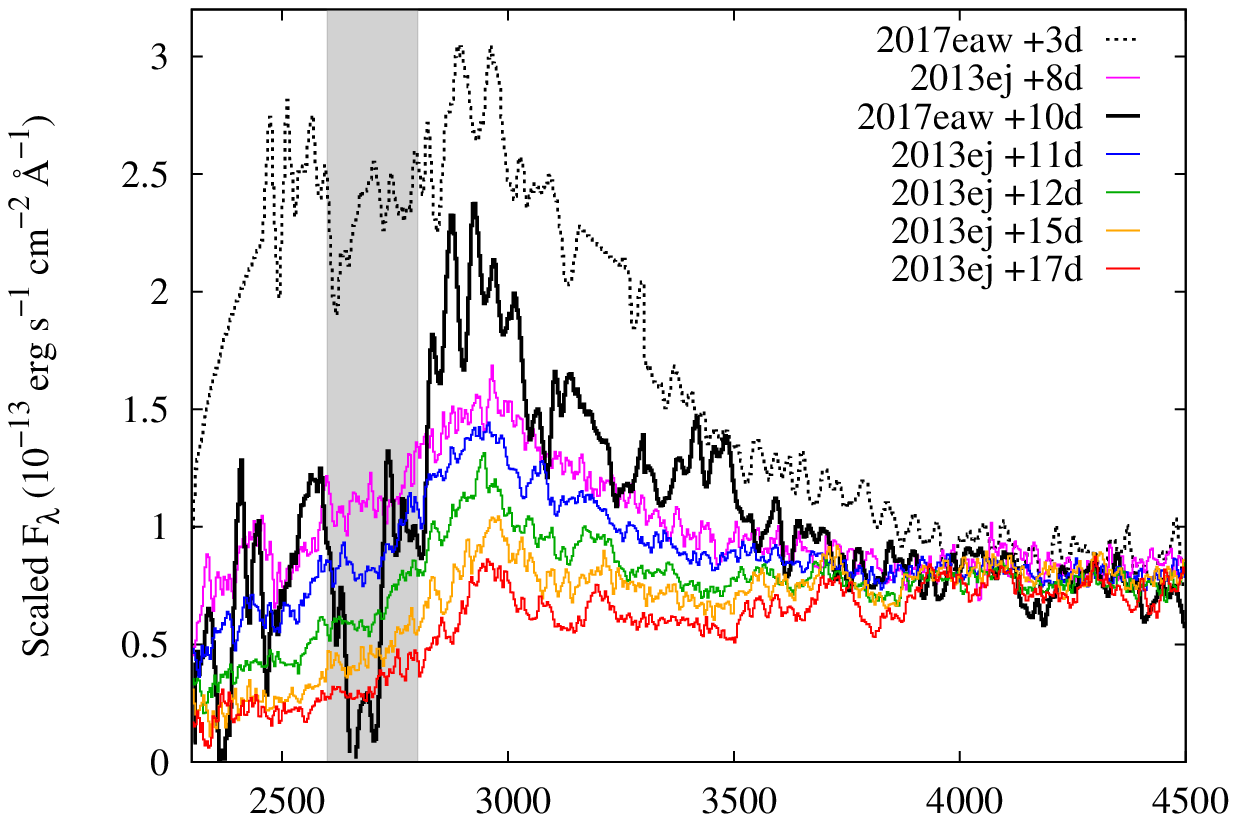}
\end{center}
\caption{Early phase near-UV spectra of SN~2017eaw taken with the {\it Swift} UVOT/UGRISM instrument (black curves) compared to those of SN~2012aw \citep[][left panel]{Bayless13} and SN~2013ej \citep[][right panel]{Dhungana16}. All spectra are dereddened and normalized to the same flux level in the 4000$-$4500\AA\ regime. Note that the deep feature in the spectrum of SN~2017eaw at $\sim$2700 \AA\ (marked by gray background color) is artificial due to contamination to the background level from a nearby stellar source.}
\label{fig:uvsp}
\end{figure*}

\begin{figure}
\begin{center}
\leavevmode
\includegraphics[width=.45\textwidth]{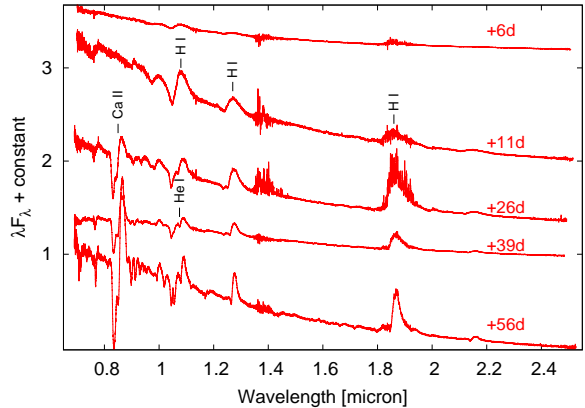}
\end{center}
\caption{Near-IR spectra of SN~2017eaw obtained with NASA IRTF SpeX.}
\label{fig:near-IRsp}
\end{figure}

\begin{figure}
\begin{center}
\leavevmode
\includegraphics[width=.45\textwidth]{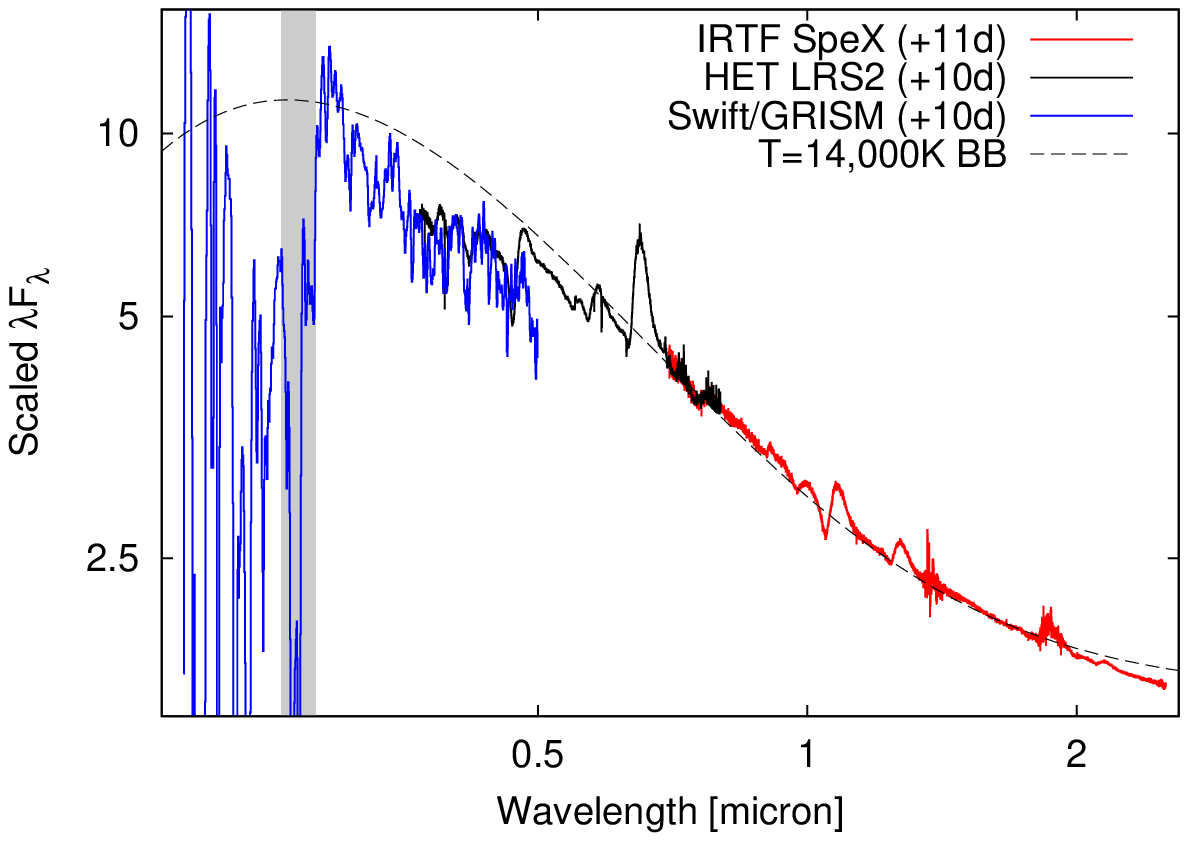}
\end{center}
\caption{Simple fitting of a $T$=14\,000K blackbody on a combined (near-UV-optical-near-IR) early-phase spectrum of SN~2017eaw. Contaminated region of the UV spectrum is marked by gray background color.}
\label{fig:bb}
\end{figure}

Comparison of the near-UV spectra of SN~2017eaw with those of two other Type II-P SNe, 2012aw and 2013ej, are presented in Fig. \ref{fig:uvsp}. Based on the findings of \citet{GY08}, Type II-P SNe look very similar in the 2000$-$3000\AA\ range; however, there are only a few objects with high-quality data. In the left panel of Fig. \ref{fig:uvsp}, we show the two near-UV spectra of SN~2017eaw together with the sequence of early-phase spectra of SN~2012aw \citep{Bayless13}. All spectra are corrected for extinction and normalized to the same flux level between 4000 and 4500\AA. Note that the strong flux depression in the +10d spectrum of SN~2017eaw is not real; it is due to contamination caused by the presence of 0th order images of nearby stars in the background region of the SN spectrum. Disregarding the contaminated region, the spectra of both SNe as well as their evolution are very similar, confirming the findings by \citet{GY08}. 

The right panel of Fig. \ref{fig:uvsp} contains the same two near-UV spectra of SN~2017eaw, but compared to those of SN~2013ej \citep{Dhungana16}. The similarity is less pronounced in this case, as SN~2017eaw appears to be relatively brighter than SN~2013ej between 2500 and 3500\AA\ at +10d. As SN~2013ej was a ``transitional'' object between the Type II-P (``plateau'') and II-L (``linear'') SNe \citep{Dhungana16}, such minor differences between the near-UV spectra are not unexpected and likely real. 

Because all three SNe showed X-ray emission shortly after explosion \citep[see][as well as Sec. \ref{csm}]{Bayless13, Chakraborti16} that are consistent with the presence of very nearby CSM, the relatively lower near-UV flux of SN~2013ej is probably not due to the lack of early CSM-interaction. 
As SN~2013ej showed a shorter plateau than 2017eaw in its optical light curves \citep{Dhungana16}, a faster spectral evolution, i.e. the faster decline of the near-UV flux in time, may be a more likely cause of the difference of its near-UV spectra with those of SNe~2017eaw.  

The five near-IR spectra of SN~2017eaw, obtained with IRTF between +6 and +39 days, are plotted in Fig. \ref{fig:near-IRsp}. During the early phases, the spectra do not show many features; they are mostly dominated by the P Cygni profiles of the \ion{Ca}{2} triplet and the hydrogen Paschen features. Nevertheless, the +6 and +11d IRTF spectra are the earliest near-IR spectra of SN~2017eaw published to date.

Moreover, the contemporaneous near-UV, optical, and near-IR spectra obtained at +10/11 days allowed us to make a well-constrained estimation for the photospheric temperature based on a wider wavelength range. We constructed a combined spectrum, which can be well fitted with a $T$=14,000K blackbody (see Fig. \ref{fig:bb}) -- this value is in a good agreement with the photospheric temperature determined by \citet{Bose13} from the spectral modeling of SN~2012aw at the same epoch.

Based on higher resolution spectra obtained with Gemini Near-Infrared Spectrograph between +22 and +205 days, \citet{Rho18} carried out a more detailed analysis on SN~2017eaw. Their most important conclusion is that the spectra show the formation of a moderate amount of CO molecules and hot dust after $\sim$120 days. We will return to this finding in Section~\ref{dust}.

\subsection{Distance estimates}\label{epm}

The distance to SN~2017eaw and its host galaxy is estimated by combining the results from various methods, as detailed below.


\subsubsection{Expanding Photosphere Method}

\begin{figure}
    \centering
    \includegraphics[width=.49\textwidth]{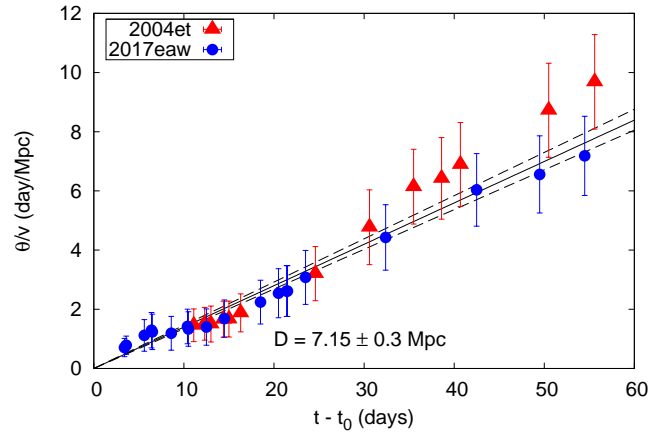}
    \caption{Distance determination from EPM applied to the combined SN~2017eaw + 2004et data (see text). The slope of the solid line gives $\sim D^{-1}$, and the dashed lines illustrate the effect of $\pm 0.3$ Mpc statistical uncertainty in the distance.}
    \label{fig:epm1}
\end{figure}

First, we apply the Expanding Photosphere Method (EPM) to the combined dataset of
SN~2017eaw (this paper) and 2004et \citep{Sahu06, Misra07, Maguire10}. The photospheric velocities are derived from the absorption minima of $H\beta$ and the standard \ion{Fe}{2} $\lambda 5169$ feature, as shown in the previous section. 
For the explosion dates we adopt JD 2,453,270.5 \citep{Li05} for
SN~2004et and JD 2,457,886.5 for SN~2017eaw (see above). 

Our method uses the combination of the light and velocity curves
of two SNe that exploded within the same host galaxy. This technique has
been applied for a number of cases recently: SN~2011dh \& 2005cs in M51 \citep{Vinko12}, and SN~2013ej \& 2002ap in M74 \citep{Dhungana16}. 
The constraint that the distance must be the same for both SNe helps
to overcome some of the issues related to the application of EPM to a single object, e.g. the sensitivity to the explosion date or stronger deviations from the modified blackbody evolution. 

After correcting for the interstellar extinction using $E(B-V)_{tot} = 0.41$ mag for the total reddening (see above) and assuming $R_V = 3.1$ for the extinction law, we construct a quasi-bolometric LC for both SNe from their
measured $BVRI$ data (see Section~\ref{bol}). Then we fit the standard equations of EPM \citep[e.g.][]{Vinko12} coupled with the dilution factors of \citet{DH05} to the quasi-bolometric LCs simultaneously. The fit is restricted to the epochs $10 < t < 50$ days after explosion for several reasons. First, the LC of SN~2017eaw shows a bump at the earliest epochs (Fig.~\ref{fig:phot_comp1}) that might be due to physical processes (e.g. CSM-interaction) that are not included in the simple physical model (an expanding blackbody) used in EPM. Also, at $t < 10$ days the contribution from the UV-band flux, which is treated only approximately when assembling the quasi-bolometric LCs, is higher than at later phases. After $\sim 10$ days these complications seem to have less effect. After $t \sim 50$ days  NLTE effects become increasingly dominant \citep[see e.g.][]{DH05}, which also cause deviations from the simple blackbody approximation used in EPM. Thus, as a compromise, we fit the equations of EPM to data taken at $10 <t <50$ days. 

The result is shown in Figure~\ref{fig:epm1}. The slope of the line gives $D = 7.15$~$\pm 0.30$ (statistical) $\pm 0.70$ (systematic) Mpc.  
The quoted systematic uncertainty comes from two main sources: a $\pm 1$ day uncertainty in the adopted explosion dates, and the sensitivity of the distance to the minimum and maximum epochs used in the EPM fitting. If we restrict the fitting to data taken in $0 < t < 30$ days, as recommended by \citet{DH05}, we get a $\sim 0.7$ Mpc higher common distance (see Table~\ref{tab:dist}).   

From Fig.\ref{fig:epm1} it is seen that the data of the two SNe, even though they are consistent within the errors, start to deviate systematically from each other after $t > +30$ days. Thus, fitting the equations of EPM to only one of them would give different distances. Indeed, fitting only to SN~2004et, but also letting the moment of explosion ($t_0$) float, would result in $D \sim 5.23 \pm 0.15$ Mpc and $t_0$ being $\sim 5$ days later than the assumed moment of explosion (see above), which is in conflict with the discovery date. Using SN~2017eaw only, the same analysis would give $D \sim 7.08 \pm 0.11$ Mpc and $t_0 \sim 1.9$ days later than assumed. These conflicting results illustrate why the fitting to the combined dataset (coupled with the constraints on the moment of explosion) can give more reliable results. Keeping in mind the uncertainties of reddening/extinction and the explosion date, this might explain why the previous applications of EPM to SN~2004et resulted in lower ($D \sim$5 Mpc) distances \citep[see e.g.][and references therein]{TV12}. 

Note that the application of the template velocity curve based on spectroscopic modeling by \citet{TV12} gives a distance 
that is only $\sim 0.1$ Mpc lower, thus, it is within the uncertainty of the fitting. 

The results detailed above also depend on the assumed reddening ($E(B-V)_{tot} = 0.41$). If we adopt only the reddening
from the Milky Way dust, $E(B-V) \sim 0.3$ \citep{SF11}, and thus ignore the reddening within NGC~6946, then the EPM distance from the combined dataset would decrease to $D \sim 6.7$ Mpc. Given that dust in the host galaxy should also contribute somewhat to the total reddening, this is probably a lower limit, and the true distance is closer to $\sim$7 Mpc. Table~\ref{tab:dist} summarizes the distances derived above and also from other methods (see below).

\begin{table}[]
    \centering
    \caption{Distance estimates to NGC~6946}
    \label{tab:dist}
    \begin{tabular}{lcccc}
    \hline
    \hline
    Method & Calibration & E(B-V) & D (Mpc) & $\sigma$ (Mpc) \\
    \hline
    EPM & A & 0.41 & 7.15 & 0.3 \\
    EPM & A & 0.30 & 6.66 & 0.3 \\
    EPM & B & 0.41 & 7.85 & 0.2 \\
    EPM & B & 0.30 & 6.93 & 0.4 \\
    SCM & C & -- & 6.69 & 0.3 \\
    SCM & D & -- & 6.69 & 0.2 \\
    SCM & E & -- & 6.02 & 0.3 \\
    TRGB & F & -- & 6.7 & 0.2 \\
    TRGB & G & -- & 7.7 & 0.3 \\
    PNLF & H & -- & 6.1 & 0.6 \\
    \hline
    average & & & 6.85 & 0.63 \\
    \hline
    \hline
    \end{tabular}
{\bf Notes.}  A) 10d$<t<$50d; B) 10d$<t<$30d ; C) \citet{Poznanski09}; D) \citet{deJaeger17}, E) \citet{Gall18}; F) \citet{Tikhonov14}; G) \citet{Anand18}; H) \citet{Herrmann08}.
\end{table}

\subsubsection{Standard Candle Method}

Second, we estimate the distance to SN~2017eaw by applying the Standard Candle Method (SCM).
This method was first proposed by \citet{HP02}, then it was refined and re-calibrated in various studies later \citep{TV06, Poznanski09, Oliva10, DAndrea10, Maguire10a, deJaeger17, Gall16}. 

For SN~2017eaw 
we measure $m_V(50) = 13.20$, $m_I(50) = 12.25$, $m_r(50) = 12.77$, $m_i(50) = 12.61$, $v_{FeII}(50) = 4600 \pm 200$ km~s$^{-1}$ and 
$v_{H\beta}(50) = 5350 \pm 200$ km~s$^{-1}$ for the $V$, $I$, $r$, and $i$-band magnitudes and expansion velocities at $t=50$d after explosion, respectively. Table~\ref{tab:dist} lists the distances of SN~2017eaw inferred from the three most recent SCM calibrations. 

Compared to other distances listed in NED, most of which are based on using SCM on SN~2004et  ($\sim 5$ Mpc), these new SCM-based distances to SN~2017eaw are all systematically higher. This is the same as found above when comparing the individual EPM-based distances of SNe~2017eaw and 2004et. 
The lower SCM-based distance to SN~2004et is due to the fact that SN~2004et showed brighter plateau, but lower expansion velocity at $t=50$d than SN~2017eaw. Indeed, from the data by \citet{Sahu06} and \citet{Maguire10} we measure $m_V(50)= 12.83$, $m_I(50) = 11.93$, $v_{FeII}(50) = 4230 \pm 200$ km~s$^{-1}$ and derive $D = 5.43 \pm 0.24$ Mpc from the calibration by \citet{Poznanski09}. If we correct the plateau brightness for Milky Way extinction ($E(B-V) = 0.3$ mag) first and use these corrected magnitudes, then the  SCM-based distance to SN~2004et decreases to $\sim 4.9 \pm 0.2$ Mpc. Overall, the SCM-distances show the same trend as the EPM-based distances: they seem to be systematically higher for SN~2017eaw than for SN~2004et. This strengthens the suspicion that SN~2004et was not a typical Type II-P, thus, the previous distance estimates based on SN~2004et are probably biased. 

\subsubsection{Other distances to the host galaxy}

NED also contains other redshift-independent distance estimates for NGC~6946 that are not related to SN data. In Table~\ref{tab:dist} we list the four most recent ones that are based on the ``Tip of the Red Giant Branch'' (TRGB) \citep{Tikhonov14, Anand18} and the ``Planetary Nebula Luminosity Function" (PNLF) \citep{Herrmann08} methods. The PNLF distance ($D \sim 6$ Mpc) is in between the ones derived for SN~2017eaw ($D \sim 7$ Mpc) and 2004et ($D \sim 5$ Mpc), while the other two are closer to that of SN~2017eaw.

We assign the average of the various distances listed in Table~\ref{tab:dist} to the final distance of NGC~6946, i.e. $D \sim 6.85$~$\pm 0.63$ Mpc (the quoted uncertainty is the rms error, but takes into account the uncertainties of the individual distances). This value disfavors the previous measurements from SN~2004et that all gave $\sim$30 percent lower distances. We use $D = 6.85$ Mpc as the distance to SN~2017eaw in the rest of this paper. 

\subsection{Progenitor mass from nebular spectra}\label{nebular}


Observations taken during the nebular phase ($\sim$200-500 days post-explosion) reveal the inner nucleosynthetic products of the progenitor star and its explosive burning.
The strength and  shape of emission lines of individual elements can be mapped back to properties of the progenitor and explosion. 
In particular, a monotonic relation exists between the intensity of the [OI] doublet ($\rm{\lambda\lambda 6300, 6364}$) and the mass of the progenitor star \citep{2012jerkstrand, 2014jerkstrand}. 
We use this relationship to find the progenitor mass of SN 2017eaw using the spectra taken during the nebular phase.

We model the oxygen emission line using the suite of models presented in \citet{2012jerkstrand} which are computed using the spectral synthesis code described in \citet{2011jerkstrand}.
Model spectra are produced for $\rm{M_{ZAMS}=12, 15, 19,}$ and 25 $\rm{M_{\odot}}$ at epochs of 212, 250, 306, 400, and 451 days for the 12, 15, and 25 $\rm{M_{\odot}}$ models and at 212,  250, 332, 369, and 451 days for the 19 $\rm{M_{\odot}}$ model. These models were generated for SN~2004et using a nickel mass of  $\rm{M_{{}^{56}Ni,mod}}$ = 0.062 $\rm{M_{\odot}}$ and $\rm{d_{mod}}$ = 5.5 Mpc.
To apply these models to SN~2017eaw, the synthetic spectra are scaled to the inferred distance and nickel mass of SN 2017eaw via the following relation:
\begin{equation}\label{eqn:NiMass}
F_{obs} = F_{mod} \times \left(\frac{d_{mod}}{d_{obs}}\right)^{2} \left(\frac{M_{{}^{56}Ni, obs}}{M_{{}^{56}Ni, mod}}\right) e^{\frac{t_{mod} - t_{obs}}{111.4}}
\end{equation}
where $\rm{F_{obs}}$ and $\rm{F_{mod}}$ are the observed and model fluxes, $\rm{d_{obs}}$ and $\rm{d_{mod}}$ are the observed and model distances, and $\rm{M_{{}^{56}Ni, obs}}$ and $\rm{M_{{}^{56}Ni,mod}}$ are the observed and model nickel masses synthesized during the explosion. Although SN 2004et and SN 2017eaw are in the same galaxy, we use the distance found in Section 3.3 as the distance to SN 2017eaw and scale the flux accordingly.

\begin{figure}
\begin{center}
\includegraphics[width=0.99\columnwidth]{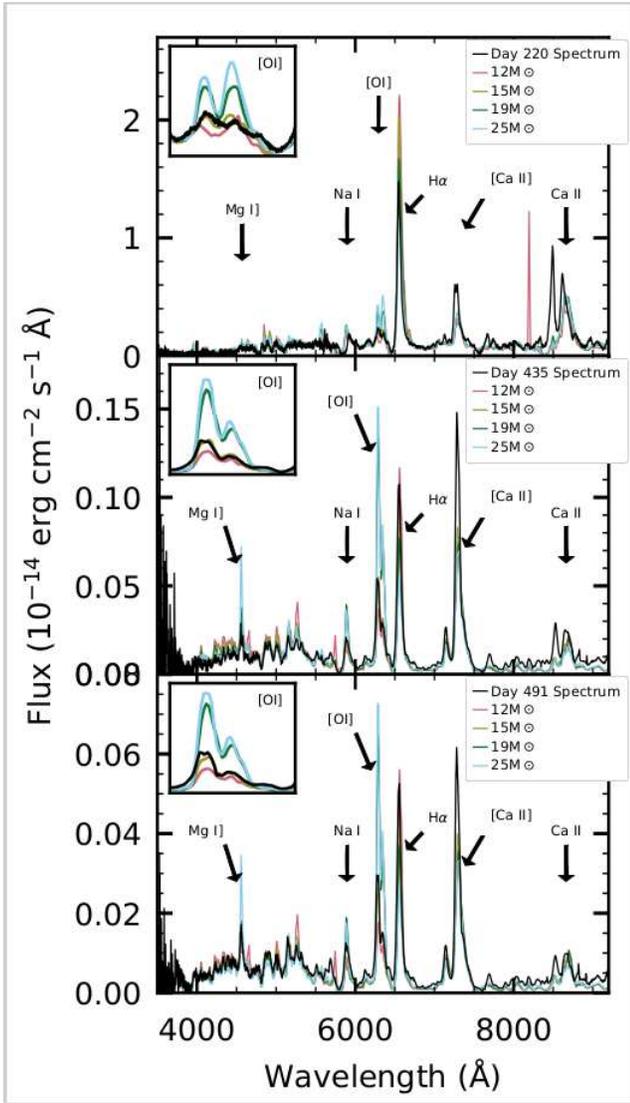}
\caption{The nebular spectra of SN 2017eaw (black) observed 220 (top panel), 435 (middle panel), and 491 (bottom panel) days after explosion.
The model spectra of \citet{2012jerkstrand} are scaled by the integrated flux at each epoch such that the continuum of the observed spectrum and model spectrum are aligned for each 12 (pink), 15 (yellow), 19 (green), and 25 (blue) $\rm{M_{\odot}}$ progenitor. 
The [OI] line doublet ($\rm{\lambda\lambda 6300, 6364}$) for the model and the observed spectra at each epoch is shown in the insets.
The observed spectrum matches the 15 $\rm{M_{\odot}}$ model at all three epochs implying that this is the $\rm{M_{ZAMS}}$ of the progenitor.}
\label{fig:nebular}
\end{center}
\end{figure}

Using this method we find the progenitor mass to be 15 $\rm{M_{\odot}}$ for the first nebular spectrum and 12 $\rm{M_{\odot}}$ for the two later spectra. 
However, we find that the blue part of the continuum in the last two observed spectra is noticeably below the continuum in the models.
For this reason, we scale the models empirically by the ratio of the integrated flux of the observed and model spectra.
This produces a much better alignment of the continuum on the blue side of the spectrum and a consistent progenitor mass of $\sim 15$ $\rm{M_{\odot}}$ for all three spectra.
The empirically scaled model spectra are plotted with the observed spectra in Figure \ref{fig:nebular}. 
The inset in each panel shows the oxygen doublet in detail.
As a sanity check, we use the empirical scale factor at each epoch and Equation \ref{eqn:NiMass} to compute the inferred nickel mass of SN 2017eaw. 
We find values of 0.025-0.036 $\rm{M_{\odot}}$ for the 15 $\rm{M_{\odot}}$ model, reasonably close to the value found in Section 3.5.

\subsection{Modeling the bolometric light curve}\label{bol}

\begin{figure*}
\begin{center}
\leavevmode
\includegraphics[width=.45\textwidth]{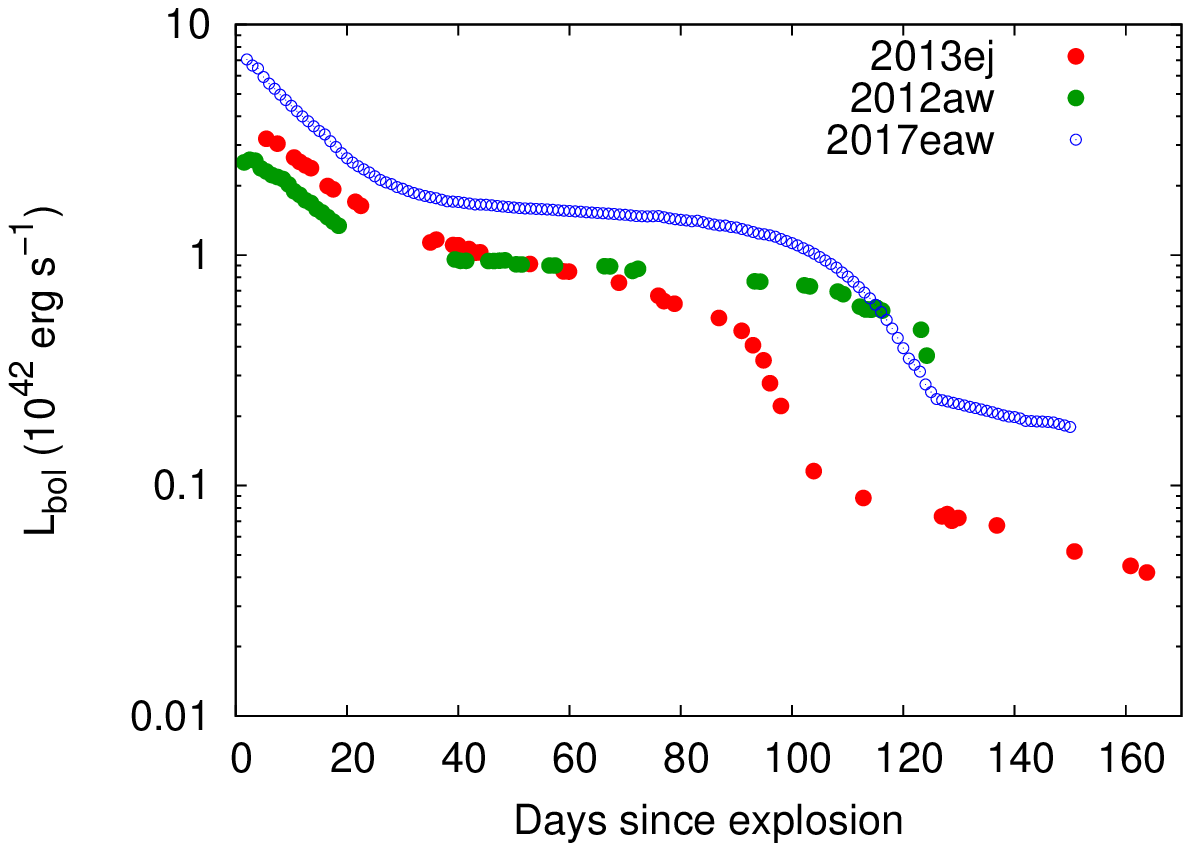}
\includegraphics[width=.45\textwidth]{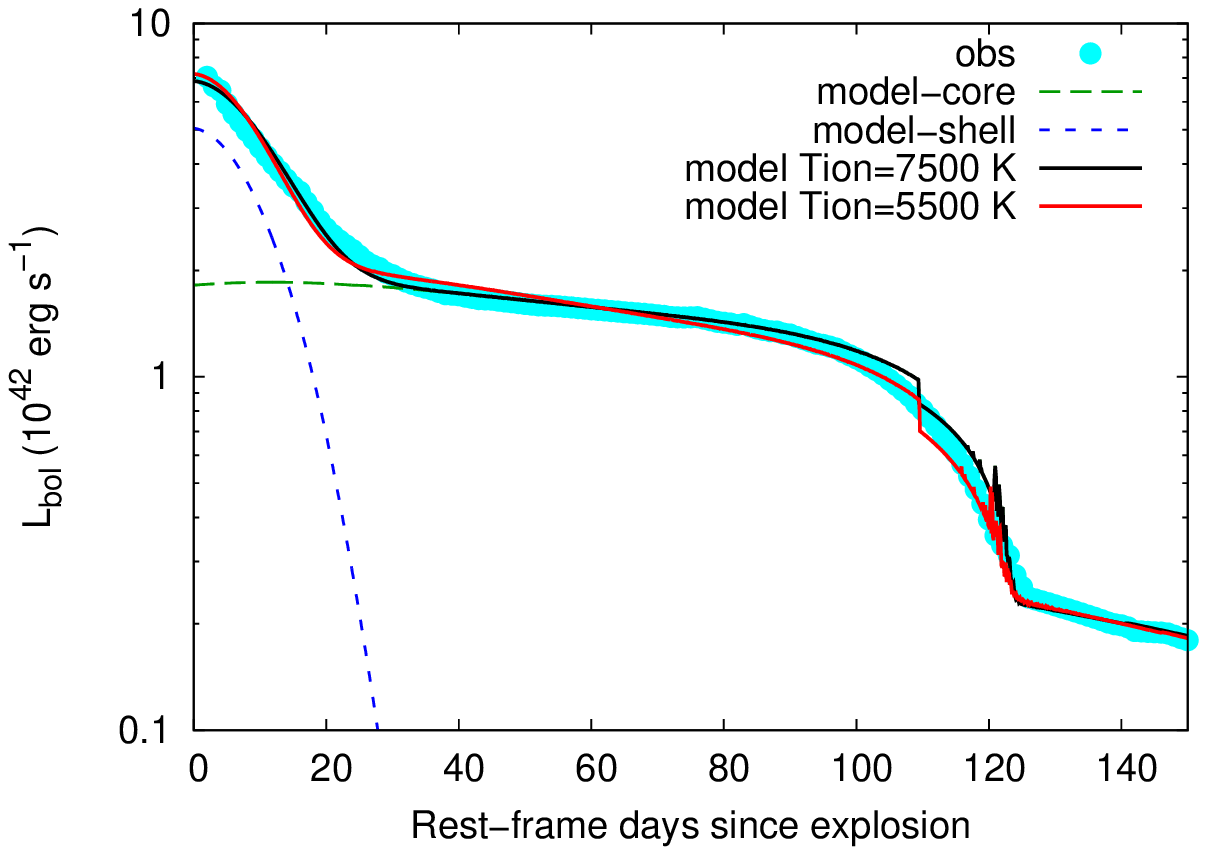}
\end{center}
\caption{Left panel: The bolometric light curve of SN~2017eaw (including the UV-contribution from {\it Swift}) compared to that of SNe 2012aw \citep{Bose13} and 2013ej \citep{Dhungana16}. Right panel: Fitting of two-component Arnett-Fu models (lines) to the bolometric light curve of SN~2017eaw (circles). The contributions from the``core'' and the ``shell'' part of the ejecta are plotted with long- and short-dashed lines, respectively.
The kink in the model light curves at the end of the plateau phase is a numerical artifact related to the finite resolution of the grid used to locate the recombination front.}
\label{fig:Lbol}
\end{figure*}


\begin{table}[]
   \caption{Parameters of the best-fit two-component Arnett-Fu models. See text for explanation. Parameters for SNe 2012aw and 2013 are adopted from \citet{Nagy16}.}
    \label{tab:Lbol-model}
    \centering
    \begin{tabular}{l|cc|c|c}
   \hline
   ~ & \multicolumn{2}{c}{SN 2017eaw} & SN 2012aw & SN 2013ej \\
    \hline
     $T_{ion}$ (K) & $7,500$ & $5,500$ & $7,500$ & $7,500$ \\
     \hline
    \multicolumn{5}{c}{''Core''} \\
    \hline
    $R_0$ ($10^{13}$ cm) & 3.3 & 2.0 & 2.9 & 2.9 \\
    $M_{\text{ej}}$ (M$_\odot$) & 14.3 & 14.6 & 20.0 & 10.0 \\
    $M_{\text{Ni}}$ (M$_\odot$) & 0.045 & 0.046 & 0.056 & 0.020 \\
    $E_{\text{tot}}$ (10$^{51}$ erg) & 2.70 & 4.87 & 2.20 & 1.45 \\
    $E_{\text{kin}}/E_{\text{th}}$ & 1.99 & 1.92 & 2.67 & 3.14 \\
    $\kappa$ (cm$^2$~g$^{-1}$) & 0.24 & 0.24 & 0.13 & 0.20 \\
    $v_{\text{sc}}$ (km~s$^{-1}$) & 4583 & 6033 & 3660 & 4290 \\
    $t_{\text{lc}}$ (d) & 98.2 & 86.7 & 95.8 & 77.6 \\
    \hline
    \multicolumn{5}{c}{''Shell''} \\
    \hline
    $R_0$ ($10^{13}$ cm) & 4.9 & 4.5 & 4.5 & 6.8 \\
    $M_{\text{ej}}$ (M$_\odot$) & 0.37 & 0.33 & 1.0 & 0.6 \\
    $M_{\text{Ni}}$ (M$_\odot$) & -- & -- & -- & -- \\
    $E_{\text{tot}}$ (10$^{51}$ erg) & 0.18 & 0.20 & 1.0 & 1.39 \\
    $E_{\text{kin}}/E_{\text{th}}$ & 1.93 & 2.22 & 9.0 & 14.4 \\
    $\kappa$ (cm$^2$~g$^{-1}$) & 0.34 & 0.34 & 0.40 & 0.40 \\
    \hline
    \end{tabular}
\end{table}

The quasi-bolometric light curve, including the contributions from the UV and IR, is constructed by applying the same technique as described in \citet{Dhungana16}. 
After correcting the data for the total interstellar extinction (assuming $E(B-V) = 0.41$ mag, see Sect.~\ref{sec:ana}) and converting the magnitudes to physical fluxes, the SEDs are integrated along wavelength using the trapezoidal rule. Note that computing proper extinction correction for the {\it Swift} UV data is not as simple as for the optical data \citep{brown10, brown16}. Here we follow a somewhat simplified procedure by assuming constant extinction coefficients for the {\it UVOT} filters as determined by \citet{brown10} for the Type II-P SN~1999em (see their Table~14).    
The optical data are integrated between the B-band and the I-band, while the {\it Swift} data are
used to compute the contribution between the B-band and 2000 \AA.
The integrated flux from the unobserved IR bands are taken into account by extrapolating the I-band fluxes with a Rayleigh-Jeans tail and integrating that curve to infinity. Finally, the integrated fluxes are corrected for distance using $D = 6.85$ Mpc (Sect.~\ref{epm}). The resulting quasi-bolometric light curve is
plotted together with those of SN~2012aw \citep{Bose13} and SN~2013ej \citep{Dhungana16} in Figure~\ref{fig:Lbol}.

Radiation-diffusion models \citep{AF1,AF2} for the bolometric light curve are computed with the LC2.2 code\footnote{\href{http://titan.physx.u-szeged.hu/~nagyandi/LC2.2/}{\tt http://titan.physx.u-szeged.hu/$\sim$nagyandi/LC2.2/}} \citep{Nagy16} that assumes a two-component ejecta having an inner, denser, more massive envelope \citep[referred to as the ``core'', following][]{Nagy16} and an outer, less massive, lower density ``shell''. The code takes into account H- or He-recombination in the same way as in \citet{AF1}. More details on the physics of these models can be found in \citep{Nagy16}. Briefly, the main difference between the two components is that the outer, low-density ``shell'' is assumed to be powered only by shock heating (and not by $^{56}$Ni-decay). 

The right panel in Figure~\ref{fig:Lbol} plots the observed bolometric light curve together with several models that are found to show similar luminosity evolution. The model parameters are listed in Table~\ref{tab:Lbol-model}: the progenitor radius $R_0$ (in $10^{13}$ cm units), the mass of the ejecta ($M_{\text{ej}}$, in M$_\odot$), the initial mass of the radioactive $^{56}$Ni ($M_{\text{Ni}}$, in M$_\odot$), the total energy ($E_{\text{tot}}$, in $10^{51}$ erg) and the ratio of the thermal ($E_{\text{th}}$) and kinetic energy ($E_{\text{kin}}$) of the ejecta, the opacity ($\kappa$, in cm$^2$~g$^{-1}$), the scaling velocity ($v_{\text{sc}}$, in km~s$^{-1}$), and the light-curve timescale ($t_{\text{lc}}$, in days) \citep[the geometric mean of the expansion and the diffusion timescales as defined by][]{AF1}. The last two parameters are derived from the previous ones listed above.

The density profiles for all models are assumed to be constant as in \citet{AF1}. The ``shell'' is assumed to be hydrogen-rich, thus, the usual $\kappa = 0.34$ cm$^2$~g$^{-1}$ (which is equal to the Thompson-scattering opacity of a fully ionized solar-like plasma) is adopted as the opacity in this component. Since the ``core'' is more abundant in heavier elements, its Thompson-scattering opacity could be somewhat lower, thus, $\kappa \sim$0.24 cm$^2$~g$^{-1}$ is adopted there \citep{Nagy18}. For the recombination temperature two different values ($T_{ion} = 5500$ K and 7500 K) are assumed as lower and upper limits that roughly bracket the recombination temperature in a hydrogen-rich and a hydrogen-depleted atmosphere, respectively.

It has to be noted that, as it is described in detail in \citet{Nagy16}, uncertainty of the explosion
date can be a serious limitation during this modeling process; at the same time, the $\pm$1 day uncertainty in $t_0$ of SN~2017eaw (see Section \ref{sec:ana}) may cause only a $\sim$5\% relative error in the derived physical parameters. Moreover, the mass estimate based on LC modeling has a well-known degeneracy with the assumed (constant) optical opacity and the kinetic energy; these parameters are correlated via the $t_{\text{lc}}$ parameter as $t_{\text{lc}} \sim \kappa M_{\text{ej}}^{3/2} E_{\text{kin}}^{-1/2}$. Thus, for the same light curve but a slightly different opacity than in Table~\ref{tab:Lbol-model} one can get different ejecta masses. For example, if using $\kappa = 0.33$ cm$^2$~g$^{-1}$ in the ``core'' one would get a factor of 0.8 lower mass, i.e. $M_{\text{ej}} \sim 11$ M$_\odot$.
Therefore, a more realistic estimate for the uncertainty of the derived ejecta mass is at least $\pm 3$ M$_\odot$, which takes into account the correlation between these key parameters.

In Table~\ref{tab:Lbol-model}, parameters for SNe 2012aw and 2013ej calculated with the same two-component model \citep[adopted from][]{Nagy16} are also shown. While slightly different opacities have been used during the modeling of the three SNe, the main parameters are similar. This suggests that the three progenitors were probably similar to each other. However, as can be also seen in Figure~\ref{fig:Lbol}, the early-time bolometric fluxes are larger in the case of SN~2017eaw, which can be modeled with a higher total energy in the ``core'' (or, can be the sign of early-time CSM interaction).
Further implications for the light curve models are discussed in Section~\ref{disc}.


\section{Discussion}\label{disc}

From the observations and models presented in the previous sections, we draw a comprehensive picture of SN~2017eaw, its progenitor and circumstellar environment. 

\subsection{Mass of the progenitor, explosion parameters} \label{prog}

The model parameters shown in Table~\ref{tab:Lbol-model} imply a relatively, but not unusually massive Type II-P SN ejecta: the total (``core'' + ``shell'') envelope mass is $\sim 14.5 \pm 3.0$ M$_\odot$. Assuming $\sim$1.4 M$_{\odot}$ for the mass of the remaining neutron star, this is in a good agreement with the progenitor mass of $\sim$15 M$_{\odot}$ inferred from our modeling of the nebular spectra (during which we used 12, 15, 19 and 25 M$_{\odot}$ models). 

\citet{Tsvetkov18} applied the multi-group radiation-hydro code STELLA \citep{Blinnikov98,Blinnikov00,Blinnikov06} to model their UBVRI light curves for SN~2017eaw. They obtained $R_0$ = 600 $R_{\odot}$ ($\sim$4.2 $\times 10^{13}$ cm), $M_{\text{ej}}$ = 23 $M_{\odot}$, $M_{\text{Ni}}$= 0.05 $M_{\odot}$, and $E_{\text{kin}}$= 2.0 foe, which are consistent with ours in Table~\ref{tab:Lbol-model}. The only exception is their $\sim$1.5 times higher total ejecta mass. It is a well-known issue that radiation hydro codes sometimes give higher envelope masses than simple semi-analytic models \citep[e.g][]{Nagy16}. Given the uncertainties of the parameters from the semi-analytic models, which use a lot of approximations, such a difference within a factor of 2 is not unexpected. Note that our derived mass is more consistent with the mass estimates for the observed progenitor of SN~2017eaw \citep[$14 \pm 3$ M$_\odot$;][]{VD17,Kilpatrick18}, as well as with the results of \citet{Rho18} who compare their near-IR spectra with the models of \citet{Dessart17,Dessart18} and conclude a progenitor mass of 15 M$_\odot$ (with M$_{\text{ej}}$ of 12.5 M$_{\odot}$ and  M$_{\text{Ni}}$ of 0.084 M$_\odot$). Note also that \citet{Williams18} give a much lower value for the progenitor mass ($\sim 8.8^{+2}_{-0.2}$ M$_\odot$) from modeling the local stellar population, but, from their Figure 2, this looks more like being a lower limit. 

The initial ``shell'' radius of $\sim 4.5 \times 10^{13}$ cm is in very good agreement with the conclusion by \citet{Kilpatrick18} that the progenitor of SN~2017eaw was a red supergiant (RSG) star. 

\subsection{Early-time circumstellar interaction, mass-loss of the progenitor}\label{csm}

\begin{figure*}
\begin{center}
\leavevmode
\includegraphics{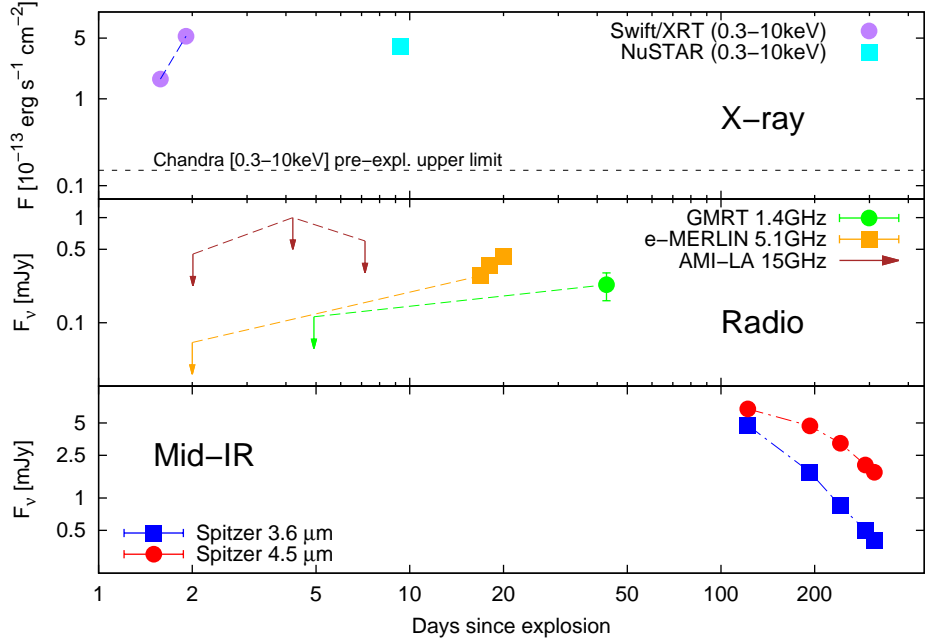}
\end{center}
\caption{X-ray, radio and mid-IR data (together with published non-detections) of SN~2017eaw. Sources of data are listed in the text.}
\label{fig:xrir}
\end{figure*}

\begin{figure}
\begin{center}
\leavevmode
\includegraphics[width=.45\textwidth]{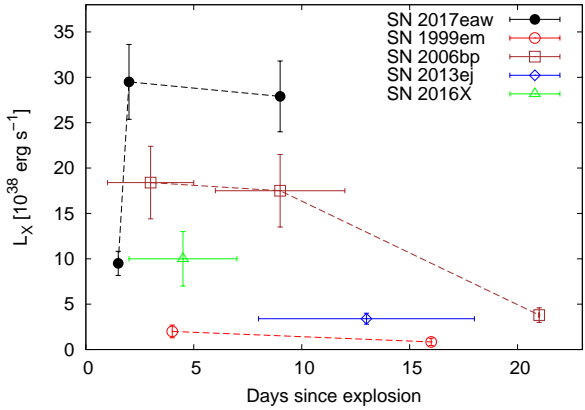}
\end{center}
\caption{Early-time X-ray luminosities measured in Type II-P SNe: 2017eaw (0.3$-$10 keV), 1999em \citep[0.4$-$8 keV, assuming a distance of $D$=12.5 Mpc,][]{Pooley02}, 2006bp \citep[0.4$-$8 keV, assuming a distance of $D$=14.9 Mpc,][]{Immler07}, SN II-P/II-L 2013ej \citep[0.5$-$8 keV, assuming a distance of $D$=9.57 Mpc,][]{Chakraborti16}, and 2016X \citep[0.3-10 keV,][]{Grupe16}. For the latter three objects, horizontal error bars indicate the (not contiguous) periods covered by the observations.}
\label{fig:xray_comp}
\end{figure}

Being one of the nearest SNe in the last decade, SN~2017eaw has been intensively followed up in both X-ray and radio bands in order to look for signs of possible early-time circumstellar interaction. Within only a day after discovery, the SN was positively detected in X-rays with {\it Swift}/X-Ray Telescope (XRT) at two different epochs, showing a significant early brightening in the 0.3-10 keV range by \citet{Kong17}, who also gave a (much lower) pre-explosion upper flux limit based on archival {\it Chandra} images of the SN site. A few days later the SN was also observed with the Nuclear Spectroscopic Telescope Array (NuSTAR) telescope \citep{Grefensetette17}, detecting a slightly lower flux between 0.3-10 keV than previously found by {\it Swift}. Moreover, the latter authors also reported the presence of a line from ionized Fe around 6.65 keV, which implies the presence of shock-heated ejecta. Unfortunately, no further X-ray observations have been published to date; however, the SN has been also detected with the AstroSat/UV Imaging Telescope (UVIT) in the far-UV channel $\sim 2$ weeks after explosion \citep{Misra17}.

The top panel of Fig.~\ref{fig:xrir} presents all the published X-ray fluxes measured for SN~2017eaw. Using the distance of $D$=6.85 Mpc (see above), we determined the integrated unabsorbed X-ray luminosities for SN~2017eaw in the 0.3$-$10 keV range as $L_X$= 9.5$\times$10$^{38}$, 29.5$\times$10$^{38}$, and 27.9$\times$10$^{38}$ erg s$^{-1}$ at +1.5d, +2d and +9d, respectively. 

In order to compare these X-ray luminosities with those of other Type II-P SNe, we have collected the available data from the literature.
There are only a few SNe II-P that were observed in X-rays at such early phases. Fig.~\ref{fig:xray_comp} shows the X-ray luminosities ($L_X$) of SN 2017eaw, together with that of SNe 1999em \citep{Pooley02}, 2006bp \citep{Immler07}, the Type II-P/II-L 2013ej \citep{Chakraborti16} and 2016X \citep{Grupe16}. It is seen that the $L_X$ for SN~2017eaw (measured in the 0.3$-$10 keV range) is a few times higher than that of SNe~2006bp and 2016X and much higher than for SNe~1999em and 2013ej (however, the latter objects were observed only in the 0.4/0.5$-$8 keV range). Note that if we use $D\sim$5 Mpc for the distance of SN~2017eaw, we get $\sim$50\% lower $L_X$ values, which are at the same level as that of SN~2006bp but are still much larger than other published values regarding Type II-P SNe.
We also note that SN~2013fs was also followed up by {\it Swift}/XRT in the first $\sim$25 days, and a combined upper limit of $L_X<$4.7 $\times$ 10$^{40}$ erg s$^{-1}$ was determined by \citet{Yaron17}; however, as those authors noted, most of the estimated flux may originate from the host galaxy instead of the SN, because of the relatively large distance.

While the level of the early-time X-ray emission measured in SN~2017eaw is much lower than usually found in Type IIn or other strongly interacting SNe \citep[see e.g.][]{Chevalier17}, its origin can be best explained by assuming a moderate interaction between the SN shock and the ambient circumstellar medium. As showed by, e.g., \citet{Immler07} in the case of SN~2006bp, other possible sources (radioactive decay products of the ejecta, or inverse Compton scattering of photospheric photons off relativistic electrons produced by the explosion) can be responsible for only a fraction of the observed X-ray emission.

Regarding radio observations, all the early notifications reported non-detections at 1.4, 5.1, and 15 GHz \citep{Nayana17a,Argo17a,Bright17,Mooley17}. Later, subsequent observations at the two lower frequencies resulted in positive detections: on three epochs between +17-20 days at 5.1 GHz \citep[using e-MERLIN,][]{Argo17b}, and at +42 days at 1.4 GHz \citep[using Giant Metrewave Radio Telescope, GMRT,][]{Nayana17b}. All of these data are shown in the middle panel of Fig. \ref{fig:xrir}.
Since there are only a few observations of SN~2017eaw (obtained at three different frequencies), detailed modeling of the radio LCs can not be accomplished. Nevertheless, the estimated radio luminosities at 5.1 GHz are $\sim$10$^{26}$ erg s$^{-1}$ Hz$^{-1}$, which agree well with the peak luminosities of other Type II-P SNe assumed to go through moderate CSM interaction \citep[see e.g.][]{Chevalier06}. 

Beyond X-ray and radio data, optical LCs may also indicate the presence of early-time CSM interaction. As has been found by \citet{Moriya11,Moriya17,Moriya18} and \citet{Morozova17,Morozova18}, the mass-loss processes of the presumed RSG progenitors may significantly affect the optical LCs of Type II(-P) SNe, especially during the first few days. As mentioned in Section \ref{lc} and seen in Fig.\ref{fig:phot_comp1}, SN~2017eaw shows a low-amplitude, early bump peaking at $\sim$6-7 days after explosion in the optical bands (most obviously in the I-band and weakening toward shorter wavelengths). This phenomenon is quite similar to the one observed in SN~2013fs, and is supposed to be caused by the interaction between the expanding SN ejecta and the ambient matter originating from the pre-explosion RSG wind. 

There is a long-term debate on the amount, density distribution and geometry of the circumstellar material surrounding SN progenitors, as well as on the pre-explosion mass-loss history of RSG stars. In the basic (perhaps simplistic) framework, RSG stars have slow ($v_{\text{w}} \sim$10$-$20 km s$^{-1}$), steady winds resulting in mass loss rates ($\dot{M}$) of 10$^{-6}-$10$^{-5} M_{\odot}$ yr$^{-1}$. At the same time, mass loss may become enhanced just before the explosion, resulting in a (more or less) compact and dense inner region in the CSM: $\dot{M} \sim$ 10$^{-4}-$10$^{-2} M_{\odot}$ yr$^{-1}$ and $R \sim 10^4 R_{\odot}$ \citep{Moriya11,Moriya17,Moriya18}, or, even $\dot{M} \sim$ 10$^{-2}-$15 $M_{\odot}$ yr$^{-1}$ and $R \sim 2000-3000 R_{\odot}$ \citep{Morozova17,Morozova18}. On the other hand, it is also possible that the shock simply breaks out from a very extended RSG atmosphere; in this case the ``superwind'' description may not be adequate \citep[see e.g.][]{Dessart17}.

In the case of SN~2017eaw, \citet{Kilpatrick18} carried out a detailed investigation on the pre-explosion environment of the assumed progenitor using archived HST and {\it Spitzer} data (see above). They suggested the presence of a low-mass ($M >2 \times 10^{-5} M_{\odot}$), extended ($R = 4000 R_{\odot}$) dust shell enshrouding the progenitor site. They also estimated the mass-loss rate by applying the method of \citet{Kochanek12} and obtained $\dot{M} \sim9 \times 10^{-7} M_{\odot}$ yr$^{-1}$. 

Applying the method described in \citet{Kochanek12} \citep[adopted from][]{Chevalier03}, and using the parameters of our two LC models in Table~\ref{tab:Lbol-model} combined with the X-ray luminosities ($L_X$) given above we can derive another constraint for $\dot{M}$ via Eq. 4 of \citet{Kochanek12},

\begin{eqnarray}
    L_X \simeq 1.63 \times 10^7 E_{51}^{27/20} M_{e10}^{-21/20} \dot{M}_{-4}^{7/10} v_{\text{w}10}^{-7/10} \nonumber \\ 
    \times t_1^{-3/10} L_{\odot}, \label{eq:k12}
\end{eqnarray}

\noindent where the total energy of the SN is $E = 10^{51} E_{51}$ ergs, the ejected mass is $M_{ej} = 10 M_{e10} M_{\odot}$, $\dot{M} = 10^{-4} \dot{M}_{-4} M_{\odot} \text{yr}^{-1}$, $v_{\text{w}} = 10 v_{\text{w}10}$ km s$^{-1}$, and $t_1$ is the elapsed time in days (+5 and +9 days in this case).

Assuming $v_{\text{w}1}$= 10 km s$^{-1}$ for the RSG wind velocity, we got $\dot{M} \sim3 \times 10^{-7}$ and $\sim$1 $\times 10^{-6} M_{\odot}$ yr$^{-1}$ for the two models listed in Table \ref{tab:Lbol-model}. Both of these values are consistent with the mass-loss rate estimated by \citet{Kilpatrick18}. On the other hand, they are orders of magnitude lower than the mass-loss rates estimated by Moriya et al. and Morozova et al. via LC-modeling, or the value of $\dot{M} \sim$10$^{-3} M_{\odot}$ yr$^{-1}$ derived by \citet{Yaron17} for SN~2013fs based on modeling the early-time spectroscopic emission features. We note that from Eq. \ref{eq:k12} it would be necessary to have $L_X \sim$10$^{41}$ erg s$^{-1}$ to get $\dot{M} \sim$10$^{-3} M_{\odot}$ yr$^{-1}$ for the mass-loss rate of SN~2017eaw. Such high-level X-ray luminosity has been measured only in strongly interacting SNe IIn to date.

Nevertheless, while it seems to be a serious contradiction, some caveats in the above analysis must be mentioned. First, the mass-loss rate we estimated from X-ray luminosities (beyond the intrinsic uncertainties of the model) is based on the assumption that the dominating counterpart of $L_X$ is the cooling of the reverse shock and its softer emission dominates the observable X-ray flux; however, as \citet{Grefensetette17} noted, the analysis of the +9d X-ray spectrum of SN~2017eaw indicates a hard X-ray spectrum having detectable flux up to 30 keV (they also mention that the contribution of the 10-30 keV counterpart to the total $L_X$ is $<$10 percent). 
Second, the radius of the dust-rich pre-explosion region ($\sim4000 R_{\odot}$) derived by \citet{Kilpatrick18} is in good agreement with the general estimation given by \citet{Morozova17,Morozova18} for the size of the cocoons of CSM around SNe II-P; the only difference is that the latter authors suggest the presence of a much denser environment. Signs of such a dense gas/dust shell are not seen in the combined optical-IR SED of the assumed progenitor of SN~2017eaw. High-resolution near-IR spectroscopy also did not detect narrow lines that may be an indication of CSM gas \citep[see][]{Rho18}. However, it is also true that these data do not cover the region of cold ($T \lesssim$ 300K) dust. Third, a common problem is the geometry; while the models generally assume a spherically symmetric CSM, it may also take the form of a thick disk, or a more complex structure of the inflated RSG envelope material \citep[see e.g.][and references therein]{Dessart17,Morozova17}. The actual shape of the CSM cloud may also have a strong influence on the estimated parameters. All of these uncertainties point toward the need for further observations and more detailed modeling in order to better understand the role of nearby CSM around Type II-P SNe as well as the mass-loss history of their RSG progenitors.
 
Moreover, it is also an intriguing question why we did not see any narrow (``flash-ionized'') emission lines in the earliest spectra of SN~2017eaw, unlike in the early ($<$5d) spectra of SN~2013fs and several other interacting Type II SNe \citep{Quimby06,Khazov16,Yaron17}. While this problem also requires further data and modeling, the geometry and/or clumpiness of the CSM may also play a role here.

\subsection{Possible signs of late-time dust formation}\label{dust}

\begin{figure}
\begin{center}
\leavevmode
\includegraphics[width=.45\textwidth]{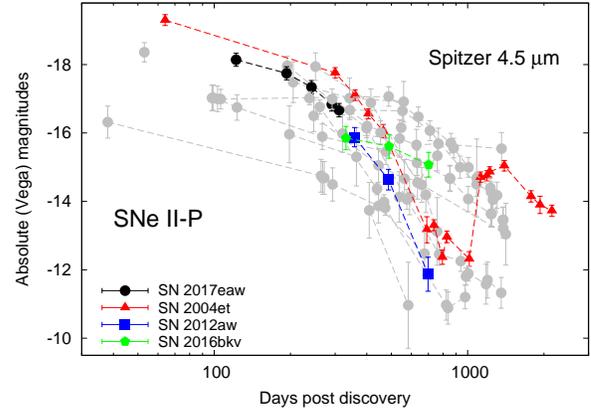}
\end{center}
\caption{Mid-IR evolution of SN~2017eaw compared to that of other normal (gray) or interacting Type II-P SNe. 4.5 $\mu$m magnitudes of SNe 2017eaw and 2016bkv come from this work, while the source of other values is \citet{Szalai18}.}
\label{fig:mir}
\end{figure}

\begin{figure*}
\begin{center}
\leavevmode
\includegraphics[width=.45\textwidth]{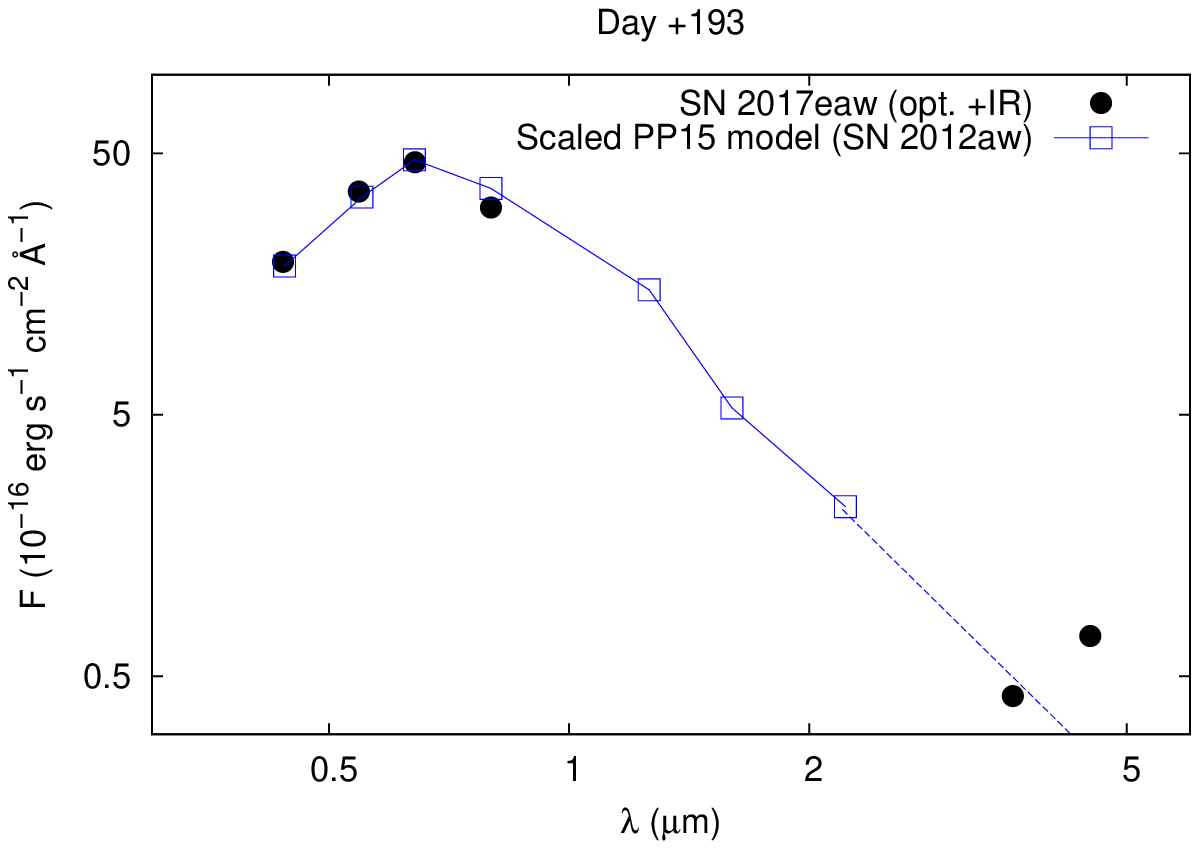} \hspace{5mm}
\includegraphics[width=.45\textwidth]{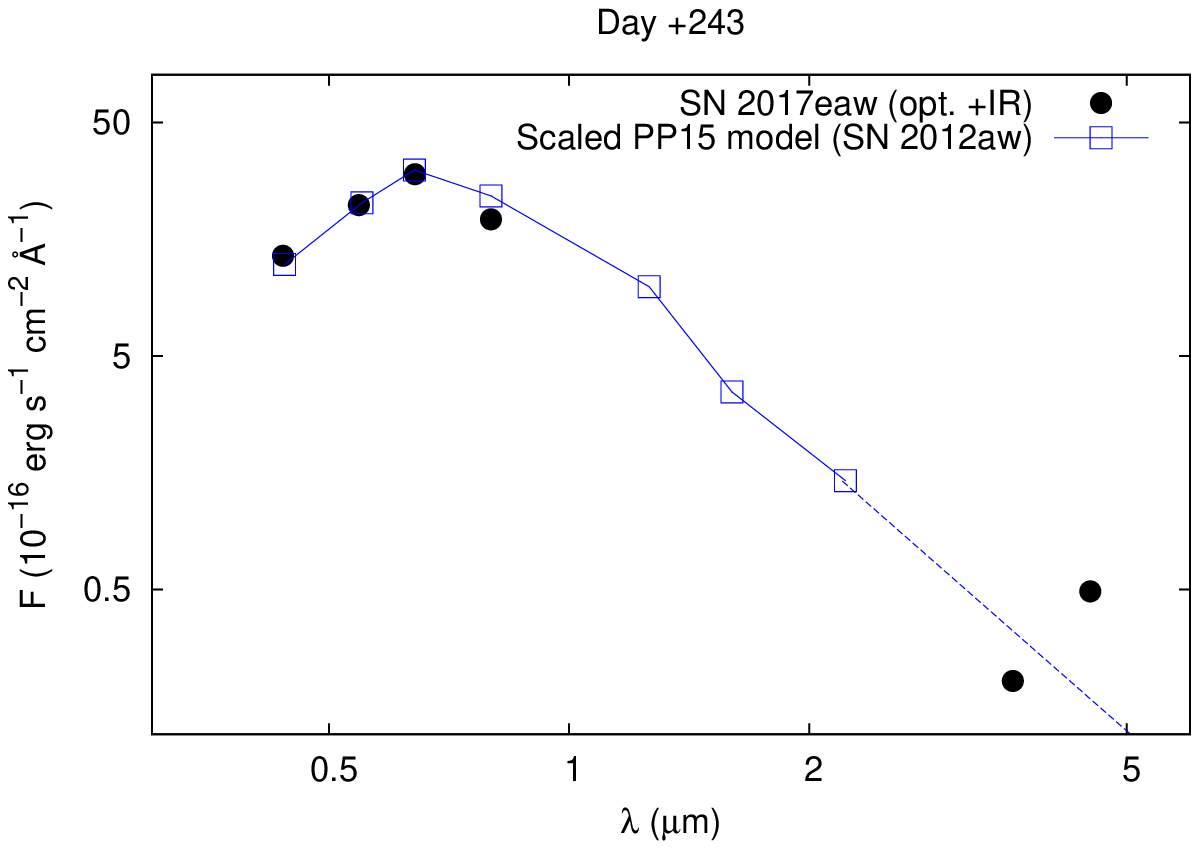}
\end{center}
\caption{Optical-IR combined SED of SN~2017eaw (filled circles) compared to the scaled model SEDs of SN 2012aw calculated from \citet{PP15} (PP15) (open rectangles).}
\label{fig:sed}
\end{figure*}

SN~2017eaw was also followed by {\it Spitzer} as the target of two different programs (PID 13239, PI: K. Krafton; PID 13053/SPIRITS, PI: M. Kasliwal). We have downloaded the public data from the {\it Spitzer} Heritage Archive (SHA)\footnote{http://sha.ipac.caltech.edu} and carried out simple aperture photometry on post-basic calibrated (PBCD) images. The SN appears as a bright, continuously fading object in both IRAC 3.6 and 4.5 $\mu$m channels. We present the results from our photometry at the bottom panel of Fig. \ref{fig:xrir}, while Fig. \ref{fig:mir} shows the 4.5 $\mu$m light curve of SN~2017eaw compared to those of other Type II-P SNe \citep[most of these data are adopted from][and references therein, while for SN~2016bkv we carried out a similar process as above]{Szalai18}.

During the observed period, the mid-IR evolution of SN~2017eaw seems to be similar to that of the highlighted Type II-P events (SNe 2004et, 2012aw, 2016bkv) that do not show strong, direct signs of dust formation (e.g. re-brightening in the mid-IR after several hundred days). At the same time, comparing the combined optical-IR SEDs of SN~2017eaw taken at +193d and +243d (Fig.\ref{fig:sed}) to model SEDs of SN~2012aw \citep{PP15} there is a clear mid-IR excess in the 4.5 $\mu$ channel on both epochs. Here we note that in Type II-P SNe the 4.5 $\mu$m flux may also be contaminated by the 1$-$0 vibrational band of CO at 4.65 $\mu$m, which can influence the SED modeling at epochs $\lesssim$500d \citep[see e.g.][]{Kotak05,S11}.

These results seem to strengthen that of \citet{Rho18}, who, based on the detailed analysis of ground-based near-IR spectra, suggest ongoing (moderate) molecule (CO) and dust formation between $\sim$125-205 days.
A more detailed study of dust and molecule formation in SN~2017eaw has been very recently published by \citet{Tinyanont19}; based on the analysis of the full {\it Spitzer} data set and near-IR photometry and spectroscopy up to $\sim$550d, these authors present similar conclusions to ours.
We also note that, as can be seen, e.g., in the case of SN~2004et in Fig. \ref{fig:mir}, dust formation can become more significant at later times ($\sim$800$-$1000d after explosion), probably due to an interaction between the forward shock and a denser CSM shell \citep[see e.g.][and references therein]{Szalai18}. 

\section{Conclusions} \label{sec:concl}

SN~2017eaw, one of the most nearby supernovae to appear in this decade, is a Type II-P explosion that shows early-time, moderate CSM interaction. We made a comprehensive study of this SN using multi-colour optical photometry and high-quality optical spectroscopy starting at very early epochs and extending into the early nebular phase, near-UV and near-IR spectra, early-time X-ray and radio detections, and late-time mid-IR photometry.

We derived a new distance to the host galaxy, NGC~6946, after combining various distance estimates including the EPM analysis of the combined data of SNe~2017eaw and 2004et. The final distance, $D \sim 6.85$ $\pm 0.63$  Mpc,  disfavors the previous measurements from SN~2004et only that all gave $\sim$30 percent lower distances. 

During the whole period covered by the observations, the evolution of SN~2017eaw seems to be similar to those of some other ``normal'' Type II-P SNe (2004et, 2012aw). However, SN~2017eaw shows a small, early bump peaking at $\sim$6-7 days after explosion in all optical bands, which resembles the early light curve of SN 2013fs and is presumably the sign of early-time circumstellar interaction. Nevertheless, it is an intriguing question why we did not see any narrow (``flash-ionized'') emission features in the earliest optical spectra of SN~2017eaw; the solution of this problem might be related to different geometries and/or clumpiness of CSM around the two objects.

We modeled the quasi-bolometric LC of SN~2017eaw using a two-component radiation-diffusion model and estimated the basic physical parameters of the explosion and the ejecta. We also carried out modeling of the nebular spectra using different progenitor masses. The results agree well with the previous findings of a RSG progenitor star with a mass of $\sim$15-16 M$_{\odot}$.

We also used these calculated explosion parameters -- together with early-phase X-ray luminosities -- to derive the mass-loss rate of the progenitor. We got $\dot{M} \sim3 \times 10^{-7} -$ $1\times 10^{-6} M_{\odot}$ yr$^{-1}$; these values agrees well with that estimated by \citet{Kilpatrick18} based on the opacity of the dust shell enshrouding the progenitor, but it is orders of magnitude lower than the generally estimated values for Type II-P SNe from from early-phase LC modeling \citep{Moriya11,Moriya17,Moriya18,Morozova17,Morozova18}. We discussed several factors that may seriously influence the various estimations of $\dot{M}$ including the limitations within the models as well as the simplifying assumptions on the geometry and clumpiness of the nearby CSM. 

Finally, we also studied the available mid-IR data of SN~2017eaw. The combined optical-IR SEDs show a clear mid-IR excess at +193 and +243 days, which are consistent with the  results of \citet{Rho18} and \citet{Tinyanont19} on the (moderate) dust formation in the vicinity of this SN.

\acknowledgments
We thank the thorough work of an anonymous referee that helped us in improving this paper.
This work is part of the project “Transient Astrophysical Objects” GINOP-2-3-2-15-2016-00033 of the National Research, Development and Innovation Office (NKFIH), Hungary, funded by the European Union, and is also supported by th New National Excellence Program (UNKP-17-4) of the Ministry of Human Capacities of Hungary. TS has received funding from the Hungarian NKFIH/OTKA PD-112325 Grant.
Research by DJS is supported by NSF grants AST-1821967, 1821987, 1813708 and 1813466. JCW is supported in part by NSF grant 1813825. The work of OP has been supported by the PRIMUS/SCI/17 award from Charles University in Prague. EYH, CA and MS acknowledge the support provided vy the National Science Foundation under Grant No. AST-1613472.
This work makes use of observations from Las Cumbres Observatory network. DAH, GH and CM are supported by NSF grant AST-1313484.
MS, EYH and DJS are visiting astronomers at the Infrared Telescope Facility, which is operated by the University of Hawaii under contract NNH14CK55B with the National Aeronautics and Space Administration.
X. W. is supported by the National Natural Science Foundation of China (NSFC grants 11325313, 11633002, and 11761141001), and the National Program on Key Research and Development Project (grant 2016YFA0400803). 
Some of the data presented herein were obtained at the W. M. Keck Observatory, which is operated as a scientific partnership among the California Institute of Technology, the University of California and the National Aeronautics and Space Administration. The Observatory was made possible by the generous financial support of the W. M. Keck Foundation.
The authors wish to recognize and acknowledge the very significant cultural role and reverence that the summit of Mauna Kea has always had within the indigenous Hawaiian community.  We are most fortunate to have the opportunity to conduct observations from this mountain.
Based on observations obtained at the Gemini Observatory (processed using the Gemini IRAF package), which is operated by the Association of Universities for Research in Astronomy, Inc., under a cooperative agreement with the NSF on behalf of the Gemini partnership: the National Science Foundation (United States), National Research Council (Canada), CONICYT (Chile), Ministerio de Ciencia, Tecnolog\'{i}a e Innovaci\'{o}n Productiva (Argentina), Minist\'{e}rio da Ci\^{e}ncia, Tecnologia e Inova\c{c}\~{a}o (Brazil), and Korea Astronomy and Space Science Institute (Republic of Korea).

\vspace{5mm}
\facilities{LCO(1m, 2m), IRTF (SpeX), Konkoly (Schmidt), W. M. Keck Observatory (Keck I LRIS), Gemini Observatory (Gemini North GMOS), Swift(UVOT and XRT), Spitzer(IRAC)}

\software{astropy \citep{astropy13},
          IRAF, HEAsoft, 
          SExtractor \citep{Bertin96},
          Spextool \citep{Cushing04},
          XTELLCOR \citep{Vacca03},
          lcogtsnpipe \citep{Valenti16}
          }




\newpage
\appendix

\section{Photometric and spectroscopic data}
\setcounter{table}{0} 
\renewcommand{\thetable}{\Alph{section}\arabic{table}}

\startlongtable
\begin{deluxetable}{cc|cccccccc}
\tablecaption{Konkoly BVRI photometry of SN~2017eaw \label{tab:phot}}
\tabletypesize{\small}
\tablehead{
\colhead{Date} & \colhead{Epoch} & \colhead{B} & \colhead{$\sigma$B}  & \colhead{V} & \colhead{$\sigma$V}  & \colhead{R} & \colhead{$\sigma$R}  & \colhead{I} & \colhead{$\sigma$I} \\
\colhead{(MJD)} & \colhead{(days)} & \colhead{(mag)} & \colhead{(mag)}  & \colhead{(mag)}  &\colhead{(mag)}  & \colhead{(mag)}  & \colhead{(mag)}  & \colhead{(mag)} & \colhead{(mag)}
}
\startdata
57887.99 & 1.99	& 13.270 & 0.028 & 13.066 & 0.022 & 12.812 & 0.016 & 12.629 & 0.013 \\
57889.83 & 3.83	& 13.202 & 0.032 & 12.886 & 0.031 & 12.567 & 0.026 & 12.340 & 0.012 \\
57890.84 & 4.84	& 13.156 & 0.030 & 12.843 & 0.026 & 12.488 & 0.047 & 12.237 & 0.048 \\
57892.00 & 6.00	& 13.204 & 0.045 & 12.848 & 0.040 & 12.448 & 0.033 & 12.184 & 0.026 \\
57892.98 & 6.98	& 13.186 & 0.022 & 12.796 & 0.024 & 12.405 & 0.028 & 12.165 & 0.023 \\
57894.95 & 8.95	& 13.170 & 0.014 & 12.830 & 0.021 & 12.398 & 0.034 & 12.165 & 0.019 \\
57897.90 & 11.90 & 13.280 & 0.018 & 12.885 & 0.019 & 12.460 & 0.016 & 12.224 & 0.017 \\
57898.96 & 12.96 & 13.305 & 0.021 & 12.915 & 0.021 & 12.470 & 0.028 & 12.245 & 0.016 \\
57900.90 & 14.90 & 13.405 & 0.026 & 12.957 & 0.026 & 12.490 & 0.024 & 12.294 & 0.025 \\
57901.90 & 15.90 & 13.431 & 0.026 & 12.951 & 0.026 & 12.502 & 0.024 & 12.294 & 0.025 \\
57902.90 & 16.90 & 13.475 & 0.026 & 12.983 & 0.026 & 12.508 & 0.025 & 12.286 & 0.025 \\
57904.90 & 18.90 & 13.531 & 0.028 & 12.976 & 0.027 & 12.507 & 0.025 & 12.277 & 0.026 \\
57905.90 & 19.90 & 13.559 & 0.007 & 12.959 & 0.006 & 12.491 & 0.005 & 12.262 & 0.005 \\
57906.90 & 20.90 & 13.623 & 0.006 & 12.962 & 0.006 & 12.490 & 0.004 & 12.263 & 0.005 \\
57907.90 & 21.90 & 13.628 & 0.008 & 12.979 & 0.007 & 12.494 & 0.005 & 12.257 & 0.005 \\
57909.90 & 23.90 & 13.765 & 0.008 & 12.989 & 0.007 & 12.491 & 0.005 & 12.254 & 0.006 \\
57912.90 & 26.90 & 13.919 & 0.010 & 13.011 & 0.007 & 12.522 & 0.005 & 12.286 & 0.006 \\
57913.90 & 27.90 & 13.905 & 0.010 & 13.049 & 0.007 & 12.538 & 0.005 & 12.258 & 0.005 \\
57915.00 & 29.00 & 13.964 & 0.008 & 13.066 & 0.006 & 12.547 & 0.005 & 12.286 & 0.005 \\
57915.90 & 29.90 & 14.006 & 0.008 & 13.071 & 0.006 & 12.557 & 0.005 & 12.269 & 0.005 \\
57916.90 & 30.90 & 14.054 & 0.011 & 13.087 & 0.008 & 12.561 & 0.006 & 12.272 & 0.006 \\
57917.90 & 31.90 & 14.070 & 0.009 & 13.105 & 0.007 & 12.566 & 0.005 & 12.279 & 0.005 \\
57918.94 & 32.94 & 14.135 & 0.063 & 13.109 & 0.035 & 12.569 & 0.036 & 12.275 & 0.035 \\
57919.99 & 33.99 & 14.116 & 0.008 & 13.123 & 0.006 & 12.590 & 0.004 & 12.274 & 0.005 \\
57922.00 & 36.00 & 14.197 & 0.011 & 13.106 & 0.010 & 12.603 & 0.007 & 12.300 & 0.008 \\
57924.00 & 38.00 & 14.252 & 0.008 & 13.187 & 0.006 & 12.610 & 0.005 & 12.297 & 0.005 \\
57925.95 & 39.95 & 14.301 & 0.010 & 13.176 & 0.006 & 12.620 & 0.005 & 12.277 & 0.005 \\
57927.90 & 41.90 & 14.327 & 0.009 & 13.194 & 0.008 & 12.620 & 0.005 & 12.286 & 0.006 \\
57929.02 & 43.02 & 14.375 & 0.010 & 13.199 & 0.007 & 12.622 & 0.005 & 12.281 & 0.005 \\
57931.00 & 45.00 & 14.425 & 0.054 & 13.209 & 0.039 & 12.611 & 0.022 & 12.258 & 0.024 \\
57937.86 & 51.86 & 14.486 & 0.023 & 13.230 & 0.022 & 12.628 & 0.015 & 12.261 & 0.023 \\
57940.90 & 54.90 & 14.558 & 0.038 & 13.222 & 0.028 & 12.622 & 0.020 & 12.239 & 0.020 \\
57942.91 & 56.91 & 14.563 & 0.044 & 13.229 & 0.039 & 12.617 & 0.023 & 12.245 & 0.021 \\
57945.90 & 59.90 & 14.605 & 0.070 & 13.231 & 0.030 & 12.627 & 0.027 & 12.238 & 0.034 \\
57947.90 & 61.90 & 14.595 & 0.059 & 13.240 & 0.027 & 12.627 & 0.033 & 12.240 & 0.031 \\
57951.92 & 65.92 & 14.632 & 0.030 & 13.262 & 0.020 & 12.638 & 0.010 & 12.235 & 0.010 \\
57952.90 & 66.90 & 14.627 & 0.010 & 13.270 & 0.007 & 12.637 & 0.005 & 12.235 & 0.005 \\
57956.90 & 70.90 & 14.717 & 0.046 & 13.297 & 0.038 & 12.626 & 0.031 & 12.235 & 0.024 \\
57957.94 & 71.94 & 14.723 & 0.063 & 13.300 & 0.041 & 12.655 & 0.034 & 12.231 & 0.028 \\
57959.87 & 73.87 & 14.738 & 0.055 & 13.299 & 0.035 & 12.643 & 0.033 & 12.238 & 0.028 \\
57961.96 & 75.96 & 14.770 & 0.023 & 13.292 & 0.020 & 12.626 & 0.023 & 12.226 & 0.018 \\
57962.86 & 76.86 & 14.810 & 0.037 & 13.297 & 0.033 & 12.636 & 0.029 & 12.233 & 0.019 \\
57964.86 & 78.86 & 14.797 & 0.008 & 13.325 & 0.005 & 12.657 & 0.003 & 12.252 & 0.004 \\
57965.94 & 79.94 & 14.818 & 0.008 & 13.333 & 0.005 & 12.675 & 0.003 & 12.245 & 0.004 \\
57966.92 & 80.92 & 14.821 & 0.008 & 13.332 & 0.005 & 12.669 & 0.003 & 12.257 & 0.004 \\
57967.87 & 81.87 & 14.858 & 0.062 & 13.376 & 0.032 & 12.660 & 0.034 & 12.257 & 0.019 \\
57968.88 & 82.88 & 14.888 & 0.025 & 13.356 & 0.026 & 12.658 & 0.027 & 12.235 & 0.039 \\
57970.89 & 84.89 & 14.905 & 0.064 & 13.382 & 0.033 & 12.693 & 0.025 & 12.268 & 0.022 \\
57972.88 & 86.88 & 14.961 & 0.038 & 13.417 & 0.021 & 12.692 & 0.014 & 12.282 & 0.020 \\
57973.86 & 87.86 & 14.977 & 0.046 & 13.410 & 0.025 & 12.692 & 0.022 & 12.278 & 0.020 \\
57974.86 & 88.86 & 14.989 & 0.041 & 13.422 & 0.030 & 12.716 & 0.018 & 12.291 & 0.025 \\
57976.00 & 90.00 & 15.011 & 0.046 & 13.422 & 0.016 & 12.727 & 0.024 & 12.294 & 0.014 \\
57980.02 & 94.02 & 15.117 & 0.041 & 13.523 & 0.032 & 12.766 & 0.028 & 12.349 & 0.030 \\
57982.84 & 96.84 & 15.192 & 0.031 & 13.555 & 0.028 & 12.778 & 0.029 & 12.367 & 0.021 \\
57983.85 & 97.85 & 15.229 & 0.032 & 13.583 & 0.019 & 12.813 & 0.029 & 12.383 & 0.023 \\
57986.86 & 100.86 & 15.340 & 0.034 & 13.648 & 0.019 & 12.874 & 0.019 & 12.437 & 0.016 \\
57987.92 & 101.92 & 15.337 & 0.012 & 13.694 & 0.007 & 12.909 & 0.004 & 12.466 & 0.004 \\
57988.84 & 102.84 & 15.418 & 0.032 & 13.718 & 0.024 & 12.923 & 0.023 & 12.484 & 0.021 \\
57993.81 & 107.81 & 15.647 & 0.042 & 13.925 & 0.024 & 13.088 & 0.020 & 12.622 & 0.025 \\
57994.88 & 108.88 & 15.768 & 0.059 & 13.980 & 0.044 & 13.136 & 0.031 & 12.673 & 0.033 \\
57995.88 & 109.88 & 15.818 & 0.058 & 14.051 & 0.043 & 13.178 & 0.022 & 12.700 & 0.018 \\
58000.83 & 114.83 & 16.202 & 0.068 & 14.384 & 0.030 & 13.470 & 0.016 & 12.963 & 0.048 \\
58006.81 & 120.81 & 16.880 & 0.056 & 15.084 & 0.032 & 14.051 & 0.035 & 13.505 & 0.025 \\
58011.80 & 125.80 & 17.385 & 0.048 & 15.530 & 0.027 & 14.434 & 0.019 & 13.858 & 0.020 \\
58022.92 & 136.92 & 17.779 & 0.048 & 15.679 & 0.031 & 14.561 & 0.019 & 14.024 & 0.011 \\
58023.92 & 137.92 & 17.540 & 0.037 & 15.698 & 0.029 & 14.603 & 0.030 & 14.048 & 0.019 \\
58024.82 & 138.82 & 17.617 & 0.037 & 15.725 & 0.026 & 14.617 & 0.022 & 14.049 & 0.019 \\
58025.80 & 139.80 & 17.564 & 0.041 & 15.754 & 0.021 & 14.620 & 0.024 & 14.053 & 0.021 \\
58026.88 & 140.88 & 17.672 & 0.015 & 15.761 & 0.011 & 14.655 & 0.015 & 14.048 & 0.012 \\
58027.83 & 141.83 & 17.648 & 0.022 & 15.816 & 0.009 & 14.677 & 0.019 & 14.076 & 0.013 \\
58032.75 & 146.75 & 17.615 & 0.025 & 15.809 & 0.024 & 14.695 & 0.008 & 14.099 & 0.006 \\
58035.75 & 149.75 & 17.661 & 0.019 & 15.886 & 0.009 & 14.736 & 0.011 & 14.143 & 0.013 \\
58040.81 & 154.81 & 17.684 & 0.044 & 15.899 & 0.025 & 14.818 & 0.026 & 14.211 & 0.020 \\
58041.84 & 155.84 & 17.679 & 0.012 & 15.916 & 0.013 & 14.781 & 0.011 & 14.181 & 0.011 \\
58044.79 & 158.79 & 17.678 & 0.056 & 15.906 & 0.027 & 14.828 & 0.029 & 14.217 & 0.022 \\
58046.72 & 160.72 & 17.798 & 0.035 & 15.944 & 0.025 & 14.815 & 0.025 & 14.258 & 0.015 \\
58050.70 & 164.70 & 17.771 & 0.047 & 15.973 & 0.024 & 14.884 & 0.017 & 14.276 & 0.021 \\
58055.95 & 169.95 & 17.768 & 0.038 & 16.069 & 0.026 & 14.871 & 0.017 & 14.314 & 0.023 \\
58063.89 & 177.89 & 17.901 & 0.018 & 16.131 & 0.015 & 14.964 & 0.010 & 14.364 & 0.008 \\
58064.81 & 178.81 & 17.841 & 0.031 & 16.139 & 0.023 & 14.948 & 0.006 & 14.368 & 0.006 \\
58075.80 & 189.80 & 17.874 & 0.072 & 16.221 & 0.031 & 15.110 & 0.031 & 14.729 & 0.064 \\
58094.79 & 208.79 & 18.083 & 0.054 & 16.407 & 0.030 & 15.235 & 0.025 & 14.674 & 0.028 \\
58182.12 & 296.12 & 18.659 & 0.077 & 17.330 & 0.046 & 16.191 & 0.031 & 15.620 & 0.027 \\
\enddata
\end{deluxetable}

\newpage
\startlongtable
\begin{deluxetable*}{cc|cccccc}
\tablecaption{LCO UBV$g'r'i'$ photometry of SN~2017eaw \label{tab:lcophot}}
\tabletypesize{\small}
\tablehead{
\colhead{Date} & \colhead{Epoch} & \colhead{U} & \colhead{B} & \colhead{V} & \colhead{$g'$} & \colhead{$r'$} & \colhead{$i'$} \\
\colhead{(MJD)} & \colhead{(days)} & \colhead{(mag)} & \colhead{(mag)}  & \colhead{(mag)}  &\colhead{(mag)}  & \colhead{(mag)}  & \colhead{(mag)}
}
\startdata
57888.85 & 2.85 & 12.523(.014) & 13.222(.012) & 12.992(.022) & 12.760(.018) & 12.903(.013) & 12.880(.023) \\
57889.91 & 3.91 & -- & 13.198(.011) & 12.888(.015) & 12.698(.010) & 12.774(.012) & 12.752(.015) \\
57891.93 & 5.93 & 12.431(.012) & 13.148(.019) & 12.807(.016) & 12.653(.018) & 12.653(.019) & 12.573(.015) \\
57893.82 & 7.82 & 12.394(.025) & 13.128(.007) & 12.774(.015) & 12.656(.022) & 12.591(.024) & 12.500(.012) \\
57895.91 & 9.91 & -- & 13.146(.009) & 12.855(.022) & 12.699(.012) & 12.612(.016) & 12.556(.035) \\
57896.89 & 10.89 & 12.467(.018) & 13.173(.014) & 12.868(.013) & 12.703(.020) & 12.580(.010) & 12.542(.010) \\
57897.81 & 11.81 & -- & 13.229(.012) & 12.869(.022) & 12.731(.011) & 12.593(.016) & 12.557(.016) \\
57898.83 & 12.83 & 12.655(.017) & 13.257(.017) & 12.888(.011) & 12.769(.015) & 12.613(.012) & 12.583(.018) \\
57902.81 & 16.81 & 12.886(.012) & 13.374(.038) & 12.933(.029) & 12.862(.015) & 12.634(.011) & 12.620(.014) \\
57906.82 & 20.82 & -- & 13.476(.022) & 12.945(.013) & 12.868(.019) & 12.582(.023) & 12.619(.046) \\
57907.86 & 21.86 & 13.344(.013) & 13.536(.019) & 12.868(.013) & 12.911(.032) & 12.583(.010) & 12.555(.023) \\
57908.87 & 22.87 & 13.491(.035) & 13.581(.020) & 12.924(.011) & 12.939(.027) & 12.597(.011) & 12.535(.018) \\
57909.85 & 23.85 & 13.601(.034) & 13.638(.011) & 12.920(.010) & 12.977(.010) & 12.608(.025) & 12.596(.011) \\
57910.88 & 24.88 & -- & -- & 12.995(.010) & 12.984(.023) & 12.643(.015) & 12.600(.019) \\
57916.89 & 30.89 & -- & -- & -- & 13.178(.011) & 12.669(.018) & 12.593(.010) \\
57917.84 & 31.84 & -- & 13.981(.014) & 13.037(.014) & 13.200(.011) & 12.673(.018) & 12.586(.012) \\
57928.79 & 42.79 & 14.879(.030) & 14.281(.017) & 13.169(.015) & 13.418(.010) & 12.782(.010) & 12.685(.010) \\
57934.82 & 48.82 & 15.146(.030) & 14.373(.026) & 13.164(.014) & 13.477(.015) & 12.769(.032) & 12.614(.029) \\
57940.84 & 54.84 15.426(.019) & 14.447(.012) & 13.159(.012) & 13.448(.010) & 12.774(.014) & 12.611(.016) \\
57944.80 & 58.80 & 15.493(.018) & 14.513(.015) & 13.193(.011) & 13.524(.012) & 12.791(.008) & -- \\
57945.84 & 59.84 & 15.522(.023) & 14.543(.016) & 13.185(.014) & 13.552(.013) & 12.805(.018) & 12.606(.025) \\
57961.83 & 75.83 & 16.014(.051) & 14.699(.018) & 13.221(.011) & 13.683(.012) & 12.821(.011) & 12.640(.015) \\
57965.93 & 79.93 & -- & 14.834(.054) & -- & -- & -- & -- \\
57971.87 & 85.87 & 16.425(.047) & 14.902(.016) & 13.340(.012) & 13.789(.013) & 12.850(.018) & 12.625(.026) \\
57972.83 & 86.83 & 16.392(.037) & 14.902(.014) & 13.345(.014) & 13.831(.020) & 12.875(.013) & 12.668(.005) \\
57976.80 & 90.80 & 16.641(.016) & 15.015(.018) & 13.408(.019) & 13.901(.036) & 12.979(.023) & 12.716(.013) \\
57995.81 & 109.81 & 17.756(.028) & 15.747(.027) & 13.907(.038) & 14.593(.012) & 13.315(.019) & 13.055(.013) \\
58001.74 & 115.74 & -- & 16.267(.018) & 14.430(.012) & 15.095(.028) & 13.716(.012) & 13.410(.026) \\
58002.84 & 116.84 & -- & 16.375(.039) & 14.489(.016) & 15.158(.017) & 13.777(.018) & 13.513(.017) \\
58004.81 & 118.81 & -- & 16.629(.017) & 14.768(.022) & 15.403(.026) & 13.974(.011) & 13.678(.016) \\
58005.82 & 119.82 & -- & 16.719(.043) & 14.868(.013) & 15.589(.015) & 14.099(.013) & 13.834(.012) \\
58008.82 & 122.82 & -- & 17.123(.022) & 15.289(.021) & 15.935(.036) & 14.405(.018) & 14.131(.048) \\
58009.80 & 123.80 & -- & 17.247(.028) & 15.379(.018) & -- & -- & -- \\
58010.83 & 124.83 & -- & 17.334(.016) & 15.395(.011) & 16.136(.012) & 14.597(.015) & 14.287(.028) \\
58018.78 & 132.78 & -- & 17.535(.026) & 15.651(.015) & 16.361(.015) & 14.793(.027) & 14.481(.013) \\
58019.77 & 133.77 & -- & 17.502(.014) & 15.669(.024) & 16.383(.017) & 14.786(.014) & 14.457(.022) \\
58028.75 & 142.75 & -- & 17.611(.023) & 15.756(.022) & 16.444(.015) & 14.838(.010) & 14.557(.006) \\
58034.76 & 148.76 & -- & 17.710(.025) & 15.823(.037) & 16.450(.030) & 14.900(.018) & 14.645(.025) \\
58035.73 & 149.73 & -- & 17.630(.028) & 15.781(.022) & 16.495(.021) & 14.884(.012) & 14.668(.021) \\
58039.75 & 153.75 & -- & 17.687(.017) & 15.831(.012) & 16.583(.010) & 14.960(.011) & 14.722(.041) \\
58043.73 & 157.73 & -- & 17.703(.013) & 15.868(.014) & 16.588(.013) & 14.992(.018) & 14.735(.012) \\
58047.60 & 161.60 & -- & 17.710(.019) & 15.929(.013) & -- & -- & -- \\
58051.62 & 165.62 & -- & 17.747(.034) & 15.970(.020) & 16.674(.010) & 15.038(.020) & 14.809(.013) \\
58055.59 & 169.59 & -- & 17.801(.029) & 16.028(.020) & 16.679(.010) & 15.096(.028) & 14.833(.014) \\
58059.66 & 173.66 & -- & 17.824(.023) & 16.061(.010) & 16.681(.011) & 15.088(.020) & 14.882(.015) \\
58060.65 & 174.65 & -- & 17.760(.029) & 16.032(.010) & 16.679(.018) & 15.114(.012) & 14.900(.012) \\
58071.62 & 185.62 & -- & 17.843(.112) & 16.045(.036) & 16.809(.016) & 15.132(.028) & 14.977(.013) \\
58082.57 & 196.57 & -- & 17.955(.023) & 16.298(.015) & 16.880(.016) & 15.302(.018) & 15.132(.020) \\
58099.57 & 213.57 & -- & 17.997(.022) & 16.449(.029) & 16.984(.015) & 15.460(.011) & 15.301(.015) \\
58231.02 & 345.02 & -- & 19.012(.033) & 17.819(.014) & 18.064(.013) & 16.933(.025) & 16.623(.017) \\
58246.05 & 360.05 & -- & -- & 17.984(.026) & 18.167(.021) & 17.136(.019) & 16.820(.046) \\
58246.83 & 360.83 & -- & 19.065(.040) & 17.958(.060) & 18.153(.029) & 17.131(.062) & 16.793(.018) \\
58255.02 & 369.02 & -- & 19.134(.014) & 18.030(.010) & 18.249(.026) & 17.245(.037) & 16.906(.031) \\
58258.81 & 372.81 & -- & 19.180(.023) & 18.086(.011) & 18.326(.022) & 17.374(.028) & 16.986(.010) \\
58267.92 & 381.92 & -- & -- & 18.253(.027) & 18.487(.038) & 17.466(.034) & 17.142(.051) \\
58284.78 & 398.78 & -- & 19.453(.018) & 18.354(.014) & 18.632(.015) & 17.666(.044) & 17.345(.041) \\
58292.77 & 406.77 & -- & 19.521(.032) & 18.506(.034) & -- & 17.769(.013) & 17.390(.016) \\
58318.75 & 432.75 & -- & 19.783(.021) & 18.784(.011) & 18.909(.048) & 18.175(.017) & 17.768(.022) \\
58335.90 & 449.90 & -- & 19.976(.059) & 19.045(.022) & 19.204(.033) & 18.381(.019) & -- \\
58351.74 & 465.74 & -- & 20.092(.086) & 19.230(.013) & 19.397(.047) & 18.631(.023) & 18.208(.012) \\
58374.81 & 488.81 & -- & 20.371(.014) & 19.434(.025) & 19.645(.043) & 18.876(.034) & 18.673(.026) \\
58385.76 & 499.76 & -- & -- & 19.659(.028) & 19.804(.032) & 19.087(.026) & 18.787(.012) \\
58392.72 & 506.72 & -- & 20.574(.042) & 19.757(.053) & 19.875(.015) & 19.233(.014) & 18.892(.041) \\
58394.63 & 508.63 & -- & 20.631(.023) & 19.699(.086) & 19.860(.047) & 19.316(.019) & 19.039(.021) \\
58399.68 & 513.68 & -- & -- & 19.771(.011) & 19.907(.048) & -- & -- \\
58405.74 & 519.74 & -- & -- & 19.959(.116) & 20.069(.031) & 19.386(.022) & 19.215(.075) \\
58417.67 & 531.67 & -- & -- & -- & -- & 19.590(.071) & 19.341(.016) \\
58426.55 & 540.55 & -- & -- & 20.159(.025) & 20.405(.021) & 19.711(.036) & 19.594(.012) \\
\enddata
\end{deluxetable*}



\begin{table*}
\begin{center}
\begin{footnotesize}
\caption{{\it Swift} photometry of SN~2017eaw}
\label{tab:swphot}
\begin{tabular}{cc|cccccccccccc}
\hline
\hline
Date & Epoch & UVW2 & $\sigma$UVW2  & UVM2 & $\sigma$UVM2  & UVW1 & $\sigma$UVW1  & U & $\sigma$U & B & $\sigma$B & V & $\sigma$V \\
(MJD) & (days) & (mag) & (mag) & (mag) & (mag) & (mag) & (mag) & (mag) & (mag) & (mag) & (mag) & (mag) & (mag) \\
\hline
57887.59 & 1.59 & 12.378 & 0.008 & 12.556 & 0.007 & 12.266 & 0.009 & 12.239 & 0.012 & 13.287 & 0.012 & 13.154 & 0.020 \\
57887.92 & 1.92 & 12.453 & 0.052 & 12.579 & 0.019 & -- & -- & -- & -- & -- & -- & -- & -- \\
57888.66 & 2.66 & 12.666 & 0.040 & 12.645 & 0.011 & -- & -- & -- & -- & -- & -- & -- & -- \\
57888.80 & 2.80 & 12.662 & 0.013 & -- & -- & 12.322 & 0.007 & -- & -- & -- & -- & -- & -- \\
57889.86 & 3.86 & -- & -- & 12.833 & 0.014 & -- & -- & -- & -- & -- & -- & -- & -- \\
57890.12 & 4.12 & 13.037 & 0.045 & 12.877 & 0.008 & -- & -- & -- & -- & -- & -- & -- & -- \\
57891.85 & 5.85 & 13.423 & 0.074 & 13.199 & 0.010 & -- & -- & -- & -- & -- & -- & -- & -- \\
57897.55 & 11.55 & -- & -- & 14.529 & 0.016 & -- & -- & -- & -- & -- & -- & -- & -- \\
57899.42 & 13.42 & -- & -- & 14.986 & 0.026 & -- & -- & -- & -- & -- & -- & -- & -- \\
57902.08 & 16.08 & 15.368 & 0.036 & 15.807 & 0.059 & 14.153 & 0.028 & 12.884 & 0.019 & 13.363 & 0.020 & 12.976 & 0.022 \\
57902.27 & 16.27 & 15.304 & 0.164 & 15.900 & 0.042 & -- & -- & -- & -- & -- & -- & -- & -- \\
57904.06 & 18.06 & 15.958 & 0.033 & 16.565 & 0.063 & 14.584 & 0.023 & 13.162 & 0.014 & 13.432 & 0.013 & 12.970 & 0.015 \\
57906.59 & 20.59 & 16.445 & 0.044 & 17.277 & 0.085 & 15.153 & 0.031 & 13.533 & 0.015 & 13.513 & 0.014 & 12.947 & 0.015 \\
57908.45 & 22.45 & 16.647 & 0.047 & 17.748 & 0.121 & 15.491 & 0.036 & 13.835 & 0.016 & 13.593 & 0.013 & 12.989 & 0.015 \\
57910.18 & 24.18 & 16.859 & 0.052 & 17.978 & 0.125 & 15.591 & 0.037 & 14.075 & 0.017 & 13.690 & 0.013 & 13.029 & 0.015 \\
57912.69 & 26.69 & 17.033 & 0.058 & 18.454 & 0.194 & 15.936 & 0.046 & 14.357 & 0.020 & 13.799 & 0.014 & 13.049 & 0.015 \\
57914.29 & 28.29 & 17.223 & 0.066 & 18.890 & 0.273 & 16.051 & 0.047 & 14.527 & 0.021 & 13.889 & 0.014 & 13.050 & 0.015 \\
57916.89 & 30.89 & -- & -- & -- & -- & 16.194 & 0.051 & 14.765 & 0.024 & 13.936 & 0.016 & -- & -- \\
57918.82 & 32.82 & 17.456 & 0.091 & 18.557 & 0.211 & 16.406 & 0.071 & 14.965 & 0.032 & 14.005 & 0.017 & 13.092 & 0.019 \\
57920.80 & 34.80 & 17.573 & 0.080 & 18.753 & 0.236 & 16.457 & 0.058 & 15.105 & 0.026 & 14.099 & 0.014 & 13.164 & 0.015 \\
57949.17 & 63.17 & 18.146 & 0.090 & 20.463 & 0.552 & 17.148 & 0.063 & 16.266 & 0.044 & 14.574 & 0.015 & 13.272 & 0.014 \\
57962.59 & 76.59 & 18.357 & 0.102 & 21.193 & 0.892 & 17.288 & 0.067 & 16.741 & 0.060 & 14.745 & 0.016 & 13.337 & 0.014 \\
57977.93 & 91.93 & 19.146 & 0.365 & -- & -- & 17.843 & 0.121 & 17.252 & 0.102 & 15.010 & 0.024 & 13.599 & 0.032 \\
57981.18 & 95.18 & 18.947 & 0.242 & 19.932 & 0.428 & 17.706 & 0.144 & 17.455 & 0.148 & 15.080 & 0.027 & 13.572 & 0.024 \\
57990.30 & 104.30 & 18.939 & 0.155 & 20.653 & 0.530 & 18.195 & 0.123 & 17.850 & 0.124 & 15.463 & 0.021 & 13.864 & 0.017 \\
58004.65 & 118.65 & 20.535 & 0.596 & -- & -- & 18.589 & 0.170 & 19.090 & 0.343 & 16.661 & 0.043 & 14.942 & 0.038 \\
58009.43 & 123.43 & 20.249 & 0.728 & -- & -- & 18.974 & 0.416 & 18.912 & 0.466 & 17.254 & 0.101 & 15.497 & 0.066 \\

\hline
\end{tabular}
\end{footnotesize}
\end{center}
\end{table*}

\begin{table}
\begin{center}
\begin{footnotesize}
\caption{Log of spectroscopic observations}
\label{tab:spec}
\begin{tabular}{ccccc}
\hline
\hline
UT Date & Phase & Instrument & Range & R \\
~ & (days) & ~ & (\AA) & ($\lambda$/$\Delta\lambda$) \\
\hline
2017-05-15 & +3 & {\it Swift} UVOT/UGRISM & 2000$-$5000 & 150 \\
2017-05-16 & +3 & HET LRS2 & 3700$-$10\,500 & 1100/1800/1900 \\
2017-05-16 & +3 & LCO FLOYDS & 3250$-$10\,000 & 400-700 \\
2017-05-18 & +5 & LCO FLOYDS & 3250$-$10\,000 & 400-700 \\
2017-05-19 & +6 & HET LRS2 & 3700$-$10\,500 & 1100/1800/1900 \\
2017-05-19 & +6 & LCO FLOYDS & 3250$-$10\,000 & 400-700  \\
2017-05-19 & +6 & IRTF SpeX & 8000$-$24\,000 & 1200 \\
2017-05-21 & +8 & LCO FLOYDS & 3250$-$10\,000 & 400-700  \\
2017-05-22 & +10 & {\it Swift} UVOT/UGRISM & 2000$-$5000 & 150 \\
2017-05-23 & +10 & HET LRS2 & 3700$-$10\,500 & 1100/1800/1900 \\
2017-05-23 & +10 & LCO FLOYDS & 3250$-$10\,000 & 400-700  \\
2017-05-24 & +11 & IRTF SpeX & 8000$-$24\,000 & 1200 \\
2017-05-25 & +12 & LCO FLOYDS & 3250$-$10\,000 & 400-700  \\
2017-05-27 & +14 & LCO FLOYDS & 3250$-$10\,000 & 400-700  \\
2017-05-31 & +18 & LCO FLOYDS & 3250$-$10\,000 & 400-700  \\
2017-06-02 & +20 & LCO FLOYDS & 3250$-$10\,000 & 400-700  \\
2017-06-03 & +21 & HET LRS2 & 3700$-$10\,500 & 1100/1800/1900 \\
2017-06-03 & +21 & LCO FLOYDS & 3250$-$10\,000 & 400-700  \\
2017-06-05 & +23 & LCO FLOYDS & 3250$-$10\,000 & 400-700  \\
2017-06-08 & +26 & IRTF SpeX & 8000$-$24\,000 & 1200 \\
2017-06-14 & +32 & HET LRS2 & 3700$-$10\,500 & 1100/1800/1900 \\
2017-06-21 & +39 & IRTF SpeX & 8000$-$24\,000 & 1200 \\
2017-06-24 & +42 & LCO FLOYDS & 3250$-$10\,000 & 400-700  \\
2017-07-01 & +49 & LCO FLOYDS & 3250$-$10\,000 & 400-700  \\
2017-07-06 & +54 & LCO FLOYDS & 3250$-$10\,000 & 400-700  \\
2017-07-08 & +56 & IRTF SpeX & 8000$-$24\,000 & 1200 \\
2017-07-08 & +56 & Lick Shane/Kast & 3250$-$10\,000 & 600  \\
2017-07-13 & +61 & LCO FLOYDS & 3250$-$10\,000 & 400-700  \\
2017-07-24 & +72 & LCO FLOYDS & 3250$-$10\,000 & 400-700  \\
2017-07-27 & +75 & LCO FLOYDS & 3250$-$10\,000 & 400-700  \\
2017-07-30 & +78 & HET LRS2 & 3700$-$10\,500 & 1100/1800/1900 \\
2017-08-04 & +83 & LCO FLOYDS & 3250$-$10\,000 & 400-700  \\
2017-08-10 & +89 & LCO FLOYDS & 3250$-$10\,000 & 400-700  \\
2017-08-16 & +95 & LCO FLOYDS & 3250$-$10\,000 & 400-700  \\
2017-08-24 & +103 & LCO FLOYDS & 3250$-$10\,000 & 400-700  \\
2017-08-28 & +107 & HET LRS2 & 3700$-$10\,500 & 1100/1800/1900 \\
2017-09-15 & +125 & LCO FLOYDS & 3250$-$10\,000 & 400-700  \\
2017-09-15 & +125 & Keck LRIS & 3115$-$10\,235 & 300-5000 \\
2017-09-21 & +131 & HET LRS2 & 3700$-$10\,500 & 1100/1800/1900 \\
2017-09-22 & +132 & LCO FLOYDS & 3250$-$10\,000 & 400-700  \\
2017-09-26 & +136 & LCO FLOYDS & 3250$-$10\,000 & 400-700  \\
2017-10-22 & +162 & HET LRS2 & 3700$-$10\,500 & 1100/1800/1900 \\
2017-12-19 & +220 & Keck LRIS & 3115$-$10\,235 & 300-5000 \\
2018-07-22 & +435 & Gemini-North GMOS-N & 3800$-$10\,000 & 500 \\
2018-09-16 & +490 & Gemini-North GMOS-N & 3800$-$10\,000 & 500 \\
\hline
\end{tabular}
\end{footnotesize}
\end{center}
\end{table}

\end{document}